%% file: ebstereo2.tex
\documentclass[a4paper,twoside,usenatbib]{mn2e}

\usepackage{natbib}
\usepackage{epsfig}
\usepackage{calc}
\usepackage{amssymb}
\usepackage{amstext}
\usepackage{amsmath}
\usepackage{multicol}
\usepackage{pslatex}
\usepackage{fancyhdr}
\usepackage{longtable}
\usepackage{pdflscape}
\usepackage{graphics}
\usepackage[left=2.6cm,top=3.3cm,right=2.6cm,bottom=4.2cm]{geometry}

\input{title}
\input{authors}

\begin{document}

\pagenumbering{arabic}

\onecolumn \maketitle \normalsize \vfill

\input{abstract}  

\input{keywords}

\twocolumn
\sloppy

\input{introduction}

\input{specs}
\input{method}
\input{results}
\input{prospects}
\input{conclusions}
\input{acks}
\input{appendix}

\bibliographystyle{mn2e}
\bibliography{ebstereorefs}

\end{document}

%% file: title.tex
\title{
{STEREO observations of stars and the search for exoplanets}
}

%% file: authors.tex
\author[K. T. Wraight, Glenn J. White, D. Bewsher and A. J. Norton]{{\textmd {\textbf }}\\
    \textup{K.T. Wraight}$^1$, \textup{Glenn J. White}$^{1,3}$, \textup{D. Bewsher}$^2$ \textup{and} \textup{A.J. Norton}$^1$\\    
    $^1$\textit{Department of Physics and Astronomy, The Open University, Milton Keynes, MK7 6AA, UK}\\
    $^2$\textit{Jeremiah Horrocks Institute, University of Central Lancashire, Preston, Lancashire, PR1 2HE, UK}\\
    $^3$\textit{Space Science and Technology Department, STFC Rutherford Appleton Laboratory, Chilton, Didcot, Oxfordshire, OX11 0QX, UK}\\
}

%% file: abstract.tex
\fontsize{9}{11}\selectfont \noindent The feasibility of using data from the NASA \textit{STEREO} mission for variable star and asteroseismology studies has been examined.  A data analysis pipeline has been developed that is able to apply selected algorithms to the entire database of nearly a million stars to search for signs of variability.  An analysis limited to stars of magnitude 10.5 has been carried out, which has resulted in the extraction of 263 eclipsing binaries (EBs), of which \textbf{122} are not recorded as such in the \textsc{Simbad} online database.  The characteristics of the \textit{STEREO} observations are shown to be extremely well-suited to variable star studies with the ability to provide continuous phase coverage for extended periods as well as repeated visits that allow both short and long term variability to be observed.  This will greatly inform studies of particular stars, such as the pre-cataclysmic variable V471 Tau, as well as entire classes of stars, including many forms of rotational variability.  \textbf{The high-precision photometry has also revealed a potentially substellar companion to a bright (\textit{R} = 7.5~mag) nearby star (HD~213597), detected with 5~$\sigma$ significance.  This would provide a significant contribution to exoplanet research if follow-up observations ascertain the mass to be within the planetary domain.  Some particularly unusual EBs from the recovered sample are discussed, including a possible reclassification of a well-known star as an EB rather than a rotational variable (HR~7355) and several particularly eccentric systems, including very long-period EBs.}

%% file: keywords.tex
\fontsize{9}{11}\selectfont Keywords: methods: data analysis -- binaries: eclipsing -- space vehicles: \textit{STEREO} -- stars: individual: HD~213597 -- stars: individual: NSV~7359 -- stars: individual: V~471 Tau 

%% file: introduction.tex
\section{\uppercase{Introduction}}
\label{sec:introduction}

\noindent The \textit{STEREO} mission consists of two satellites in heliocentric orbit, each observing the solar environment from the Sun's atmosphere out to the Earth's orbit and together producing 3D images of this region \citep{2008SSRv..136....5K}.  The HI-1 imager on the Ahead spacecraft (HI-1A) has been used to conduct a survey for variability of stars down to magnitude 12 in its field of view (within about 20 degrees of the Ecliptic Plane, see Figure \ref{fig44}).  The lightcurves produced have been analysed to look for exoplanet transits and eclipsing binaries.  The value of eclipsing variables in studies of stellar properties and evolution is well-known \citep{de2010binaries} and the trends from this sample, as well as some particular stars of interest, will prove informative.  \textbf{The detection of many new eclipsing binaries around comparatively bright stars, previously undetected in many cases due to shallow eclipses, allows for the detailed follow-up required to improve upon existing models of stellar evolution and behaviour.}  Also of note is a comparison of the effectiveness of different algorithms in extracting the signatures of eclipsing binaries of different types.

The rest of this paper is divided into sections giving an overview of the \textit{STEREO}/HI-1A characteristics, describing the data analysis pipeline, the results obtained, prospects for further study, the conclusions that can be drawn and the table showing all 263 EBs is given at the end.

Further research is being undertaken to identify and characterise other kinds of variability, which will be the subject of future papers.  Some HI-1B data has been extracted (for a region between 50 and 60 degrees right ascension and 10 and 20 degrees declination, which includes a variety of interesting objects, e.g. V~471 Tau - see Section \ref{sec:results}) but this has not yet been analysed on a large scale.  The HI-1B imager is on the side of the spacecraft facing the direction of travel and the lightcurves show artefacts resulting from micrometeorite impacts, complicating the task of signal extraction.

%% file: specs.tex
\section{\uppercase{\textbf{\textit{STEREO}/HI-1A} instrumentation and data preparation}}
\label{sec:specs}

\noindent The Heliospheric Imager on the \textit{STEREO}-Ahead spacecraft (HI-1A) provided the data for the large-scale analysis.  A summary of the data analysis pipeline is given in the Appendix.  This instrument is described in detail in \citet{eyles2009heliospheric} but for convenience a summary is given here.  The HI-1A has a field of view of 20 degrees \textbf{by 20 degrees} and is centred on a point 14 degrees away from the centre of the Sun, where the solar F-corona is the dominant large-scale structure.  The aperture of the HI-1A lens system is 16~mm, with a focal length of 78~mm.  The CCD has a pixel size such that the plate scale becomes 35 arcsec on a 2048x2048 chip, with images binned 2x2 on-board, resulting in 1024x1024 images being returned with an image bin size of 70 arcsec.  A filter limits the spectral bandpass to 630-730~nm with some contribution around 400~nm and 950~nm.  The characteristics of the CCD on the HI-1B are similar, with only slight differences in the spectral bandpass \citep{eyles2009heliospheric}.  Each image is the result of 30 summed exposures, each of 40 seconds, with one image produced every 40 minutes.  This prevents saturation by cosmic rays, although planetary and cometary incursions do cause saturation, and also allows for very faint structures in Coronal Mass Ejections (CMEs) and the solar environment to be detected.  Stars down to 12th magnitude are detectable and about 650000 stars have been recorded with listed magnitudes brighter than 11.5, with almost 75000 of these being brighter than magnitude 9.5 (Figure \ref{fig44}).  Only the very brightest stars, those brighter than about 3rd magnitude, saturate the detector but useful lightcurves can nevertheless still be extracted.  Since these stars are mostly within 20 degrees of the ecliptic plane, STEREO/HI-1A and HI-1B are therefore searching a region of the sky not well observed by dedicated exoplanet searches, such as \textit{CoRoT} \citep{CorotAASpecEd1} \textbf{(\textit{V} magnitudes from 11 to 16)}, \textit{Kepler} \citep{koch2010kepler} \textbf{(\textit{R} magnitudes from 7 to 17)} and SuperWASP \citep{pollacco2006wasp} \textbf{(\textit{V} magnitudes from 9 to 13)}, and it is also able to view much brighter stars than are normally observed.

Before the data from \textit{STEREO}/HI-1A and HI-1B can be analysed, it must be passed through a pipeline (shown in the Appendix, Figure \ref{fig102}) in order to
\begin{itemize}
\item Make a shutterless correction \citep{eyles2009heliospheric}
\item Apply a large-scale flat field \citep{bewsher2010determination}
\item Apply the pointing calibration to determine the positions of stars \citep{brown2009calibrating}
\item Flag out saturated pixels/columns and missing data blocks \citep{eyles2009heliospheric}
\item Carry out a daily background subtraction
\end{itemize}

\subsection{Shutterless correction}

\noindent The \textit{STEREO}/HI-1A and HI-1B cameras have no shutter and remain exposed during the readout process and also during the clear process prior to each exposure.  Consequently, the top of the image is exposed for longer than the bottom of the image, resulting in image smearing during readout.  The effect of this can be expressed as a matrix relationship and a correction for the effects of the shutterless operation can be made by a matrix multiplication.  If saturation occurs anywhere in a column, then the signal in the affected pixels cannot be corrected in this way and the entire column is flagged as bad data.  Further details can be found in \citet{eyles2009heliospheric}.

\subsection{Flat fields}

\noindent The sensitivity of the \textit{STEREO}/HI-1A and HI-1B cameras varies spatially in three distinct ways.  Firstly, the optical properties mean that the solid angle corresponding to each pixel may vary across the field of view, leading to large variations in scale across the CCD.  Secondly, the sensitivity of the CCD varies for individual pixels, independently of the sensitivity of adjacent pixels.  Thirdly, debris and manufacturing variations and defects on the CCD generate variations in sensitivity over the scale of tens to hundreds of pixels.

It is unknown to what extent the sensitivity of the CCD will vary over time but it is anticipated that the largely stationary F-corona in the field of view may be enough to cause some degradation.  The flat field calibration is updated on a regular basis, however, which should prevent this becoming a problem.

The flat field calibrations are performed on \textit{STEREO}/HI-1A and HI-1B images after the removal of the CCD DC bias \citep{eyles2009heliospheric} and the application of the shutterless correction.

Images with significant optical ghosts due to saturated objects in the data are not used for flat fielding.  The methodology is detailed in full in \citet{bewsher2010determination}.  Observations of stars near a region of the frame close to the edge of the solar disk are discarded due to additional noise of solar origin.

\subsection{Pointing calibration}

\noindent Calibrated instrument pointing solutions have been derived by comparing the locations of stars identified in the \textit{STEREO}/HI-1A and HI-1B images with known star positions given in the NOMAD astrometric dataset \citep{zacharias2004naval}.  In the raw (L0.5) data, nominal pointing information is provided, however once the data has been processed to L1 (by running \textit{secchi\underline{ }prep}) the calibrated pointing information is provided.  Full details on the calibration and pointing procedure can be found in \citet{brown2009calibrating}.

For the \textit{STEREO}/HI-1A and HI-1B cameras, the average root mean square distance between stars as seen in the image and the location given in the catalogue is about 0.1 pixels, i.e. 7 arcsec.  Thus any pointing offsets have a negligible effect on the efficacy of the stellar lightcurves.

\subsection{Photometric calibration}

\noindent The methodology used to determine the photometric calibration is described in full in \citet{bewsher2010determination}, which includes more up-to-date flat fields and calibration constants than were available for the analysis presented in this paper.  Aperture photometry of well-isolated stars down to 12th magnitude is used, that have no other stars with a difference of $\leq$1 magnitude within 1000 arcsec.

Whilst future large-scale analyses will use the most up-to-date calibration of \citet{bewsher2010determination}, the photometric calibration of the \textit{STEREO}/HI-1A used in the analysis presented herein is described by the following formula

\begin{equation}
\texttt{\textit{STEREO} magnitude} = \textit{d}~ log(\frac{\texttt{counts}}{\textit{c}})
\label{eq:1}
\end{equation}

\noindent where $\textit{d} = -2.49501 $ and $\textit{c} = 18337.7$ for \textit{STEREO}/HI-1A for a 3.5 pixel radius aperture.  The number of counts is expressed as a flux intensity in units in data numbers, DN$s^{-1}$ as provided by \textit{secchi\underline{ }prep}.  For analysing HI-1B, the following formula is used (explained more fully in \citet{bewsher2010determination})

\begin{equation}
\texttt{\textit{STEREO} magnitude} = d~ log(\frac{a * \texttt{counts}}{b * c})
\label{eq:2}
\end{equation}

\noindent where $\textit{d} = -2.5 $, $\textit{a} = 4 $ is a correction factor , $\textit{b} = 0.98 $ and $\textit{c} = 97026$ are the calibration constants for \textit{STEREO}/HI-1B for a 3.5 pixel radius aperture.

%% file: method.tex
\section{Data analysis pipeline and methods}
\label{sec:method}

\begin{figure}
\centering
\resizebox{7cm}{!}{\includegraphics{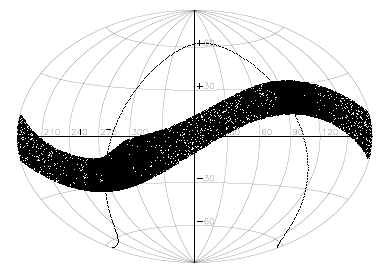}}
\caption{The locations of stars in the \textbf{\textit{STEREO}/HI-1A} with measured mean magnitude of 9th magnitude and brighter.}
\label{fig44}
\end{figure}

\begin{figure}
\centering
\resizebox{7cm}{!}{\includegraphics{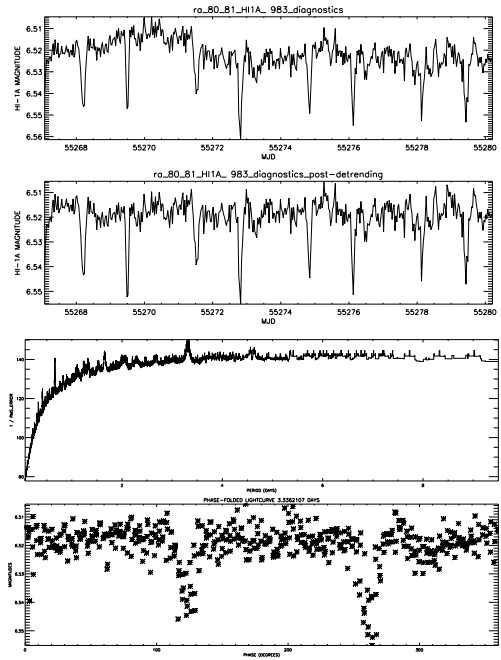}}
\caption{The main stages of the detrending and analysis, demonstrated by the star HR~1750.  Starting with the top graph showing the raw light curve for one epoch of observations, the next graph down shows the results of detrending with a 7th order polynomial, the \textbf{third graph shows the periodogram derived from the detrended lightcurve and the bottom graph is the lightcurve phase-folded on the period of the strongest signal found from the periodogram}.  Note that it is unknown whether the trend removed by the polynomial fitting is artificial or genuine variability - this potential long term variability is not the type of variability being searched for.}
\label{fig50}
\end{figure}

\noindent Three methods were used to quantify the variability of each star in the \textit{STEREO}/HI-1A database with a magnitude of 10.5 or greater, thus eliminating the faintest stars with the lowest signal-to-noise ratio (Figure \ref{fig44} shows the locations of stars with measured mean magnitude of 9th magnitude and brighter with more than twenty photometric points on their lightcurve).  They were: the Box-Least-Squares (BLS) \citep{bls2}, the Lomb-Scargle periodogram \citep{1982ApJ...263..835S} and Stetson's variability index \citep{1993AJ....105.1813W}.  IDL code to produce a Lomb-Scargle periodogram was obtained from Armagh Observatory, whilst the others were custom implementations.  Figure \ref{fig50} shows the steps in the analysis as produced by the programs doing the analysis.

Before being passed to the algorithms for analysis, a 7th order polynomial fit was carried out for each lightcurve to remove both natural and artificial trends that might obscure the faintest transit signals (the top two graphs of Figure \ref{fig50} show an example lightcurve before and after this polynomial detrending).  It should be emphasized that whether or not there is genuine long term variability that would be removed by this is not a concern as the focus is on recovering eclipsing variables and exoplanet transits whilst retaining the capacity to detect other short term variables, such as RR Lyrae and $\delta$ Scuti stars.  The IDL built-in routine \verb+poly_fit+ was the basis of this.  Each lightcurve was then passed through a combined low and high pass filter in an attempt to remove some remaining noise.  This was done by transforming into the frequency domain using the IDL built-in routine \verb+fft+ and applying a mask that blocked all features within one standard deviation of the mean before using the inverse transform to return to the time domain.  This process was repeated for each epoch of observations for each star and each epoch was analysed separately by the different algorithms (The third graph in Figure \ref{fig50} shows the BLS spectrum).  The strongest signals found from analysing all the epochs for each star were recorded and stored in an output file associated with each field of view, as were a variety of other statistics (q.v. Section \ref{sec:tables}).

The reasons for processing each epoch separately are the presence of noise and artefact features that exist within one or more epochs but seldom all, such as a planetary incursion into the field of view by Venus or Mercury,  the existence of some features near the beginning or end of some epochs as a result of streamers or other solar activity and the presence of occasional de-pointing events, typically from micrometeorite impacts, all of which would produce a false signal if all epochs were analysed together.  As the highest values from all epochs are stored, these features still give some false signals but by analysing data from unaffected epochs a genuine value may sometimes supersede the false one.  Nevertheless, the severity of some noise and artefacts precludes detection of faint signals, which is only partially alleviated in some cases by a direct visual inspection of lightcurves for signs of genuine variability.

Once a star had been identified as an eclipsing binary candidate on the basis of the statistics generated, its detrended, filtered lightcurve and phase-folded lightcurves on the periods of the strongest signals found by the BLS and Lomb-Scargle analysis were examined visually to determine if the variability detected was likely to be genuine.  An initial classification of the type of variability was made at this stage.  Those that were considered likely to be eclipsing binaries were imported into the software package \textsc{Peranso} \citep{peranso231} for more detailed analysis.  There is the possibility that many contact binary systems might have been excluded at this stage due to an inability to distinguish between different types of variability.  Indeed, there is some evidence that a significant number of fainter contact binaries may have been excluded in this manner \textbf{(Section \ref{sec:conclusions})}.

Within \textsc{Peranso}, a detailed analysis of the BLS spectrum would be used to establish as accurate a period \textbf{as} possible using as many epochs as possible, using the software's ability to manually exclude observation points that are likely to be noise or artefacts.  The lightcurve phase-folded from this period would be used to finally determine if the star was likely to be an eclipsing binary.  If it was still considered an eclipsing binary, the phase-folded lightcurve would provide estimates of the primary eclipse depth and minimum eccentricity.  The minimum eccentricity would be estimated by visually determining the offset of secondary eclipses, where these could be distinguished as such, from a phase of $0.5$ away from the primary eclipse.  The estimated errors in the estimate were taken as the extremes of the possible ranges of where the secondary minimum might be, thus the errors are larger for smoother, shallower eclipses as in many contact binaries.  The values obtained were then, along with the errors, multiplied by two, giving a value that is actually $|e \times \cos \omega|$ or effectively a minimum eccentricity that could be significantly higher depending on the value of $\omega$, the argument of periastron.

The values of the period, primary eclipse depth and $|e \times \cos \omega|$ found from this detailed examination were recorded, along with the output of the different algorithms, the star's Celestial co-ordinates, its identity (from \textsc{Simbad}) and spectral type (from \textsc{Simbad}) as well as the identity of any nearby star (within 6 arc-minutes, from \textsc{Simbad}) that could potentially have contaminated the signal and whether or not either the star or this neighbour had previously been identified as an eclipsing binary.  This dataset of 263 eclipsing binaries (Table \ref{tab:1}) was then explored for interesting trends and patterns.

\subsection{Trends}

\begin{figure}
\centering
\resizebox{7cm}{!}{\includegraphics{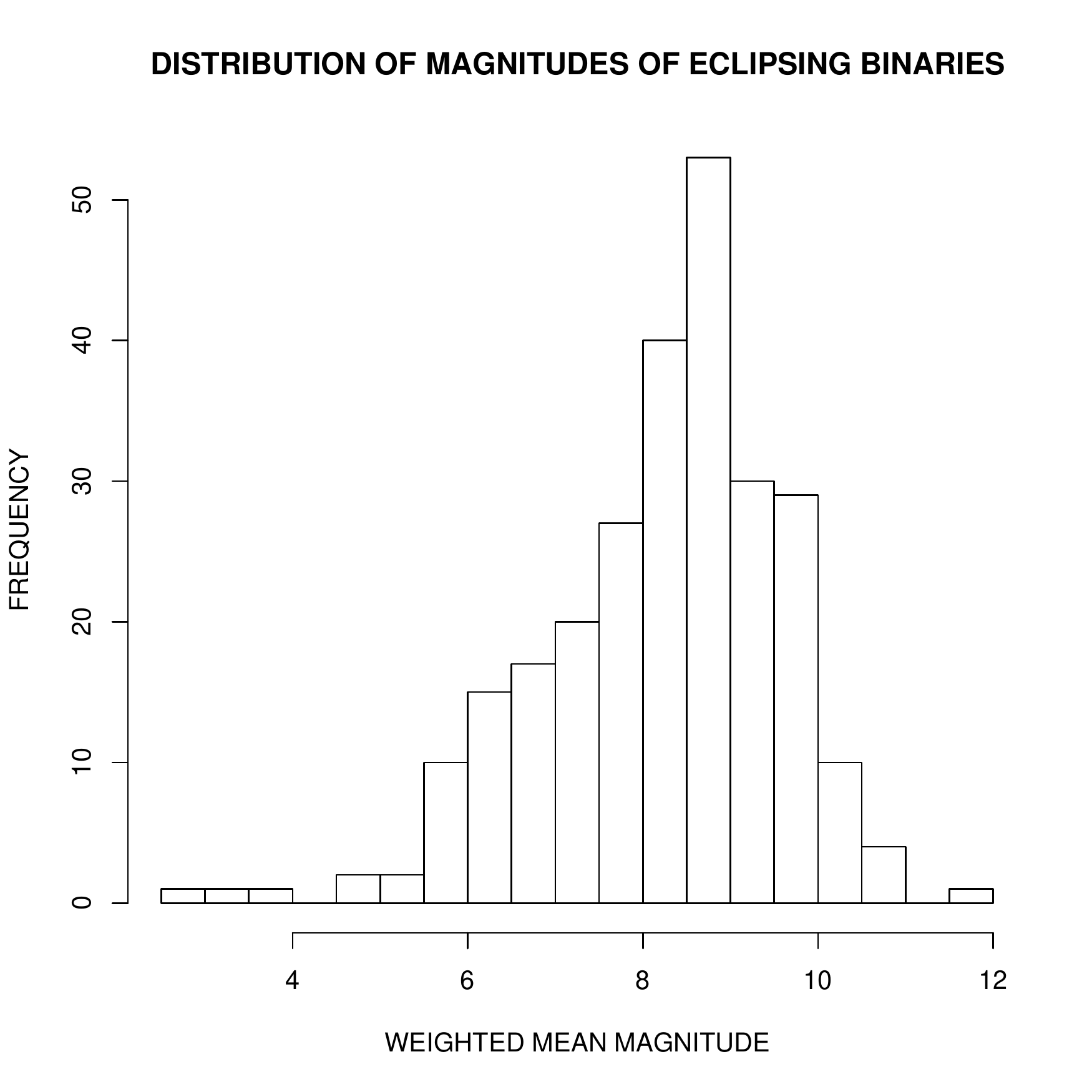}}
\caption{The distribution of the 263 eclipsing binaries observed by \textit{STEREO}/HI-1A by weighted mean magnitude.  The fall-off after magnitude 9 is the result of the decreasing signal quality with magnitude increasingly preventing the detection of eclipse signatures.}
\label{fig1}
\end{figure}

\noindent The distributions of the recorded characteristics show some trends that apply to the sample as a whole.  Most significantly, the distribution of eclipsing binaries by weighted mean magnitude (Figure \ref{fig1}) shows a peak between magnitude 8.5 and 9.0 after which the numbers extracted fall off.  This implies that, due to a combination of sensitivity and noise, eclipses are not being detected around fainter stars since the distribution of eclipsing binaries is expected to match the distribution of stars in the database, the numbers of which continue to increase with increasing magnitude.

\begin{figure}
\centering
\resizebox{7cm}{!}{\includegraphics{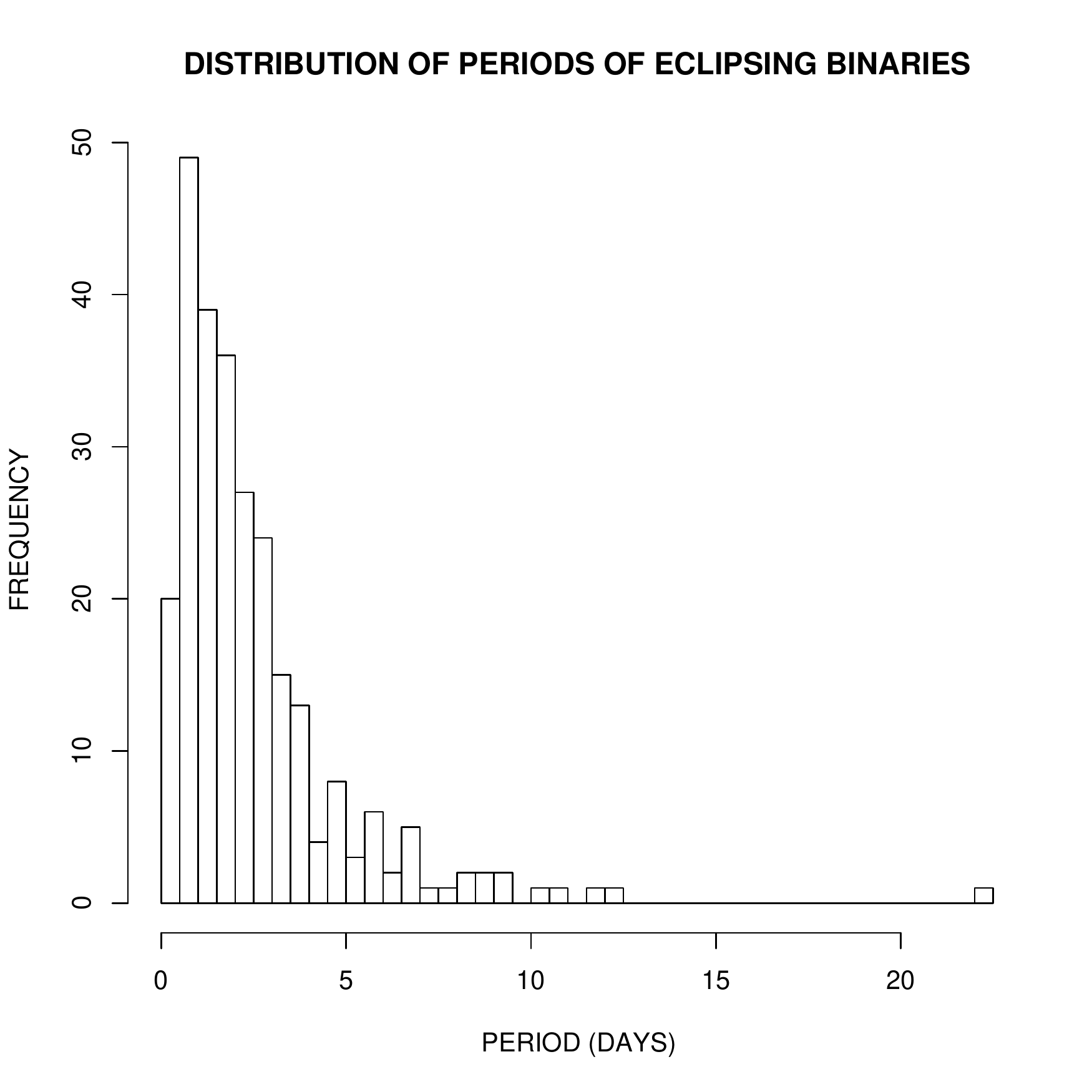}}
\caption{The distribution of the 263 eclipsing binaries observed by \textit{STEREO}/HI-1A by period.}
\label{fig2}
\end{figure}

The periods given in the statistics are the primary-to-primary period, although difficulty in distinguishing primary from secondary eclipses may cause the resulting value to be out by a factor of 2.
The distribution of the periods of the eclipsing binaries extracted reflects, as expected, the bias due to the length of individual epochs (Figure \ref{fig2}).  In order to get a good signal, three transits must be observed within an epoch and thus there is a tendency to find periods of less than this threshold.  Through the presence of secondary eclipses longer periods can be found but visual examination is required for detecting the very longest period eclipsing binaries, especially if the system is eccentric.  The star with the longest period confirmed of the 263 is HD 72208 with $22.013$ days (see Section \ref{sec:results} for details), which cannot be found through a separate analysis of each epoch but only through all three epochs presently available, owing to the significant eccentricity ($|e \times \cos \omega| = 0.391 \pm 0.006$) and the fact that no more than two transits are observed in any epoch.  There is a suspected eclipsing binary with a potentially longer period showing only one primary eclipse in each epoch (HD 173770, see Section \ref{sec:results}) but a definite period determination cannot be made and it was not included in the sample.

\begin{figure}
\centering
\resizebox{7cm}{!}{\includegraphics{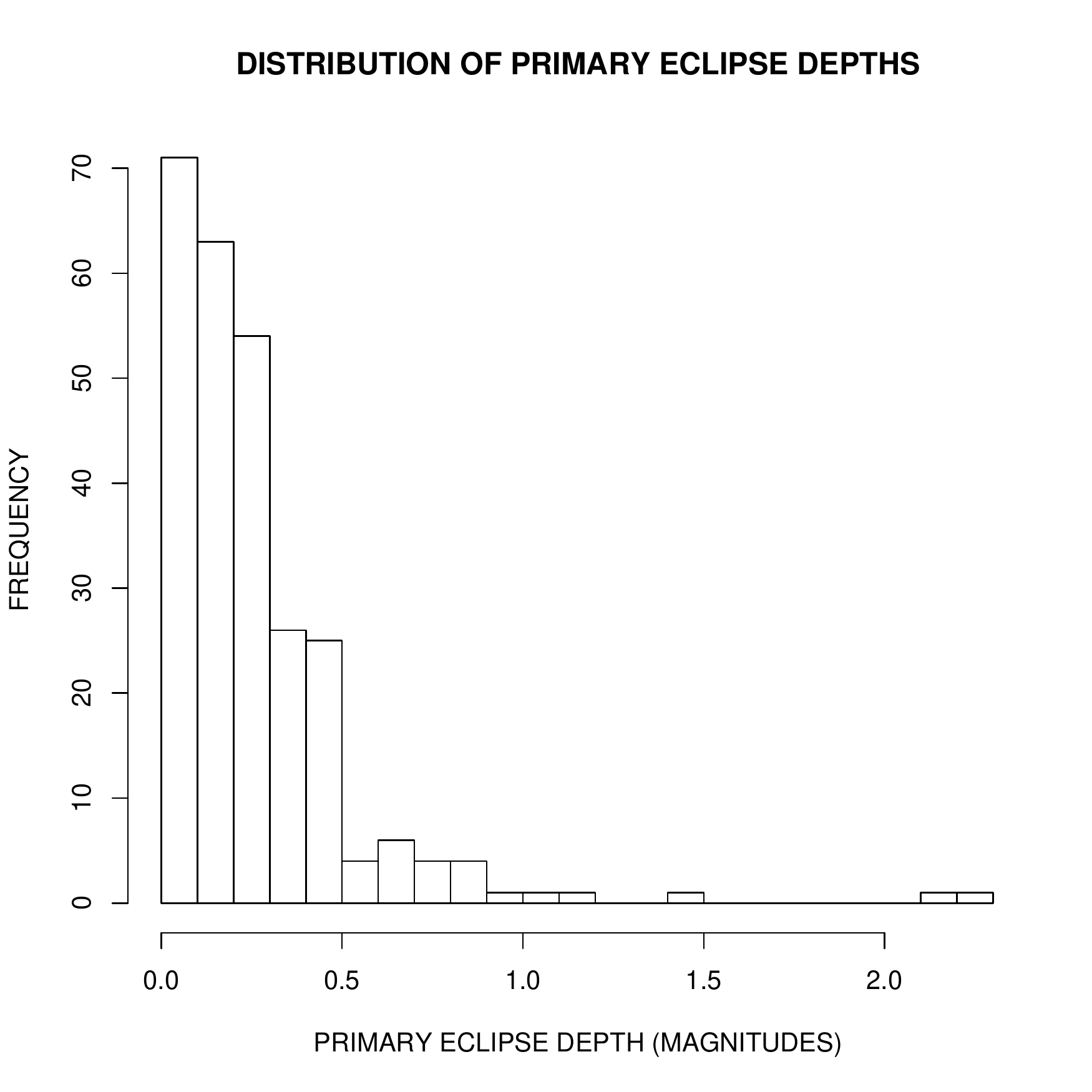}}
\caption{The distribution of the 263 eclipsing binaries observed by \textit{STEREO}/HI-1A by primary eclipse depth.}
\label{fig3}
\end{figure}

The distribution of the depths of primary eclipses reveals a trend that is interpreted as the presence of different populations of eclipsing binary (Figure \ref{fig3}).  Only \textbf{detached and semi-detached EBs} appear to have a primary eclipse depth greater than $0.5$ magnitudes, whereas contact binaries predominate at depths of $0.3$ magnitudes and below.

The estimates of $|e \times \cos \omega|$ reveal that 24 of the 263 eclipsing binaries have measurable eccentricity (see the Appendix for their phase-folded lightcurves).  Other EBs in the sample may have eccentricity, as the value of the argument of periastron, $\omega$, is unknown and there are some that show a hint of eccentricity but the margin of error does not preclude the possibility that they might have zero eccentricity, including a couple that are exactly on the line where the error margin equals the estimate of $|e \times \cos \omega|$.  Thus a small but significant fraction of the sample are eccentric, with more than 5\% showing sigificant eccentricity.  The sample of eccentric systems is not quite large enough to show statistically valid trends, unfortunately, although the indications are that small eccentricities dominate with larger eccentricities being widely distributed.  Nevertheless, this result is broadly in agreement with recent findings from the \textit{Kepler} mission \citep{2010arXiv1006.2815P}, however the sample of stars observed by \textit{STEREO}/HI-1A is not limited to main sequence stars or by other selection criteria intended to maximise the chances of finding an exoplanet transit.  It is therefore a useful result that it appears to find a similar distribution of $|e \times \cos \omega|$ to the \textit{Kepler} sample as it implies that eccentricity is not dependent upon spectral type.  Eccentric systems are expected to circularise over time and thus these systems are much-studied in the field of stellar evolution to model interactions between multiple stars as an interaction with a third body is usually required for them to be formed \citep{claret2002new}.

\begin{figure}
\centering
\resizebox{7cm}{!}{\includegraphics{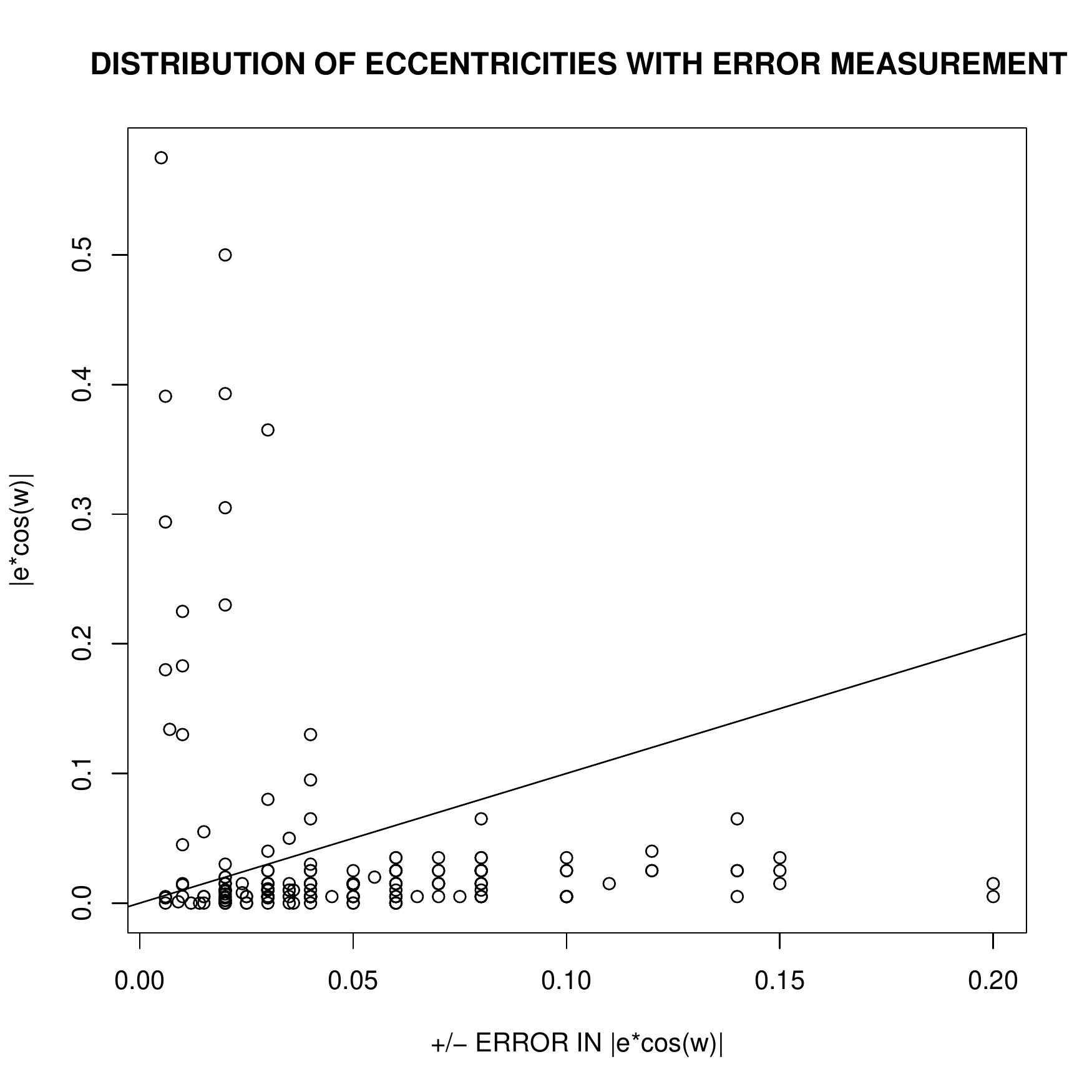}}
\caption{The estimates of the minimum eccentricities ($|e \times \cos \omega|$) of the 263 eclipsing binaries observed by \textit{STEREO}/HI-1A, where measurable, against the errors for each measurement.  Each open circle represents a single eclipsing binary.  Those points above the line represent eclipsing binaries for which an eccentricity of zero is outside the range of the errors in the measurement.}
\label{fig4}
\end{figure}

\begin{figure}
\centering
\resizebox{7cm}{!}{\includegraphics{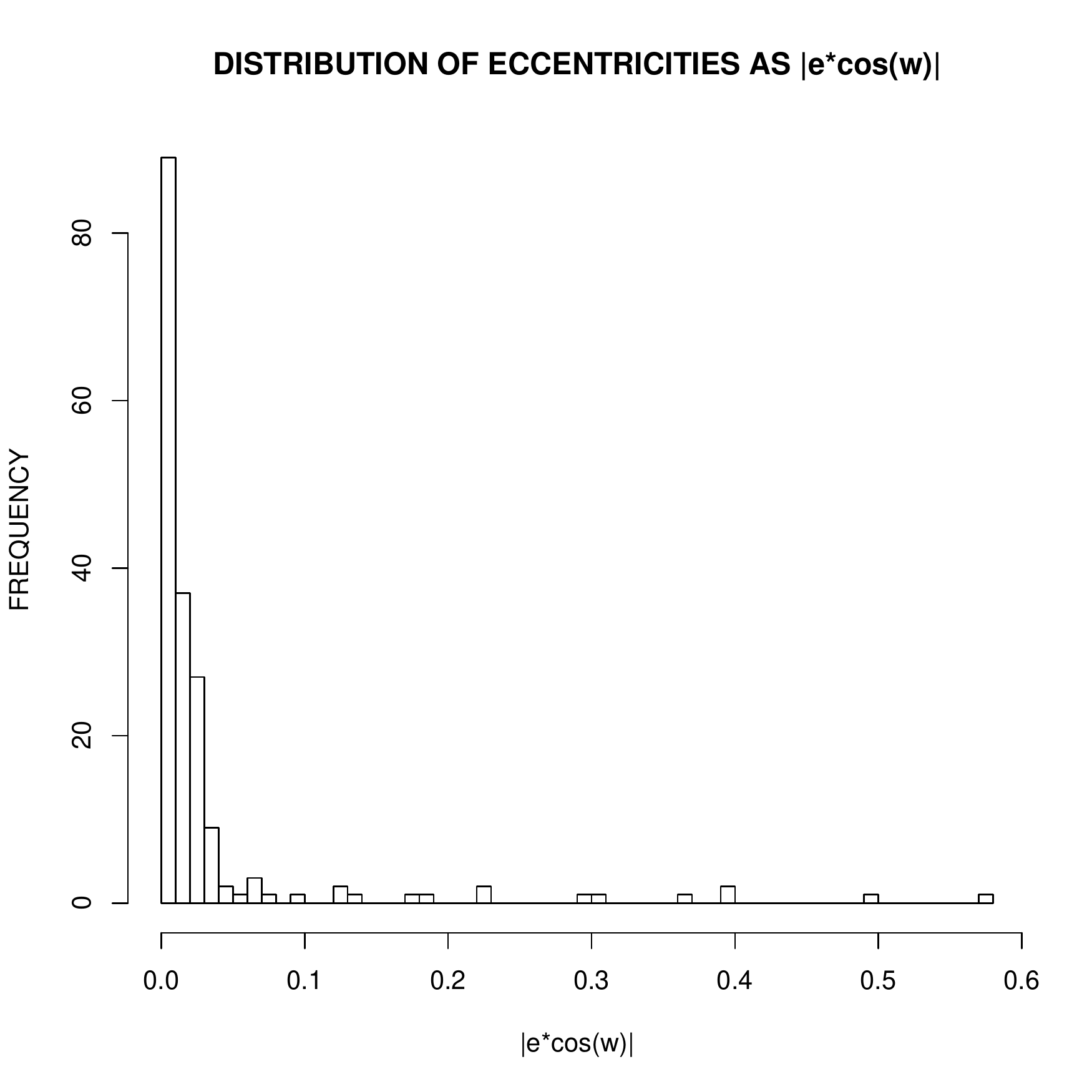}}
\caption{The estimates of the minimum eccentricities ($|e \times \cos \omega|$) of the 263 eclipsing binaries observed by \textit{STEREO}/HI-1A, where measurable.}
\label{fig4b}
\end{figure}

\begin{figure}
\centering
\resizebox{7cm}{!}{\includegraphics{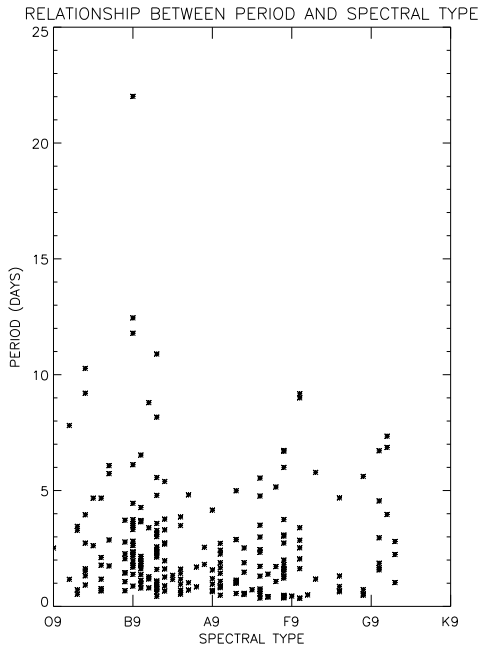}}
\caption{The 232 eclipsing binaries observed by \textit{STEREO}/HI-1A with a known spectral type, showing the relationship between spectral type and period.  \textbf{The most common spectral type is \texttt{A2}, with 22 stars in the sample}.  There are no clear patterns connecting the spectral type with period, however there are some cases where the specific star causing the eclipses has not been identified and the spectral type is of the star closest to the target co-ordinates (see Appendix \ref{subsec:eeb}).}
\label{fig5}
\end{figure}

The distribution of spectral types of the eclipsing binary primary stars shows a predominance of younger, hotter stars of earlier spectral types (Figure \ref{fig5}).  The main limitation of this distribution is that the spectral type of the closest star to the co-ordinates of the perceived eclipsing binary may not always be the star that is actually eclipsing, due to the low spatial resolution and the possible contamination of lightcurves by nearby stars.  The distribution will also be affected by any bias in the \textit{STEREO}/HI-1A database, with only stars brighter than magnitude $10.5$ being analysed en masse (although one or two fainter examples were identified visually).

\begin{figure}
\centering
\resizebox{7cm}{!}{\includegraphics{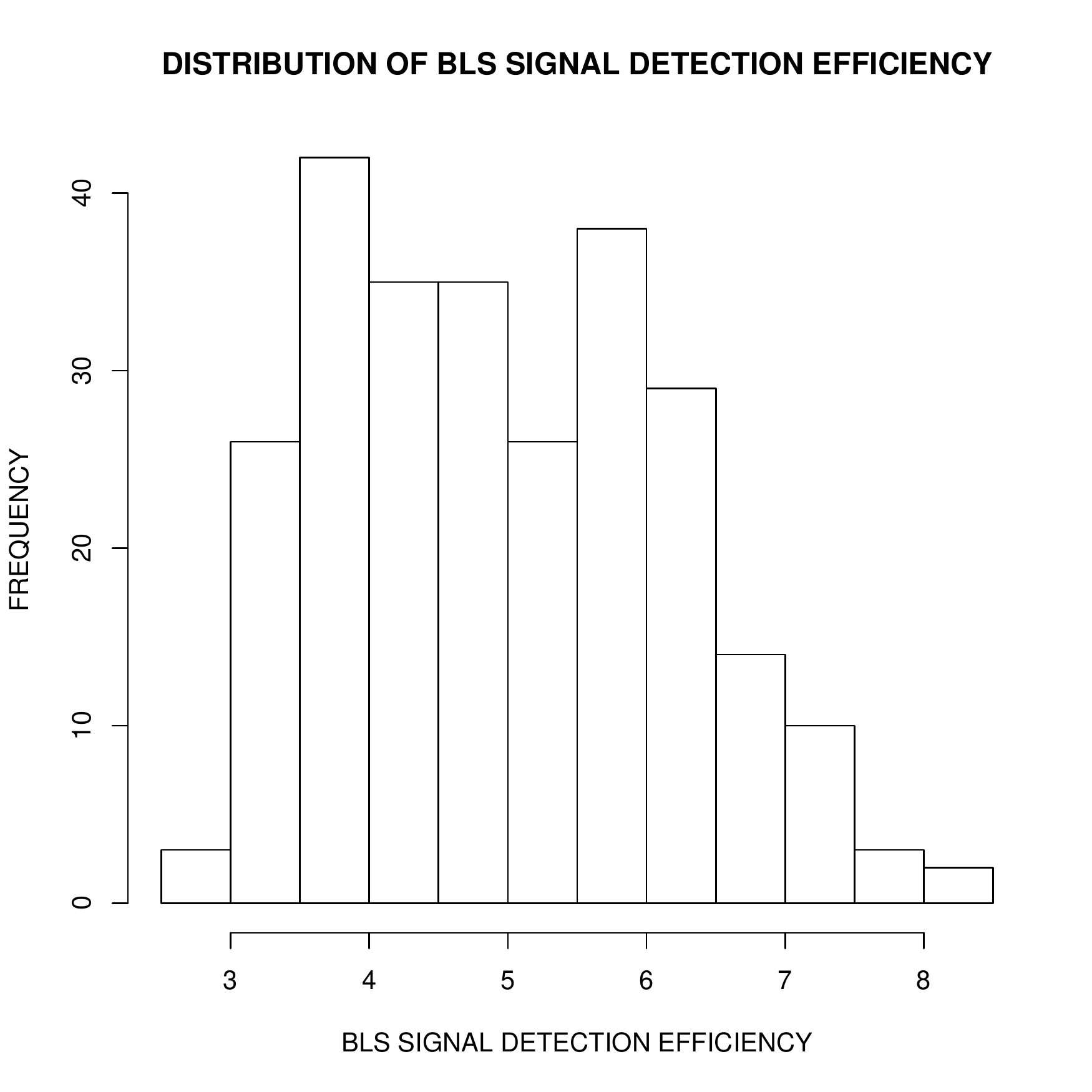}}
\caption{The distribution of the 263 eclipsing binaries observed by \textit{STEREO}/HI-1A by BLS signal detection efficiency.}
\label{fig6}
\end{figure}

\begin{figure}
\centering
\resizebox{7cm}{!}{\includegraphics{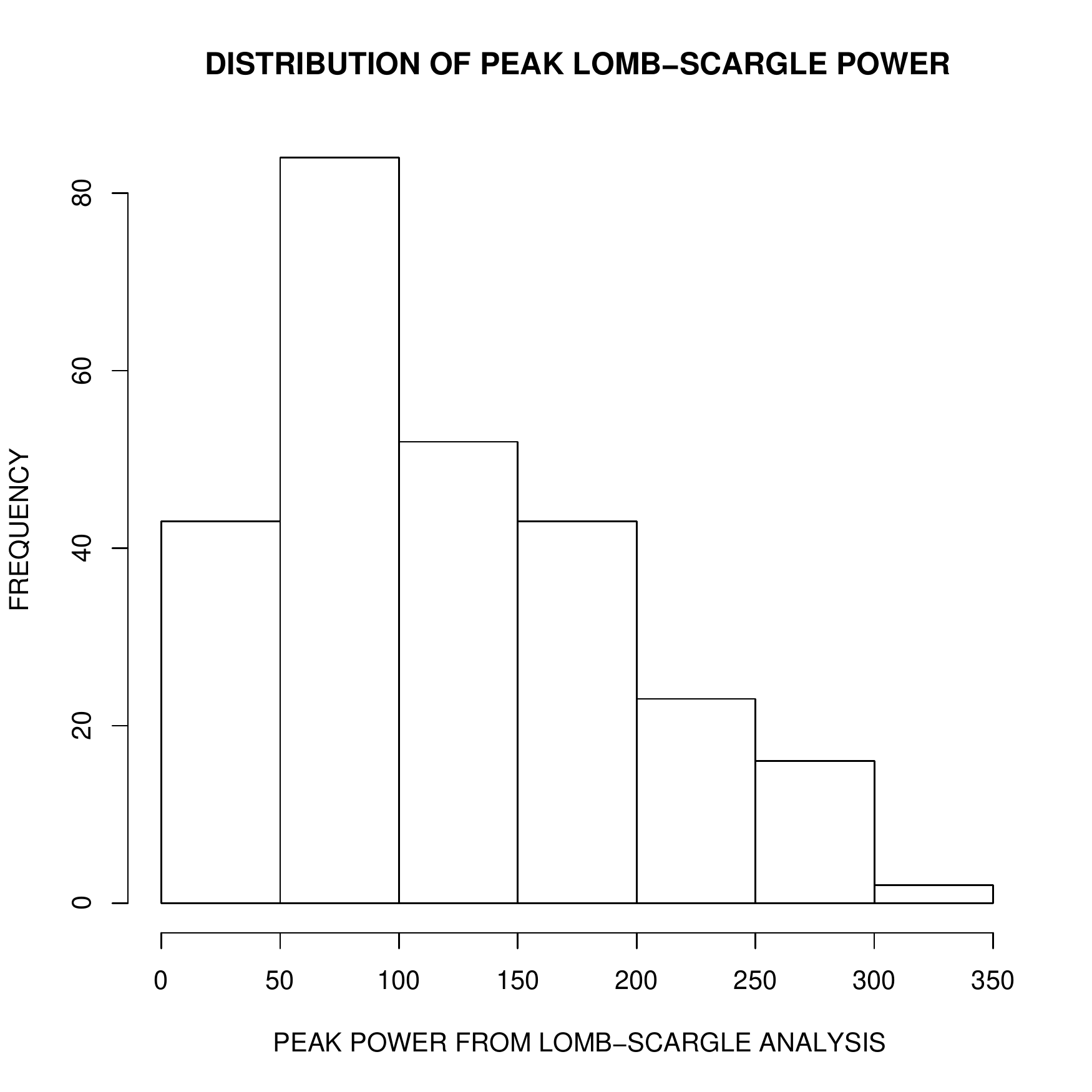}}
\caption{The distribution of the 263 eclipsing binaries observed by \textit{STEREO}/HI-1A by peak power from a Lomb-Scargle periodogram analysis.}
\label{fig10}
\end{figure}

\begin{figure}
\centering
\resizebox{7cm}{!}{\includegraphics{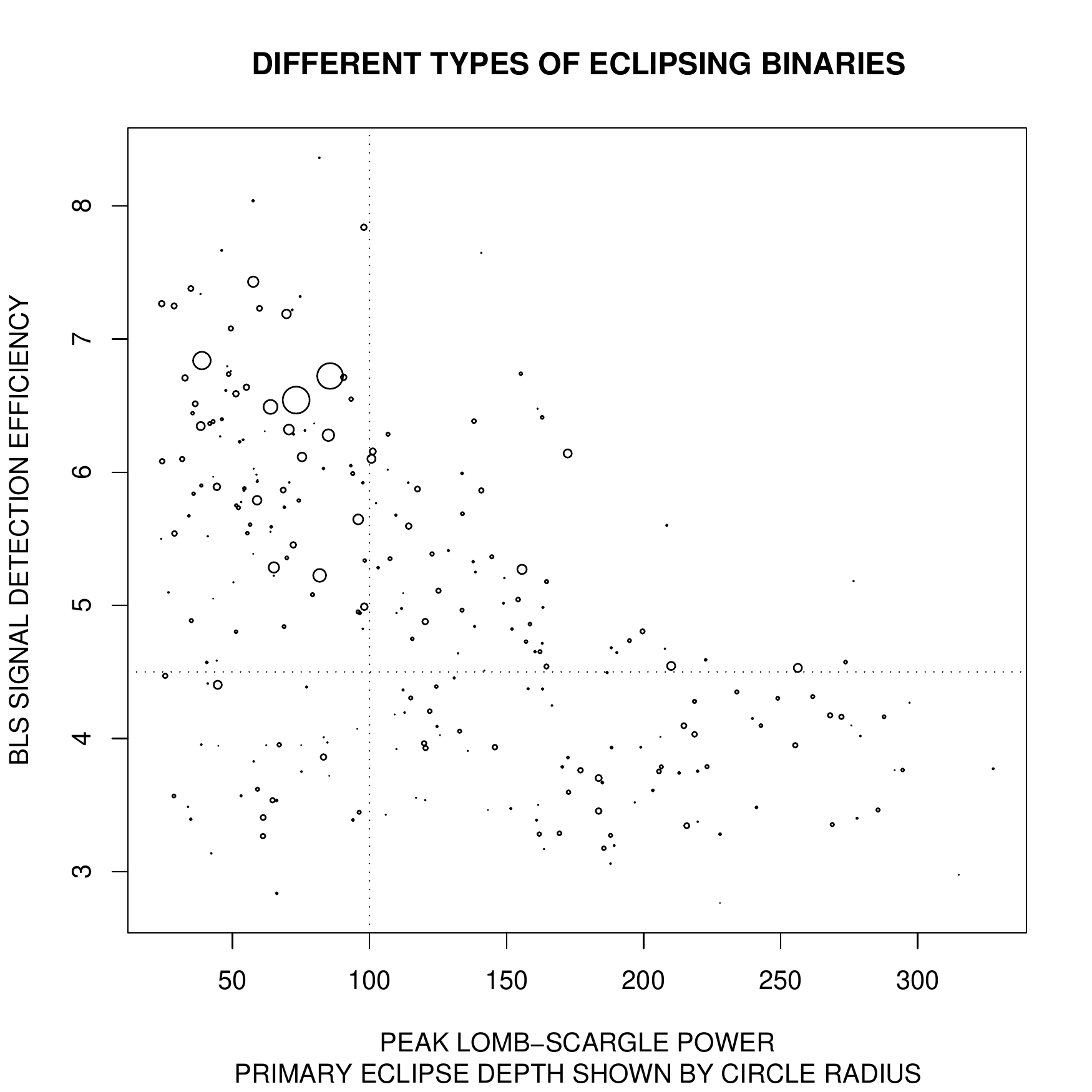}}
\caption{Comparison of the ability of the BLS and Lomb-Scargle algorithms to extract the signals of eclipsing binaries.  Each eclipsing binary is represented by a circle with a radius proportional to the depth of the primary eclipse.  Points above the horizontal line have a SDE high enough to be considered a detection by the BLS algorithm, whilst those to the right of the vertical line may similarly be considered to have been detected by the Lomb-Scargle algorithm, in the absence of artefacts and noise producing large numbers of similar signals.  Those rated poorly by both were typically found through visual examination.}
\label{fig7}
\end{figure}

The distribution of eclipsing binaries by their signal detection efficiency (SDE) from the BLS algorithm (Figure \ref{fig6}) shows two indistinct peaks, interpreted as corresponding to two different types of eclipsing binary.  The peak with the highest SDE corresponds mostly to detached binaries, whose lightcurves are a closer match to the box-like shape the BLS algorithm checks for.  The broader peak with the lower SDE corresponds to contact binaries, which have smoother, more sinusoidal lightcurves.  An Algol type binary might be the highest-rated by the BLS algorithm in its field but contact binaries can sometimes be near the bottom of such a list and thus are not reliably detected by this algorithm.  The Lomb-Scargle periodogram, (distribution shown in Figure \ref{fig10}), preferentially reveals contact binaries, although it typically does give a signal for many detached binaries as well, albeit increasingly less so for shallower, more box-like eclipses (Figure \ref{fig7}).  Combining the BLS and the Lomb-Scargle methods thus produces a positive signal for the vast majority of the eclipsing binaries, with those rated poorly by both being stars with fewer data points (both algorithms give higher scores for more data points, thus stars with less data available are rated less highly), those with very long periods and those with comparatively low signal-to-noise that were detected by visual examination.

\begin{figure}
\centering
\resizebox{7cm}{!}{\includegraphics{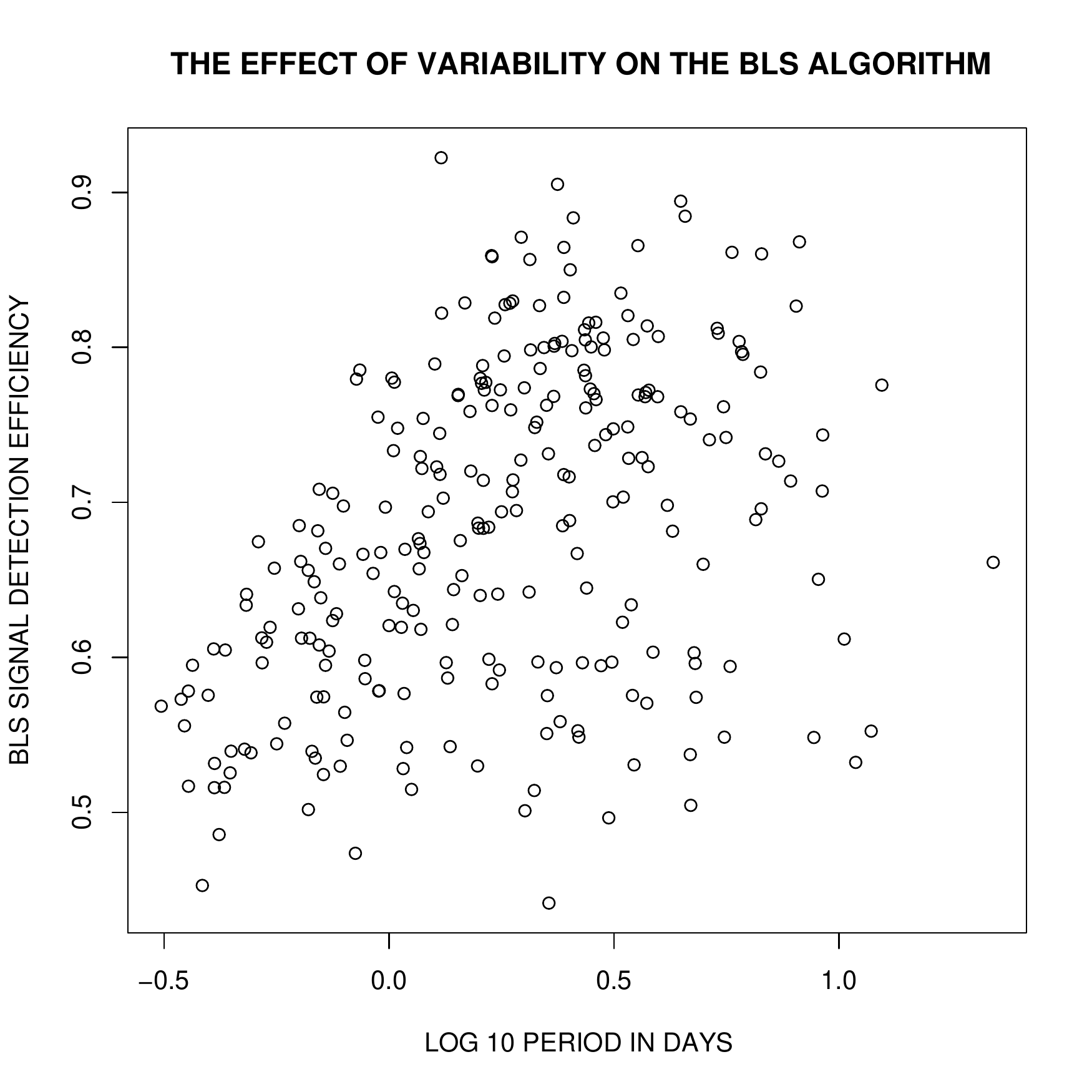}}
\caption{The dependence of the BLS algorithm on the period of a signal.  The influence is believed to be the shape of the light curves of variables with increasingly short periods, rather than the period itself.  Each open circle represents a single eclipsing binary.}
\label{fig8}
\end{figure}

\begin{figure}
\centering
\resizebox{7cm}{!}{\includegraphics{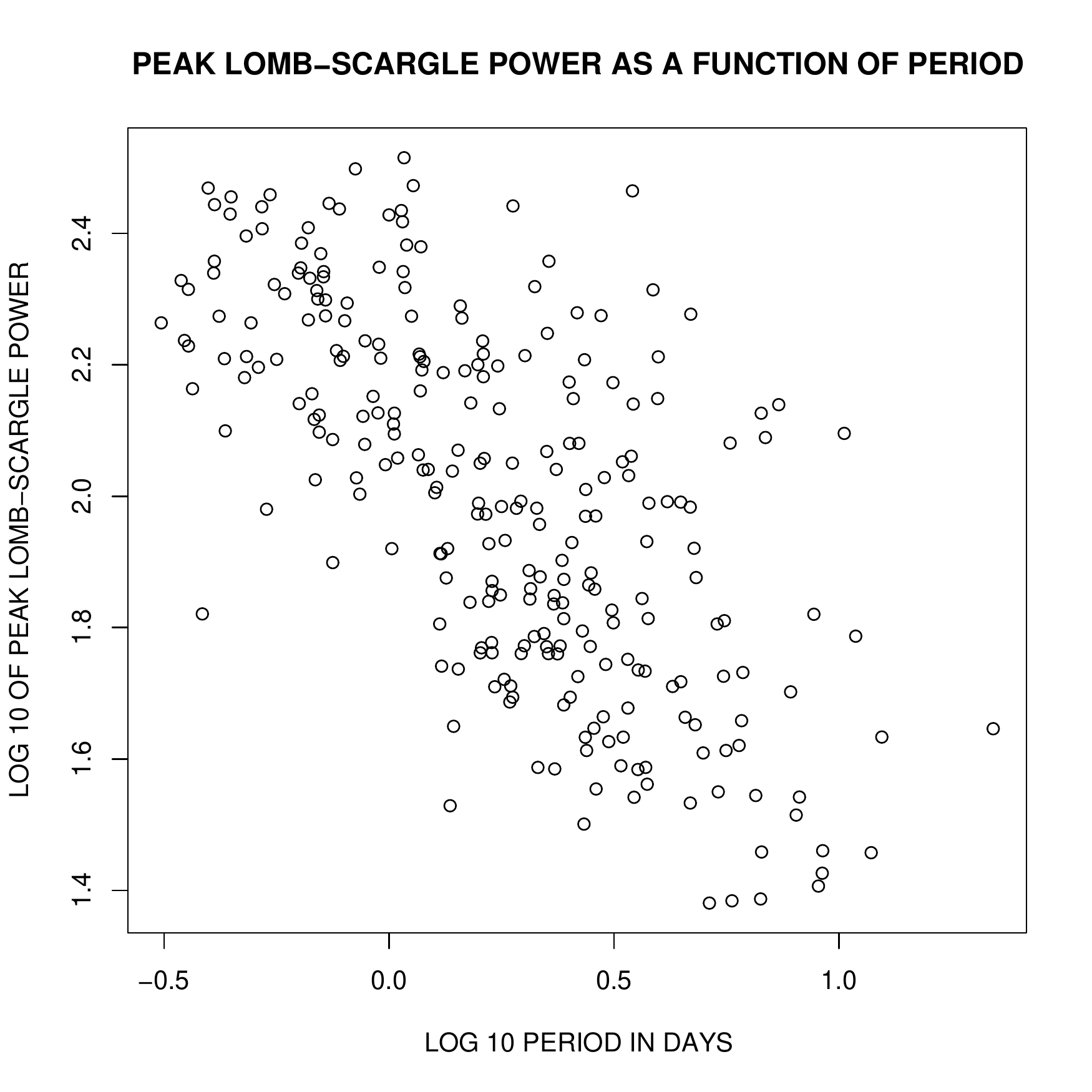}}
\caption{The dependence of the Lomb-Scargle method on the period of a signal.  Each open circle represents a single eclipsing binary.}
\label{fig9}
\end{figure}

The effect of the period on the output of the BLS and Lomb-Scargle methods was also investigated (Figures \ref{fig8} and \ref{fig9}).  The dependence discovered is again likely representative of the shape of the transits.

\begin{figure}
\centering
\resizebox{7cm}{!}{\includegraphics{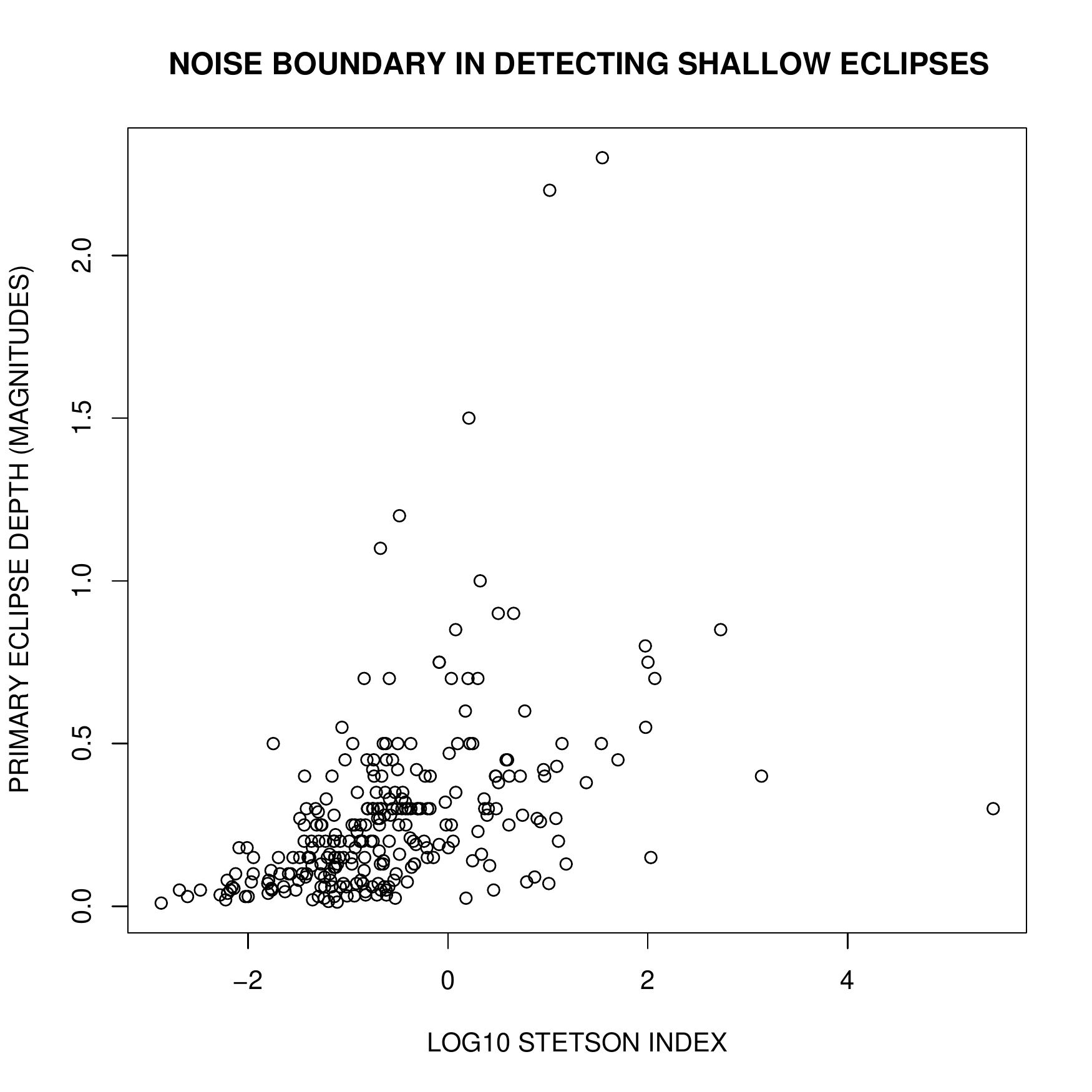}}
\caption{The dependence of primary eclipse depth on the Stetson index.  The sharp cut-off at about -0.5 shows that eclipses with a primary eclipse depth less than 0.3 magnitudes are considerably more difficult to detect through noise or variability.  Each open circle represents a single eclipsing binary.}
\label{fig11}
\end{figure}

\begin{figure}
\centering
\resizebox{7cm}{!}{\includegraphics{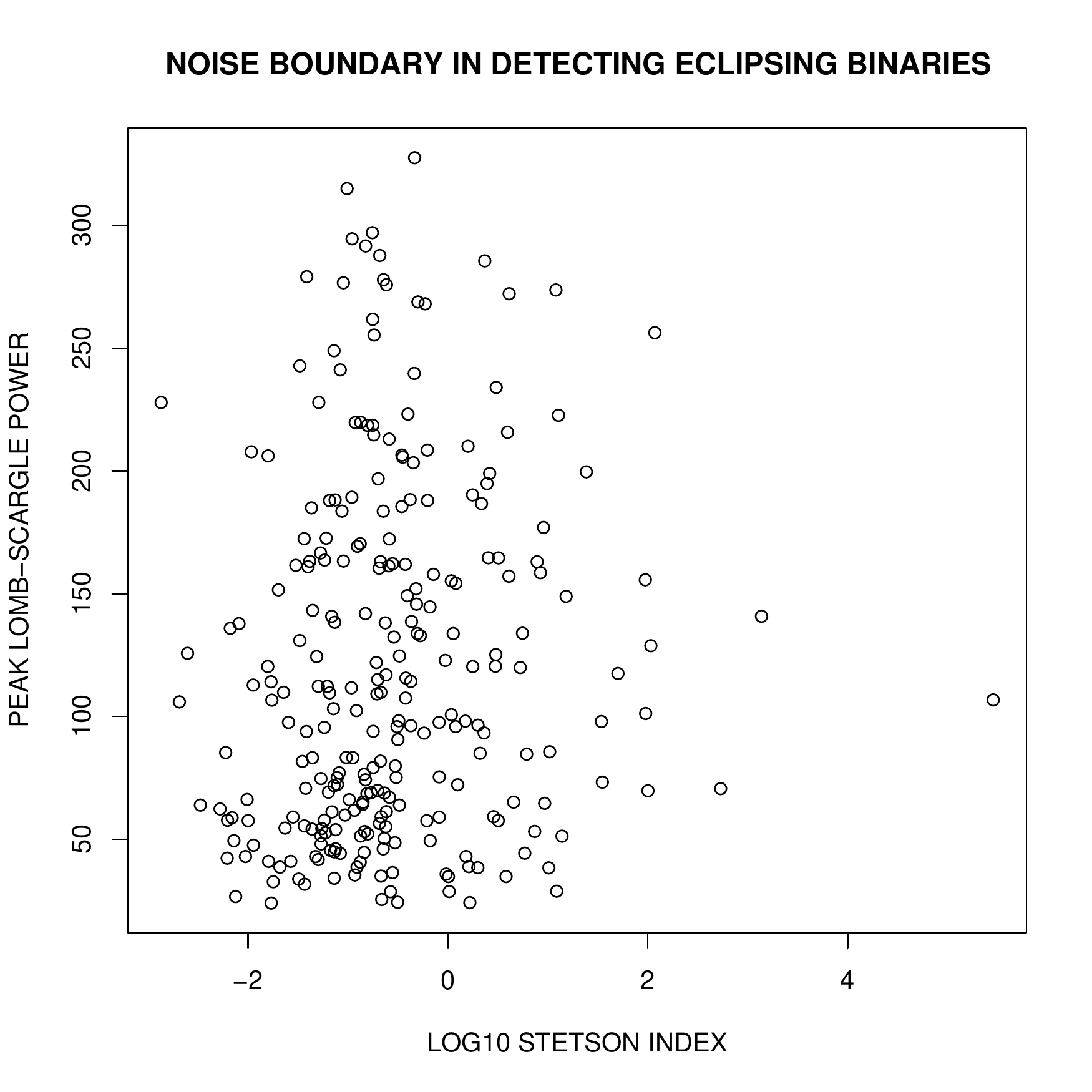}}
\caption{The dependence of peak power from a Lomb-Scargle periodogram analysis as a function of Stetson index.  The cut-off point at about -0.5 shows that variability or noise above a certain level is very effective at obscuring the signals of eclipsing binaries, regardless of their peak power.  Each open circle represents a single eclipsing binary.}
\label{fig12}
\end{figure}

\begin{figure}
\centering
\resizebox{7cm}{!}{\includegraphics{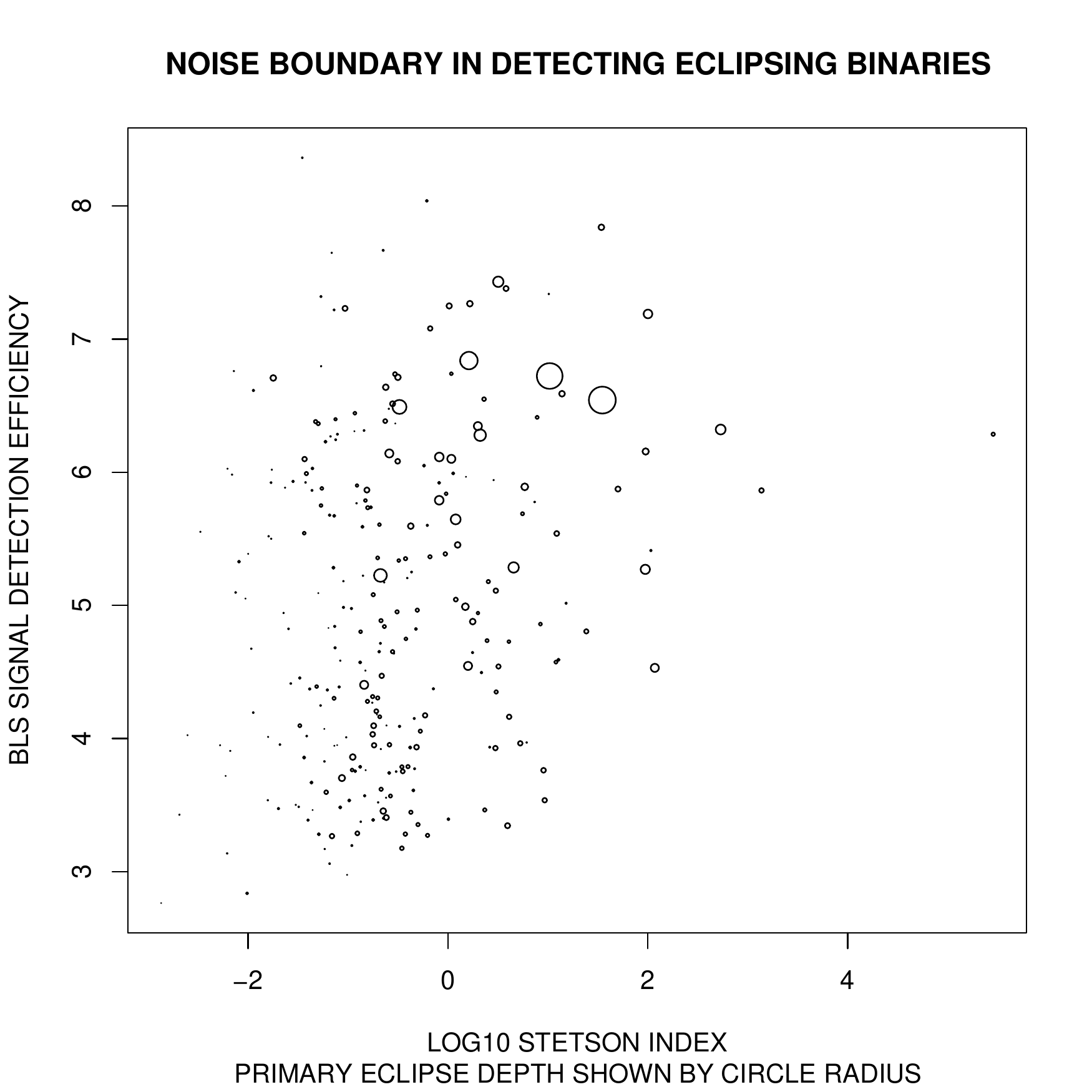}}
\caption{The dependence of signal detection efficiency as a function of Stetson index.  The cut-off point at about -0.5 shows that variability or noise above a certain level is obscuring the signals of some eclipsing binaries, although the apparent clustering of signals around a moderate value and especially the higher ratings for deeper eclipses shows that the Stetson index is also to some extent reflecting the variability due to eclipses.  Each eclipsing binary is represented by a circle with a radius proportional to the depth of the primary eclipse.}
\label{fig13}
\end{figure}

\begin{figure}
\centering
\resizebox{7cm}{!}{\includegraphics{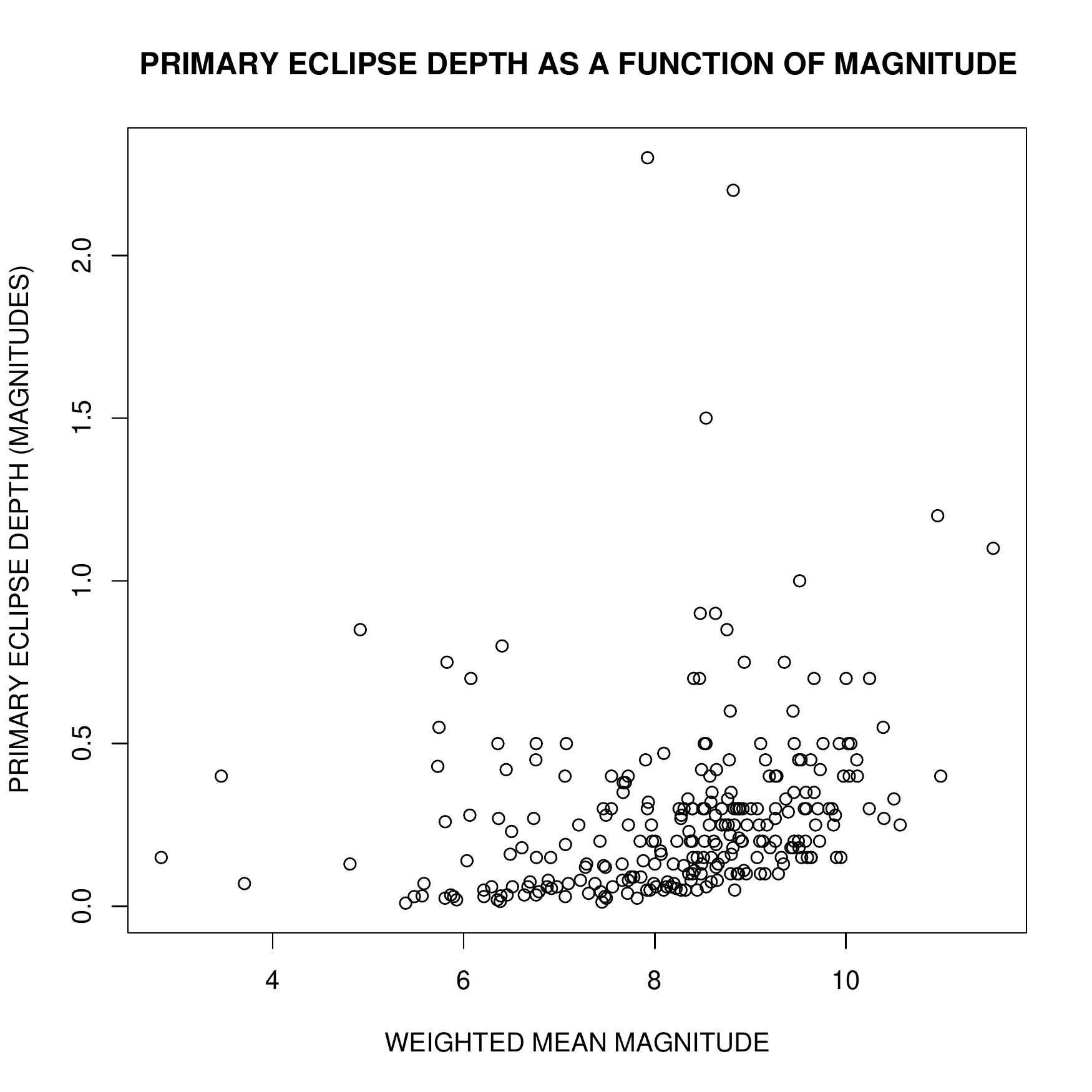}}
\caption{The dependence of primary eclipse depth as a function of weighted mean magnitude.  After about magnitude 8.5 shallower eclipses are not detected, with the threshold rising nearly linearly.  Each open circle represents a single eclipsing binary.}
\label{fig14}
\end{figure}

As a variety of noise signatures were encountered in the dataset, some very clearly obscuring genuine signals, producing both false positives and false negatives, an attempt to evaluate the effects of noise was made (Figures \ref{fig11}, \ref{fig12}, \ref{fig13} and \ref{fig14}).  Together these appear to show a tendency to overlook shallow eclipses.  This may be partly due to a mixture of red and white noise and also due to problems classifying contact binaries as such when their \textbf{lightcurves are not clear enough to distinguish them from sinusoidal variations}.  \textbf{Blending also has a tendency to obscure faint signals, with the brightest object in the photometric aperture dominating but showing additional signals overlaid, or with stars of similar magnitudes showing constructive or destructive interference resembling an amplitude modulation.  This has the effect of lowering, broadening and sometimes splitting peaks in a periodogram analysis, which affects the ability of an algorithm to identify a low-amplitude signal.}

\begin{figure}
\centering
\resizebox{7cm}{!}{\includegraphics{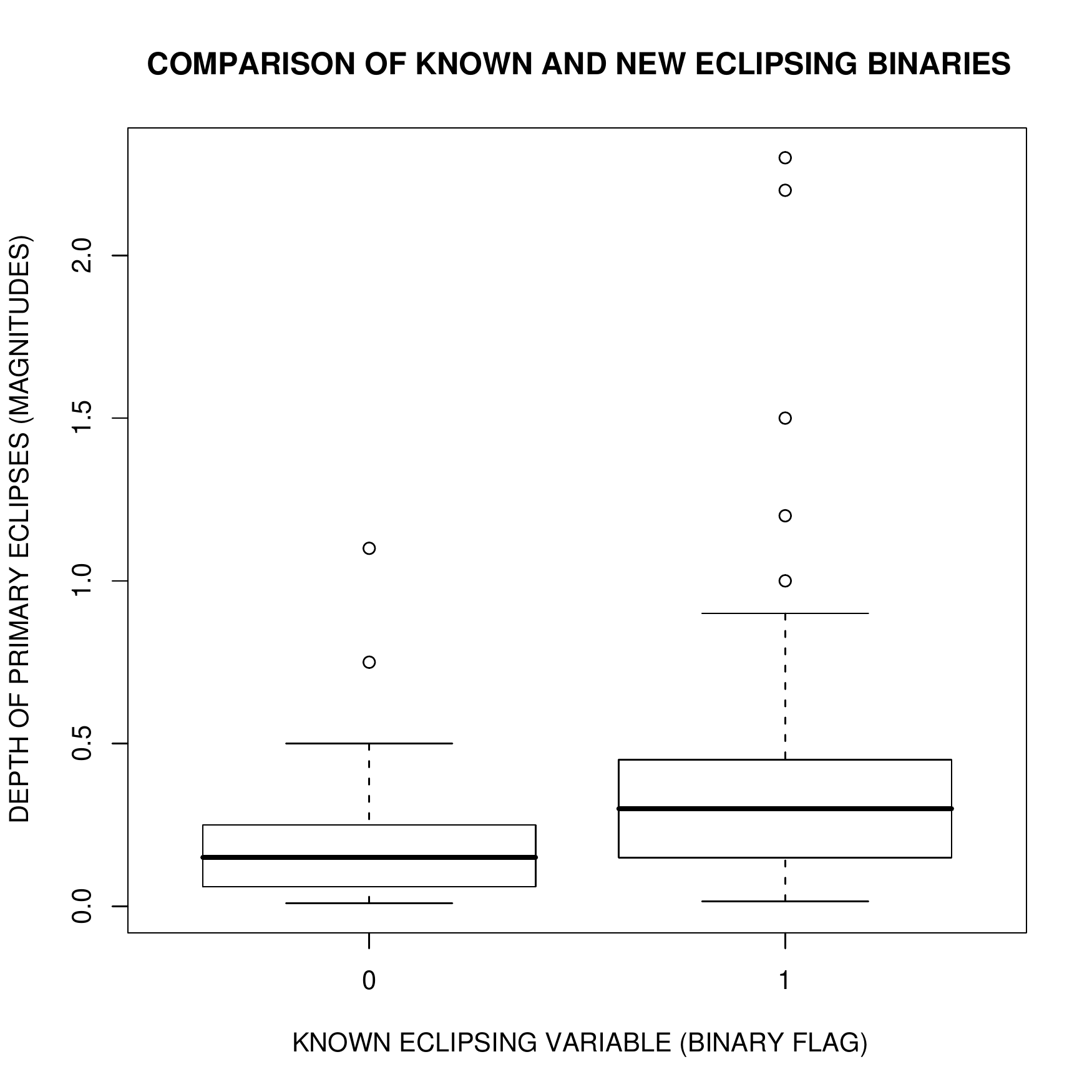}}
\caption{Comparison of the distribution of known and new eclipsing binaries with respect to the depth of their primary eclipses.  The tendency of the newly-discovered systems to show fainter eclipses reflects both the sensitivity of the \textit{STEREO}/HI-1A and the expected bias of previous surveys in discovering deeper eclipses.  The centre box contains those EBs in the second and third quartiles with the median represented by a thick line within, whilst the outer bars represent an approximately $95\%$ confidence level and outliers are individual open circles.}
\label{fig16}
\end{figure}

Since a large proportion of the sample of eclipsing binaries appear to be previously undetected or, at least, unconfirmed as such by previous observations, a search was made for characteristics in common to understand what, if anything, might be different about the new cases that might explain why they had previously gone undetected (Figures \ref{fig16}).  As might be expected, there is a tendency for the new cases to have shallower eclipses than those previously known.  There does not appear to be any significant difference in either the periods or magnitudes found, however.

\begin{figure}
\centering
\resizebox{7cm}{!}{\includegraphics{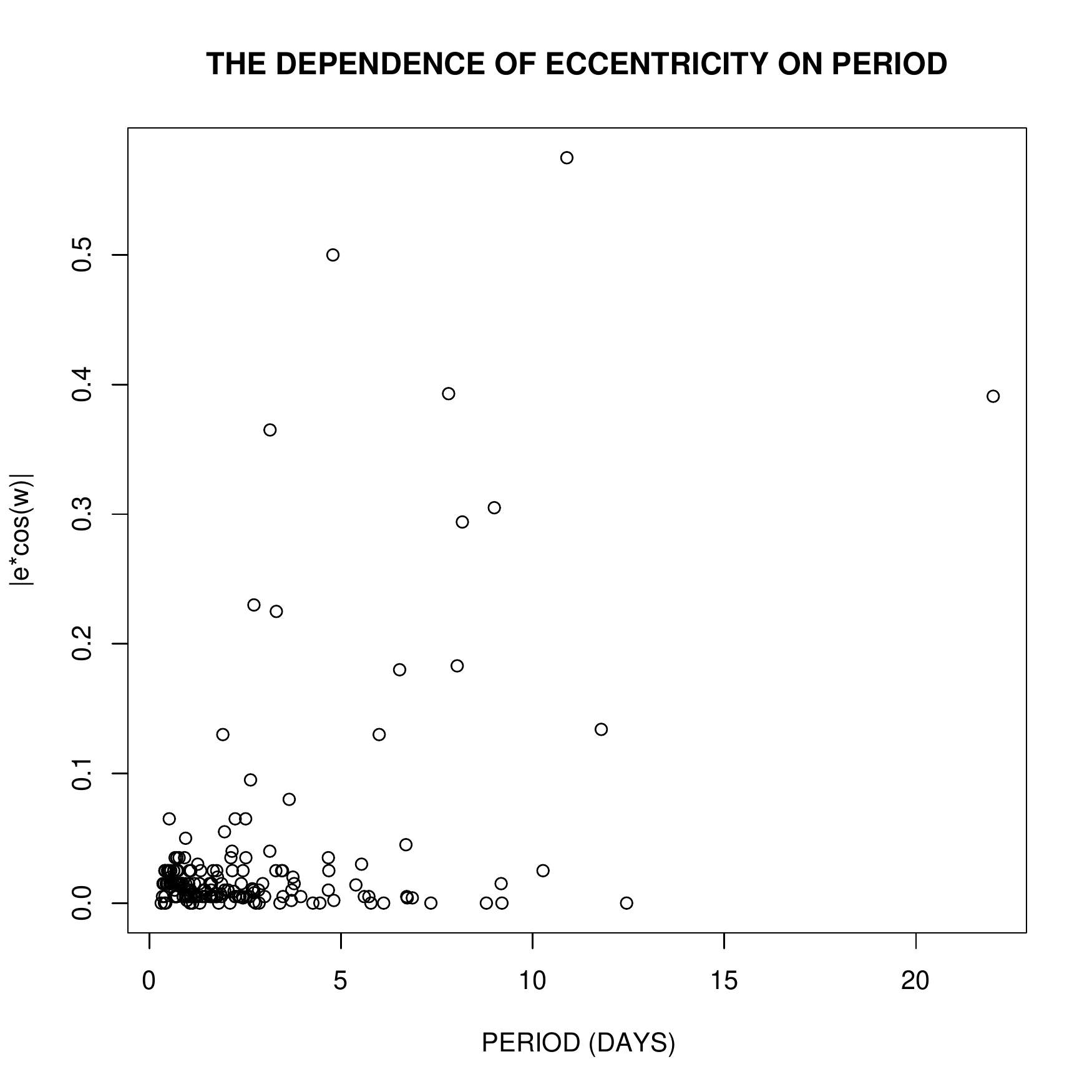}}
\caption{The dependence of the minimum eccentricity of the 263 eclipsing binaries observed by \textit{STEREO}/HI-1A, as measured by $|e \times \cos \omega|$, on the period of eclipses.  Each open circle represents a single eclipsing binary.}
\label{fig19}
\end{figure}

A previously known tendency for eclipsing binaries with long periods to be more likely to be eccentric \textbf{\citep{1988ApJ...326..256M}} is confirmed by the \textit{STEREO}/HI-1A observations (Figure \ref{fig19}).

To summarise briefly, the BLS is the algorithm of choice for extracting box-like eclipses, e.g. Algol binaries or exoplanet transits.  The Lomb-Scargle algorithm is best at extracting sinusoidal-like signals but also has some limited capability to extract other regular signals.  The Stetson index is able to extract large amplitude variables of all types.

%% file: results.tex
\section{\uppercase{Results}}
\label{sec:results}

\noindent Of the 263 eclipsing binaries extracted, \textbf{122} are not recorded as such in \textsc{Simbad}.  Some of these are, however, known variables of other types or are suspected of variability.  Owing to the purely photometric data available from \textit{STEREO}/HI-1A, there remains the possibility of mis-classification, however care was taken to include only stars whose phase-folded lightcurves appeared to be clear cases.  Thanks to the excellent phase coverage resulting from the \textit{STEREO}/HI-1A observational cadence of 40 minutes maintained for up to $19.44$ days, with three or four such epochs for most stars, the new observations are in many cases clearer than those previously available and in some cases the existing classification may be in need of revision.  A total of 24 of the eclipsing binaries have measurable eccentricity and their lightcurves are shown in the Appendix (Figure \ref{fig101}).

\subsection{Individual stars of interest}

\subsubsection{HD 213597: substellar transit candidate}

\begin{figure}
\centering
\resizebox{7cm}{!}{\includegraphics{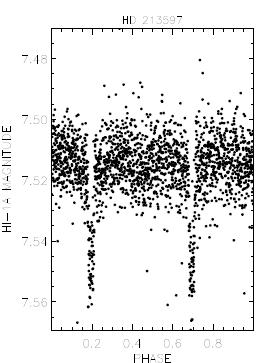}}
\caption{Lightcurve of the star HD 213597, phase-folded on a period of $4.8476$ days.}
\label{fig20}
\end{figure}

\noindent \textbf{The best potentially substellar transiting candidate is HD 213597}.  This star is recorded in \textsc{Simbad} as being of spectral type \texttt{F0} and is not suspected of any variability.  The observed eclipses are box-like, with very sudden ingress and egress phases, and are uniformly about 25~mmag deep, with a period of $2.4238$ days between successive eclipses (Figure \ref{fig20}).  Secondary eclipses are not evident.

\begin{figure}
\centering
\resizebox{7cm}{!}{\includegraphics{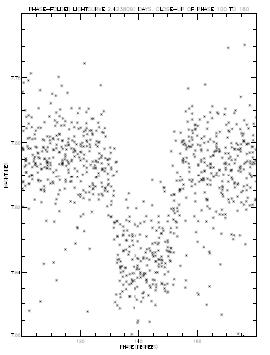}}
\caption{\textbf{Close-up of the transit of HD~213597, phase-folded on a period of $2.4238$ days.  This lightcurve was constructed using the latest data available, including updated flat-fields as well as data from \textit{STEREO}/HI-1B}.}
\label{fig21}
\end{figure}

\subsubsection{V~471 Tau: cataclysmic variable progenitor}

\begin{figure}
\centering
\resizebox{7cm}{!}{\includegraphics{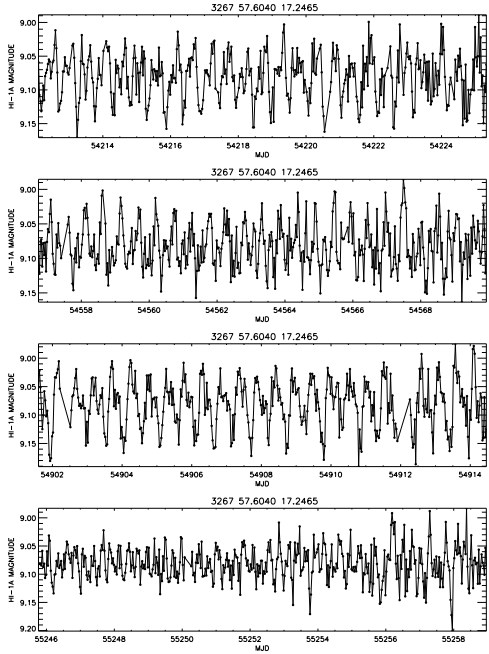}}
\caption{Lightcurve of the star V471 Tau, as observed by the \textit{STEREO} HI-1A imager.  3267 is the catalogue number of the star in the \textit{STEREO} HI-1A field of view for right ascension: 50-60 degrees and declination: 10-20 degrees.}
\label{fig471a}
\end{figure}

\begin{figure}
\centering
\resizebox{7cm}{!}{\includegraphics{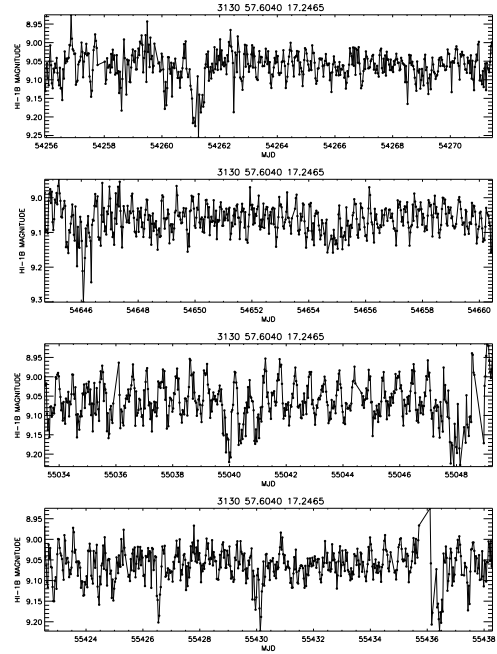}}
\caption{Lightcurve of the star V471 Tau, as observed by the \textit{STEREO} HI-1B imager.  3130 is the catalogue number of the star in the \textit{STEREO} HI-1B field of view for right ascension: 50-60 degrees and declination: 10-20 degrees.  The scale of the y-axis has been adjusted to fit the data range.}
\label{fig471b}
\end{figure}

Although it does not feature in the sample, as the eclipses by the white dwarf secondary are too short to be observed, V471 Tau nevertheless is an important star and the \textit{STEREO} observations in both HI-1A (Figure \ref{fig471a}) and HI-1B (Figure \ref{fig471b}) have the potential to inform studies of the magnetic activity cycle of the red dwarf primary (eg. \citep{o2001hubble}, \citep{hussain2006spot} and \citep{kaminski2007most}).  As a cataclysmic variable progenitor, with an unexpectedly massive red dwarf primary and the most massive white dwarf in the Hyades \citep{o2001hubble}, studies of this star may have a considerable effect on understanding the mechanism behind type 1a supernovae, thereby influencing the calibration of the extragalactic distance scale and much of modern cosmology.  Note that the spectral bandpass of the HI-1A and HI-1B imagers both allow a component of blue light through \citep{bewsher2010determination} and the white dwarf secondary is estimated to contribute $6~\%$ of the flux in this part of the spectrum \citep{o2001hubble}.  Similar behaviour can be seen in \textit{STEREO} observations of RS CVn variables, e.g. SZ Psc, and shows that \textit{STEREO} is a useful resource for studies of similar variability that require photometry with both long and short baselines.

\subsubsection{Highly eccentric eclipsing binaries}

\begin{figure}
\centering
\resizebox{7cm}{!}{\includegraphics{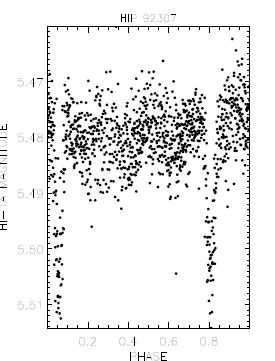}}
\caption{Lightcurve of the star HIP 92307, phase-folded on a period of $4.7911$ days.  \textbf{$|e \times \cos \omega| = 0.500 \pm 0.020$}.}
\label{fig24}
\end{figure}

The most eccentric undiscovered eclipsing binary in the sample is HIP 92307, with \textbf{$|e \times \cos \omega| = 0.500 \pm 0.020$} (Figure \ref{fig24}).  Both primary and secondary eclipses are about 30~mmag deep, indeed it is not clear which are the primary and secondary eclipses.  This star is listed as spectral type \texttt{A2V}.

\begin{figure}
\centering
\resizebox{7cm}{!}{\includegraphics{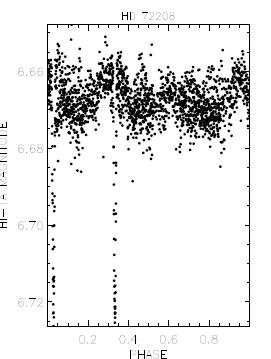}}
\caption{Lightcurve of the star HD 72208, phase-folded on a period of $22.0130$ days.  $|e \times \cos \omega| = 0.391 \pm 0.006$.}
\label{fig25}
\end{figure}

The longest period eclipsing binary in the sample is HD 72208, with $22.0130$ days between primary eclipses and $|e \times \cos \omega| = 0.391 \pm 0.006$ , making it one of the more eccentric as well (Figure \ref{fig25}).  This star, spectral type \texttt{B9p}, is a spectroscopic binary.  Both primary and secondary eclipses are about 60~mmag deep.

Possibly the longest period eclipsing binary so far observed in the \textit{STEREO}/HI-1A data may be HD 173770.  As there are multiple possible periods that can fit the lightcurve, it was not included in the sample.  The best-fitting period that has been found with the data available shows a period of $25.145$ days, which implies a minimum eccentricity $|e \times \cos \omega| = 0.108 \pm 0.012$.  The star shows some intrinsic regular variability as well, although the exact type has not been classified.  The host star is listed in \textsc{Simbad} as being of spectral type \texttt{B6V} and is a suspected triple system.  With the separation given for these components, it is unlikely either are responsible for the eclipses and this may instead be a quadruple system.  The lightcurve is included in Figure \ref{fig101} for comparison with the other measurably eccentric eclipsing binaries.

\subsubsection{NSV~7359: $\beta$ Cepheid showing the Kozai effect}

\begin{figure}
\centering
\resizebox{7cm}{!}{\includegraphics{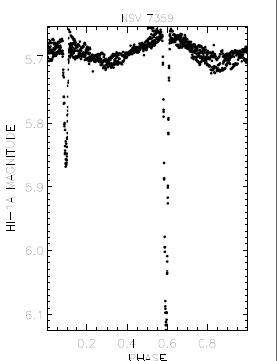}}
\caption{Lightcurve of the star NSV 7359, phase-folded on a period of $9.1999$ days.}
\label{fig28}
\end{figure}

A well-observed bright star, NSV 7359 shows relatively deep eclipses with a period of 9.1999~days that have not been recorded in \textsc{Simbad} (Figure \ref{fig28}).  This star is a known $\beta$ Cepheid, with an amplitude listed as 30~mmag.  Given the volume of observations, it is very surprising this variability has not previously been recorded.  It is in a fairly crowded area with many stars close by in the \textit{STEREO}/HI-1A field of view, yet none of these are sufficiently bright to contaminate its lightcurve to the extent of eclipses as deep as those observed.  The eclipses are therefore genuine and should help to further inform the parameters previously determined for this star.  Interestingly, this star is also shown in \textsc{Simbad} to be a spectroscopic binary with a period of $0.2872$ days, which would mean the eclipsing companion is a tertiary component and not the secondary.  The Kozai effect \citep{kozai1962secular} may have caused the inclination of the system to change so that eclipses are now visible.

\subsubsection{HR~7355: rotational variable or contact binary?}

\begin{figure}
\centering
\resizebox{7cm}{!}{\includegraphics{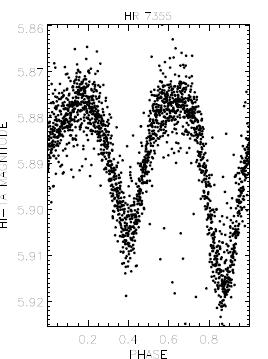}}
\caption{Lightcurve of the star HR 7355, phase-folded on a period of $0.5214$ days.}
\label{fig29a}
\end{figure}

One of the better-observed stars in the sample is HR 7355 (a.k.a. HIP 95408 and HD 182180).  \textit{STEREO}/HI-1A shows what appears to be a clear-cut case for an eclipsing binary as the phase-folding on 0.5214 days indicates in Figure \ref{fig29a}.  Other studies have noted this periodicity and associated it with a rotation period, however, rather than attributing it to a secondary companion, even the most recent which show the variability to be eclipse-like \citep{2010arXiv1009.4083O}.  We suggest that the photometry from \textit{STEREO} is enough to warrant a re-appraisal of the nature of this system.  It has been left in the sample as an eclipsing binary since from the photometry alone that is what it appears to be.  In the wider context it is clearly an unusual and fascinating, chemically peculiar star that continues to attract further study.

\subsection{False positives}

\noindent A small number of stars showing variability on a similar scale to that expected of an exoplanet or brown dwarf transit could be ruled out as such from photometric data alone, or in combination with other information known about the host star.

\textbf{Total eclipses just 30mmag deep were observed around either HIP 248 or HIP 247 (near-identical signals and in the same or adjacent pixels in \textit{STEREO}/HI-1A).  These stars are both of spectral type \texttt{F0}, however the eclipses include a substantial amount of time in the ingress and egress phases.  The eclipses are consistently about 30~mmag deep and $2.2604$ days apart (Figure \ref{fig21}).  This is a transit by a larger, stellar body rather than a brown dwarf but a low-mass star remains a possibility.}

\textbf{The star HD 222891 has similar features to HIP 248, however the eclipses are also too deep, at 40~mmag, with the host star of spectral type \texttt{F8}, to be due to a brown dwarf companion.}  It may nevertheless be a low-mass star of some interest.

Of all eclipse-like features extracted from the \textit{STEREO}/HI-1A database, the shallowest primary eclipses were 7~mmag deep around the star 18 Sgr.  This lightcurve appears more like a contact binary, although the raw lightcurve is also somewhat irregular and the behaviour might instead be due to rotational microvariability.  It was not considered conclusive enough to be labelled as an eclipsing binary and does not feature in the sample, although more data may help to clarify the lightcurve and its classification.

\textbf{NSV 1321 is the most marginal detection in the sample.  In order to verify whether the 10~mmag eclipses observed were genuine}, differential photometry was carried out, however this appeared to introduce variability from a neighbouring star - it is located in a crowded field with several stars showing clear variability and others blended.  Although the star itself is classified as variable, the NSV catalogue as shown in \textsc{Simbad} indicates this variability was considered doubtful by the catalogue's compilers.  With the host star of spectral type \texttt{B8V} this would not be an exoplanet \textbf{signal and is most likely a grazing eclipse or stellar companion}.  Although the possibility of microvariability or a blend with a neighbouring star cannot be wholly ruled out, none of the neighbouring stars show a periodicity of $2.2663$ days so this was included in the sample.

Another star showing shallow eclipses that was ruled out from being a brown dwarf is II Cnc \textbf{(period from \textit{STEREO}/HI-1A of $1.0684$ days, eclipses 25~mmag deep)}.  This is listed in \textsc{Simbad} as a BY Draconis type variable of spectral type \texttt{G8V}, however there is no mention of it showing eclipses.  The shape of the lightcurve is indicative of either a contact binary or grazing stellar transits with secondary eclipses present.  The period is more representative of a contact binary.  The secondary may nevertheless be a low-mass star, making this a relatively unusual system for a contact binary.

%% file: prospects.tex
\section{\uppercase{Survey prospects}}
\label{sec:prospects}

\noindent Although the analysis thus far done has been restricted to recovering eclipsing binaries and searching for exoplanets, the prospects for recovering variables of different types has been investigated.

The prospects for exoplanet extraction are summarised in Figure \ref{fig27} for the declination strips away from the Galactic Plane for the ranges of periods and magnitudes shown.  This indicates reasonable prospects for finding a signal, if any exist around stars of about 7th magnitude or brighter but with a sharp fall-off for fainter stars, such that the probability of extraction is negligible for stars fainter than 8th magnitude from \textit{STEREO}/HI-1A data alone.  Note that the most significant \textbf{substellar candidate so far extracted is} magnitude 7.5.  For reference, the signal of HIP~247 and HIP~248 (magnitude 7.5, period $2.2604$ days) is 30~mmag deep detected with a significance of 5~$\sigma$ and the 30~mmag deep signal of HR~1750 (magnitude 6.5, period $3.3150$ days) is detected with a significance of 8~$\sigma$.  With the inclusion of \textit{STEREO}/HI-1B data the number of data points will more than double (this satellite is in an Earth-trailing orbit and stars remain in the field of view for longer) and the significance of extracted signals will increase, as will the potential for extracting faint signals.

For eclipsing binaries, Figures \ref{fig11}, \ref{fig14} and \ref{fig43} indicate that more may yet be found and Figure \ref{fig7} shows that the combination of the BLS and Lomb-Scargle methods is a good way of extracting their signals.  The difficulty here is that, from photometric data alone, it has not been possible to conclusively classify them as eclipsing binaries.  This problem largely affects contact binaries, which have lightcurves that might easily be confused with elliptical variables and short period pulsating stars.  The ability of the Lomb-Scargle method to extract contact binaries suggests it would be a useful tool for extracting other variables also.

Asteroseismological studies are likely to require that low-level variability be detected.  If a signal of 10~mmag, comparable to an exoplanet transit, is required, then Figure \ref{fig27} might also be an indicator of the prospects for extracting microvariable behaviour, however this would assume that the algorithm chosen to extract the microvariable signal is equally effective at discerning faint signals as the BLS.  Nevertheless, the prospects would appear good for the brighter stars (7th magnitude or brighter) and these are also the most amenable to high-resolution spectroscopic follow-up by other observatories.  Investigations of the signals extracted by the Lomb-Scargle method show that the presence of systematic sources of noise inhibits the detection of a test signal, however, so additional cleaning would be required.  \textbf{Whilst the observing cadence of 40~minutes will impair detailed studies of very short period variables, the ongoing examination of the data that is including \textit{STEREO}/HI-1B observations is nevertheless able to obtain very clear photometry from a number of known $\delta$ Scuti stars (e.g. DX Cet) and potentially detect new variables down to magnitude 11.}

%% file: conclusions.tex
\section{\uppercase{Conclusions}}
\label{sec:conclusions}

\noindent The \textit{STEREO}/HI-1A observations are clearly a valuable resource suitable for a wide range of variability studies.  With 263 eclipsing binaries extracted from the database, \textbf{122} of which have not previously been identified as such, there is the potential for the discovery of a number of new and interesting objects.  The area of the sky being covered and the brightness of the stars being observed are features unique to \textit{STEREO}/HI-1A and simultaneously allow for high quality follow-up observations of interesting objects.  There is the potential to discover transiting exoplanets (Figure \ref{fig27}), however this is at the very limit of the survey's sensitivity and would only be possible for the stars with the cleanest signals.  A 30~mmag transit around a magnitude 7.5 star is detectable with 5~$\sigma$ significance and a 30~mmag transit of a magnitude 6.5 star is detectable with 8~$\sigma$ significance from \textit{STEREO}/HI-1A data alone.  With the inclusion of data from the second satellite, \textit{STEREO}/HI-1B, this will improve, e.g.  from \textit{STEREO}/HI-1A data alone the 50~mmag transits of the magnitude 8.5 star BD~+03 ~263p are detected with 7~$\sigma$ significance but with the inclusion of \textit{STEREO}/HI-1B data they are detected with nearly 13~$\sigma$ significance.

  The analysis presented here may have some impact on stellar evolution studies, in particular relating to the proportion and distribution of eccentric eclipsing binaries.  As the newly-discovered eclipsing binaries are comparatively bright, the masses, radii and effective temperatures can be determined more accurately in follow-up observations and can therefore provide more stringent tests of stellar evolution models (e.g. \citep{torres2006eclipsing}, \citep{vandenberg2006victoria}, \citep{claret2002new} and \citep{stassun2006discovery}).  The new eclipsing binaries will also help to produce more accurate distances to their host stars and any stars those are associated with \citep{southworth2005eclipsing}.

  The potential to improve upon existing phase coverage, especially relating to that obtained during eclipses, and times of maxima and minima in a variability cycle can help to refine parameters for these important stars.  In some cases, the capability of \textit{STEREO} to deliver good photometry over timescales of days and weeks in a single pass, combined with repeated observations after about a year, are ideal for observing long-term changes in variability.  This is of particular importance in studies of magnetic activity cycles (eg. V471 Tau) and particular classes of variables, such as RS CVn stars (eg. SZ Psc).

\begin{figure}
\centering
\resizebox{7cm}{!}{\includegraphics{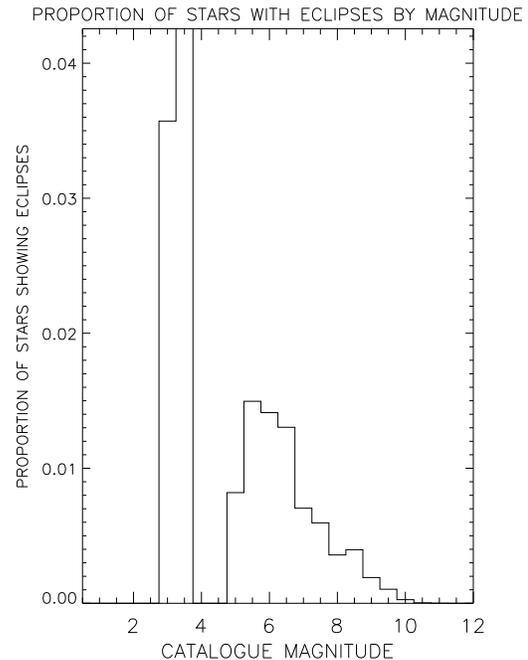}}
\caption{The proportion of stars of different magnitudes that have been found to be eclipsing.  \textbf{The small number of stars brighter than magnitude $4.5$ in the \textit{STEREO}/HI-1A field of view is responsible for the shape at the bright end of this graph, with only three EBs observed.}}
\label{fig43}
\end{figure}

\begin{figure}
\centering
\resizebox{7cm}{!}{\includegraphics{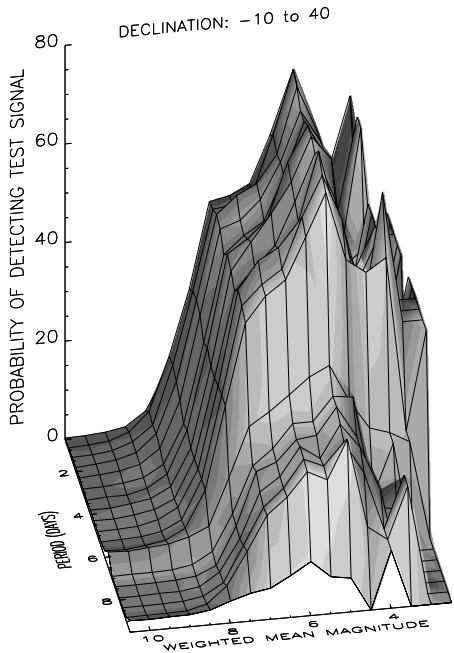}}
\caption{The percentage of stars where a 10~mmag test eclipse superimposed on the real lightcurve was successfully returned as the strongest signal by the BLS algorithm.  Each epoch of each star was examined separately and only those that returned the test signal within $0.04$ days were considered successes.  Detection of a harmonic was not considered a success.  This graph thus represents the chance of detecting a transiting `Hot Jupiter' for stars of a given magnitude for the range of periods shown.  The stars tested were those with a declination between -10 degrees and 40 degrees and having a weighted mean magnitude within $0.05$ magnitudes of the points shown.  The trial periods ranged from 0.5 to 9.5 days in half-day steps.  The dip in success rates at 6.5 days is the result of many stars having parts of their lightcurve masked out to avoid contamination by solar activity, whilst the dip at 9.5 days is the result of fewer stars remaining in the field of view for long enough for a period of this length to be checked.  The small increases in success rates for fainter stars at these periods are due to artificial trends associated with the length of an epoch, which also contributes to an increase in the number of false positives for brighter stars although this number is small compared to the decreases observed for the above reasons.  The best \textbf{substellar candidate so far extracted has a period of 2.4238~days and a host star of magnitude 7.5: a 25~mmag transit} is detectable with 5~$\sigma$ significance at this level.}
\label{fig27}
\end{figure}

The eclipsing binaries newly discovered by this analysis show a tendency to possess shallower transits than those previously known (Figure \ref{fig16}), partly due to the quality and cadence of the photometry.  With many of the host stars being of early spectral types, late \texttt{B} and early \texttt{A}, it is possible that many of these systems may be comparatively young (Figure \ref{fig5}).  A small but significant proportion are in eccentric orbits (Figures \ref{fig4} and \ref{fig4b}), supporting recent observations by \textit{Kepler} \citep{2010arXiv1006.2815P}.  Some particularly interesting systems have been identified for potential follow-up observations, including one good brown dwarf candidate (HD 213597, Figure \ref{fig20}).  The proportion of stars of different magnitudes found to be eclipsing indicates that more may yet be found as Figure \ref{fig43} shows decreasing numbers observed with fainter stars, whereas the real number should be a constant proportion of the population.

Evaluation of the algorithms used to extract eclipsing binaries suggests that, if exoplanets alone are the target, then the \textbf{Box-Least-Squares} (BLS) algorithm is capable of effectively extracting them (Figure \ref{fig27}).  For a more general search for eclipsing binaries, however, the combination of the BLS along with the Lomb-Scargle periodogram analysis will be required in order to extract a high proportion of these objects (Figure \ref{fig7}).  The Stetson index can be used as a measure of noise, as well as variability, although it does rate some Algol type binaries very highly so extreme values are not necessarily false positives (Figure \ref{fig11}).

Further research being done includes the development of a matched filter algorithm and follow-up observations of key targets has begun.  A further two eclipsing binaries showing eccentricity have been discovered (TYC~1422-1328-1, magnitude in \textit{STEREO}/HI-1A of 9.5, period 3.0999~days, $|e \times \cos \omega| = 0.175 \pm 0.025$ with primary and secondary eclipses both 0.1~mag deep and BD~+09~485, magnitude in \textit{STEREO}/HI-1A of 9.15, period 4.5293~days, $|e \times \cos \omega| = 0.125 \pm 0.025$ with primary eclipses 0.15~mag deep and secondary eclipses 0.1~mag deep), amongst numerous other short-period variables.  Periods have been extracted for a variety of stars showing sinusoidal-like variations in their lightcurve and some stars showing changes in behaviour on a year-by-year basis have been observed (eg. V~471 Tau, SZ Psc).  Additional findings will be reported in due course.

%% file: acks.tex
\section{\uppercase{Acknowledgements}}
\label{sec:acks}

\noindent The Heliospheric Imager (HI) instrument was developed by a collaboration that included the Rutherford Appleton Laboratory and the University of Birmingham, both in the United Kingdom, and the Centre Spatial de Li\'ege (CSL), Belgium, and the US Naval Research Laboratory (NRL), Washington DC, USA.  The \textit{STEREO}/SECCHI project is an international consortium of the Naval Research Laboratory (USA), Lockheed Martin Solar and Astrophysics Lab (USA), NASA Goddard Space Flight Center (USA), Rutherford Appleton Laboratory (UK), University of Birmingham (UK), Max-Planck-Institut f\"{u}r Sonnensystemforschung (Germany), Centre Spatial de Li\'ege (Belgium), Institut d'Optique Th\'eorique et Appliqu\'ee (France) and Institut d'Astrophysique Spatiale (France).  This research has made use of the \textsc{Simbad} database, operated at CDS, Strasbourg, France.  This research has made use of NASA's Astrophysics Data System.  This research has made use of the statistical analysis package \textsc{R} \citep{rproject}.  This research has made use of version 2.31 \textsc{Peranso} light curve and period analysis software, maintained at CBA, Belgium Observatory http://www.cbabelgium.com.  This research is funded by the Science and Technology Facilities Council (STFC).  This research made use of an IDL program to carry out a Lomb-Scargle periodogram analysis from Armagh Observatory at http://www.arm.ac.uk/\verb+~+csj/idl/PRIMITIVE/scargle.pro.
K. T. Wraight acknowledges support from an STFC studentship.  Many thanks to the referee for numerous thoughtful and constructive comments that have improved the quality of the paper.

%% file: appendix.tex
\appendix

\section{Eccentric eclipsing binaries}
\label{subsec:eeb}

\noindent The measurably eccentric eclipsing binaries are shown in Figure \ref{fig101}, along with the lightcurve of HD 173770, which although not included in the sample owing to an inability to uniquely determine its period due to the small number of eclipses observed, nevertheless appears to be an eccentric eclipsing binary.  The $|e \times \cos \omega|$ measured for this star is $0.108 \pm 0.012$ and the lightcurve shown is a phase-folding on a period of 25.145 days.  The values of $|e \times \cos \omega|$ measured for the other stars can be found in Table \ref{tab:1}.  The identities of the stars are as given in the table, however in the cases where nearby stars may have contaminated the signal or be the real source of eclipses it was often the case that the eclipses would feature in multiple lightcurves and the ones selected for analysis were those that were clearer and had deeper eclipses and the higher SDE.  Where this occurred, the table gives the name of a star that may be contaminating the lightcurve or be the real source of eclipses.

\section{Tables and figures}
\label{sec:tables}

\noindent The sample of 263 eclipsing binaries, with the various statistics that have been collected for each is given here.  Where data is not available, for example because secondary eclipses were not visible or the spectral type has yet to be determined, the entry is given as NA.  Right ascension and declination are given in degrees.  Primary eclipse depth and weighted mean magnitude are both in units of magnitudes.  Periods are in units of days.  The Box-Least-Squares (BLS) signal detection efficiency (SDE), the peak power from a Lomb-Scargle periodogram analysis and the Stetson variability index are all arbitrary scalar units.  The estimates of minimum eccentricity ($|e \times \cos \omega|$) and the estimated errors in those estimates are all scalar values.  The spectral type given is for the closest star to the listed co-ordinates from \textsc{Simbad}, the identity of which is given in the next column.  The alternative nearby star is the identity of a nearby star that might be the source of the eclipses, rather than the closest star (this frequently occurs due to the 70 arc-second per pixel resolution of \textit{STEREO}/HI-1A, as well as being due to nearby bright eclipsing binaries).  The Known EB is a binary flag set to 1 if either star is recorded as showing eclipses and 0 otherwise.  

\onecolumn

\begin{figure}
\centering
\resizebox{16cm}{!}{\includegraphics{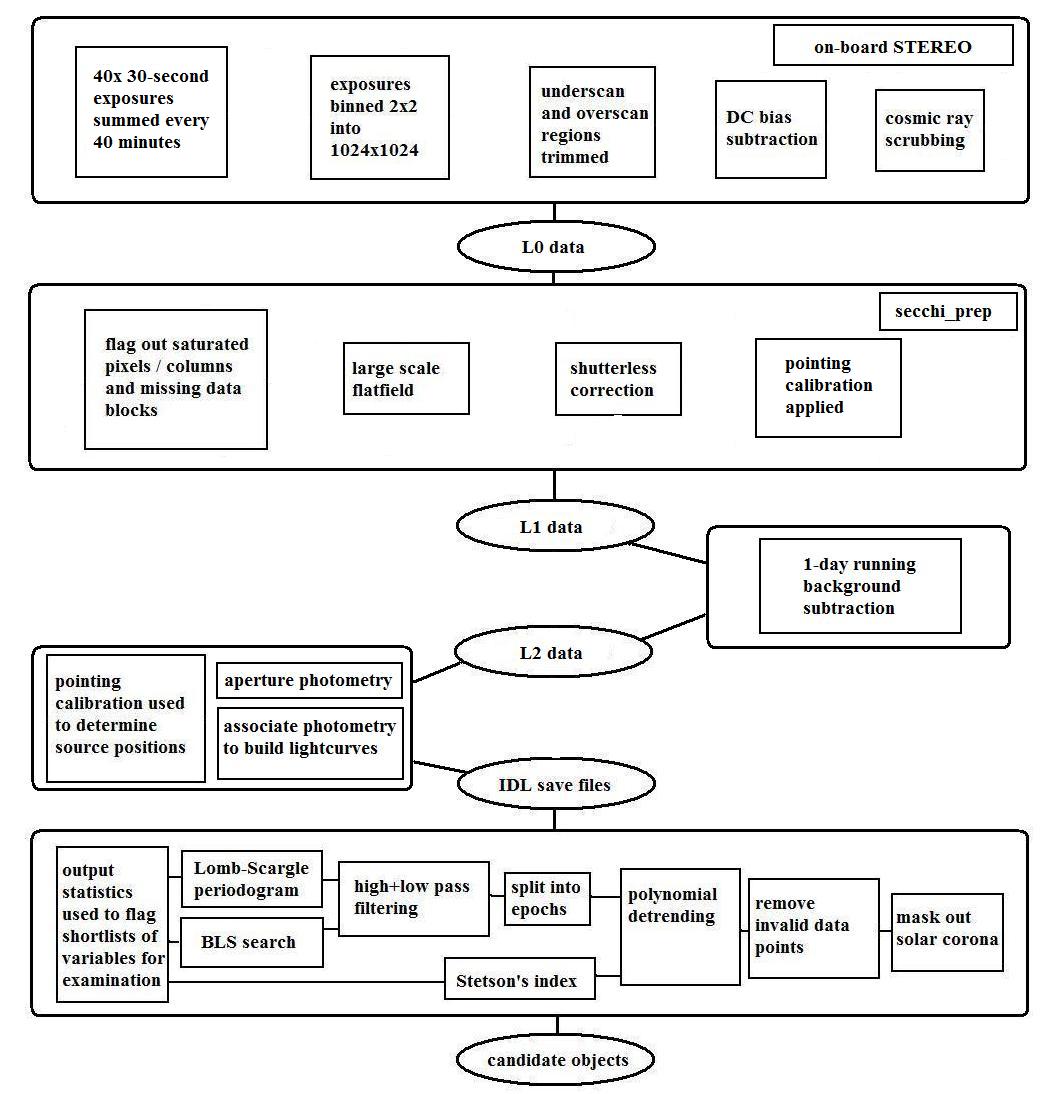}}
\caption{The \textit{STEREO}/HI-1A data analysis pipeline.}
\label{fig102}
\end{figure}

\begin{figure}
\begin{minipage}{16cm}
\centering
\resizebox{15cm}{!}{\includegraphics{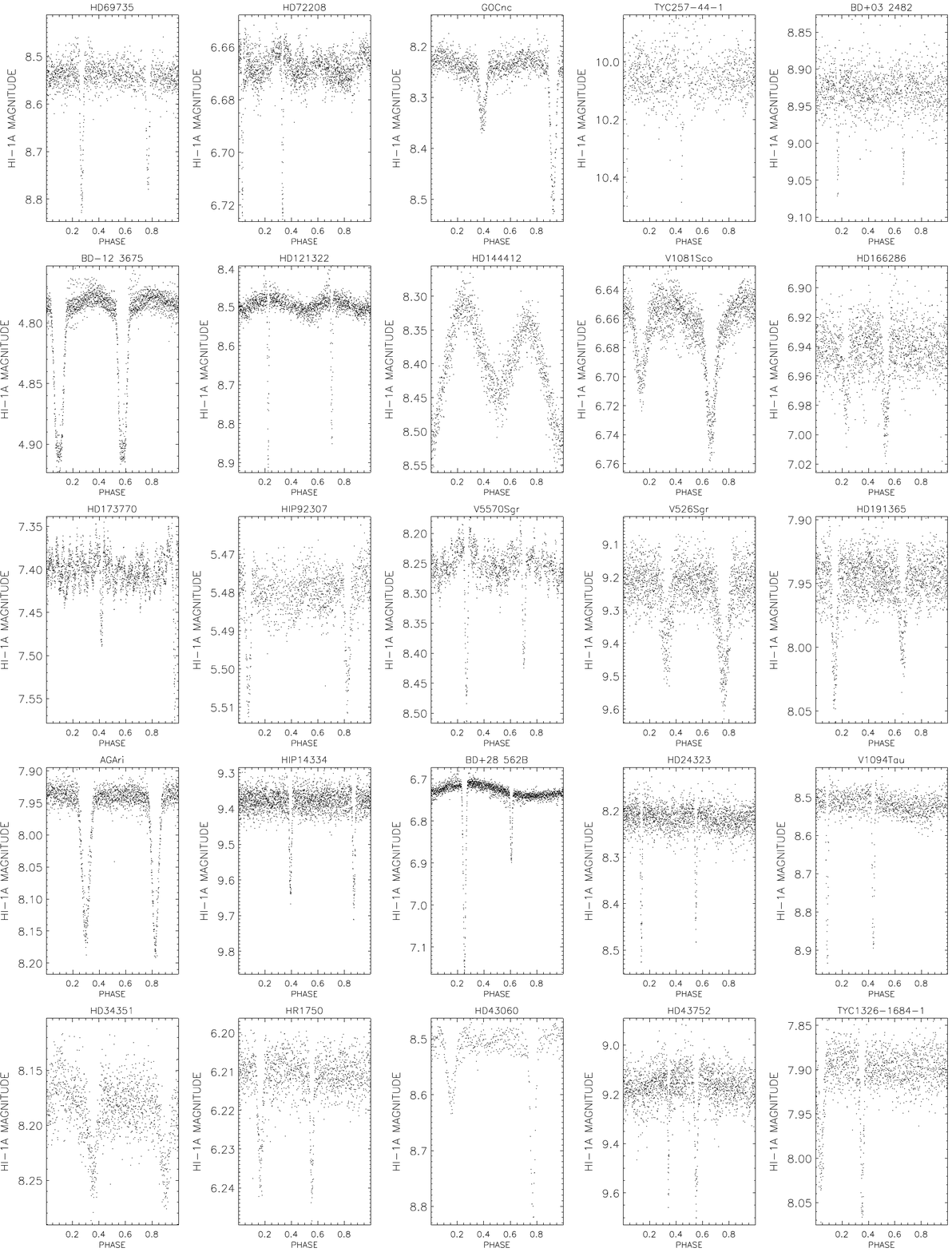}}
\caption{Lightcurves of all measurably eccentric EBs in the sample and the best-fitting lightcurve of HD~173770 (see Section~\ref{sec:results}).}
\label{fig101}
\end{minipage}
\end{figure}

\begin{figure}
\begin{minipage}{16cm}
\centering
\resizebox{15cm}{!}{\includegraphics{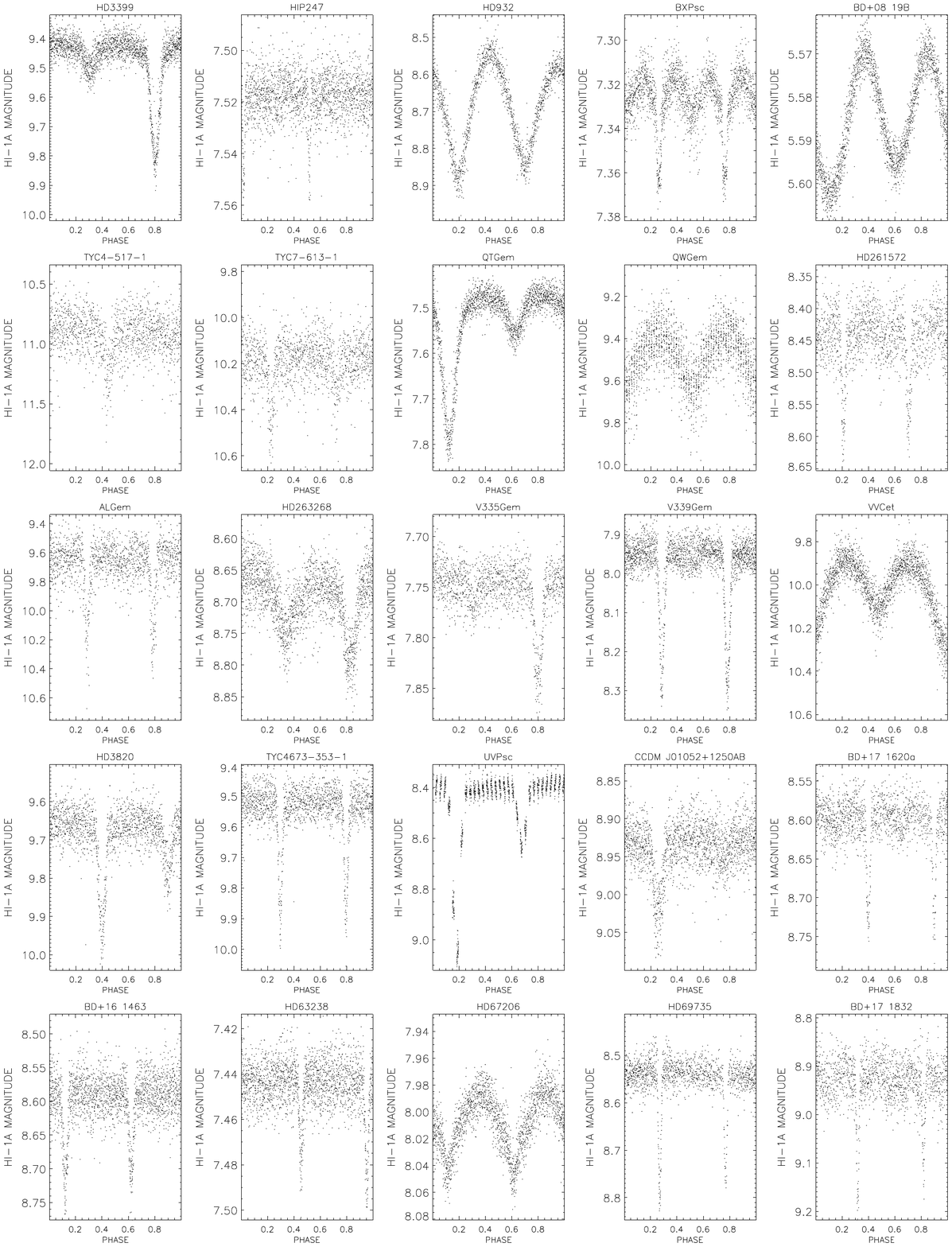}}
\caption{Lightcurves of all EBs in the sample including HD~173770 (see Section~\ref{sec:results}).  For Supporting Information only.}
\label{figB3}
\end{minipage}
\end{figure}

\begin{figure}
\begin{minipage}{16cm}
\centering
\resizebox{15cm}{!}{\includegraphics{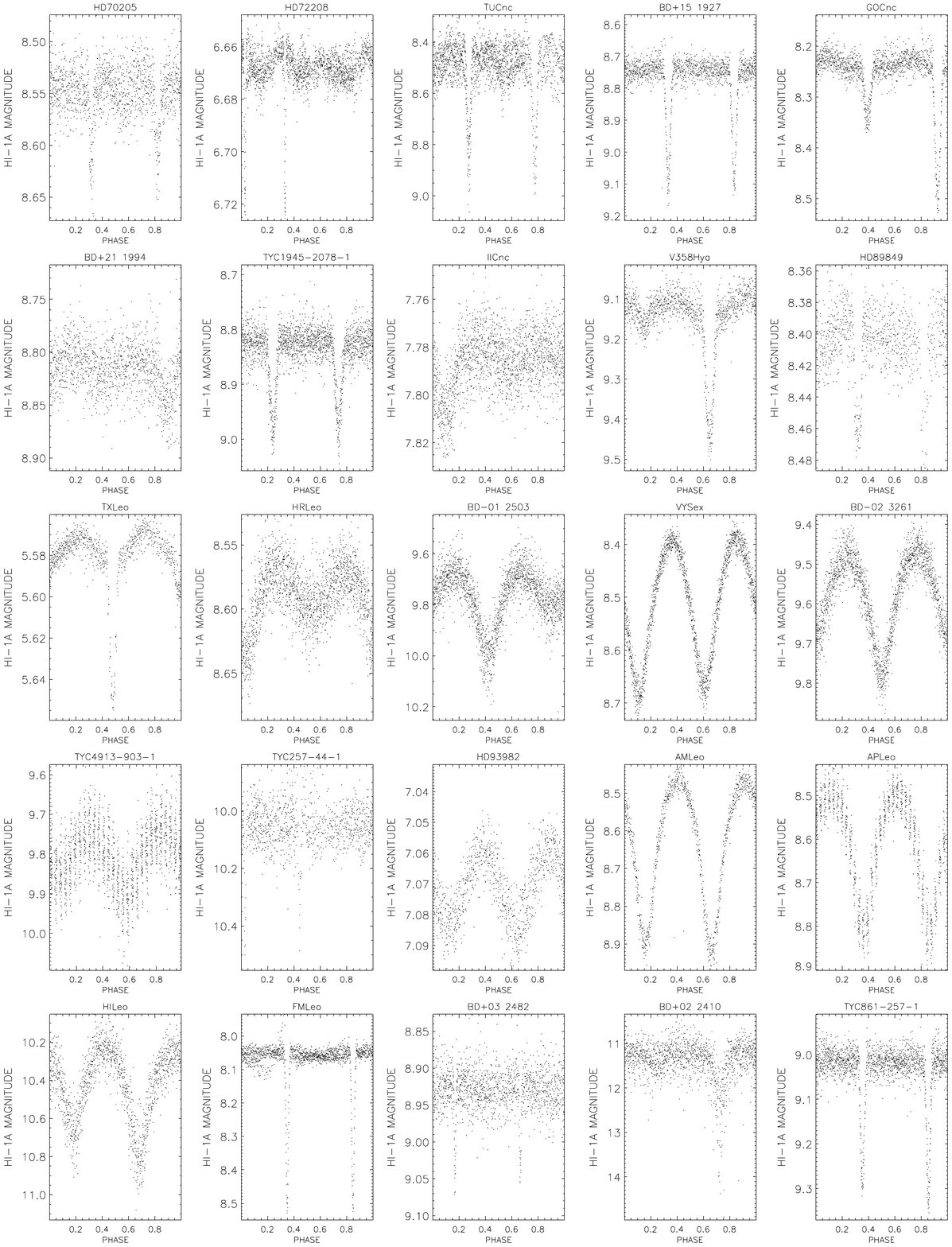}}
\caption{Lightcurves of all EBs in the sample including HD~173770 (see Section~\ref{sec:results}).  For Supporting Information only.}
\label{figB4}
\end{minipage}
\end{figure}

\begin{figure}
\begin{minipage}{16cm}
\centering
\resizebox{15cm}{!}{\includegraphics{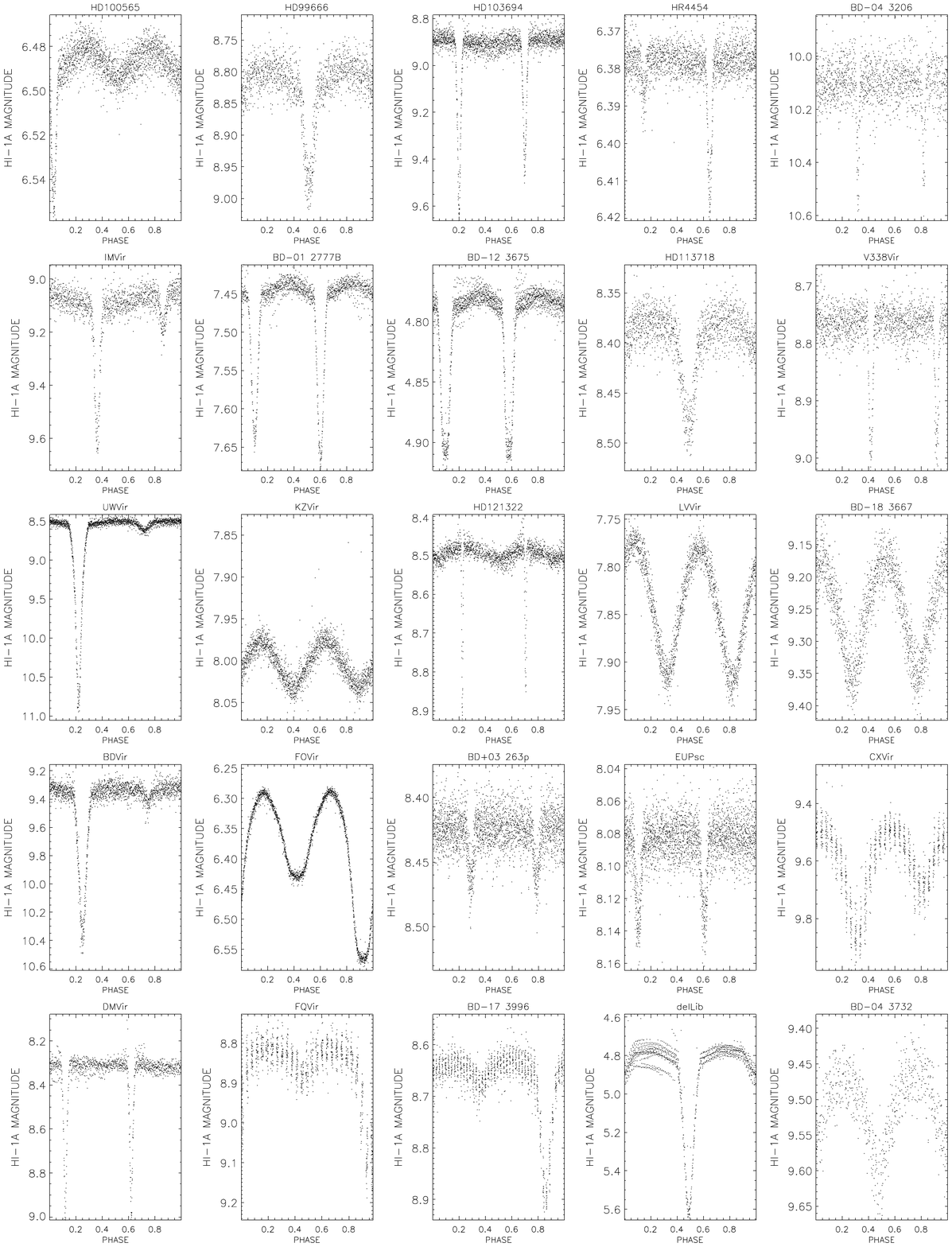}}
\caption{Lightcurves of all EBs in the sample including HD~173770 (see Section~\ref{sec:results}).  For Supporting Information only.}
\label{figB5}
\end{minipage}
\end{figure}

\begin{figure}
\begin{minipage}{16cm}
\centering
\resizebox{15cm}{!}{\includegraphics{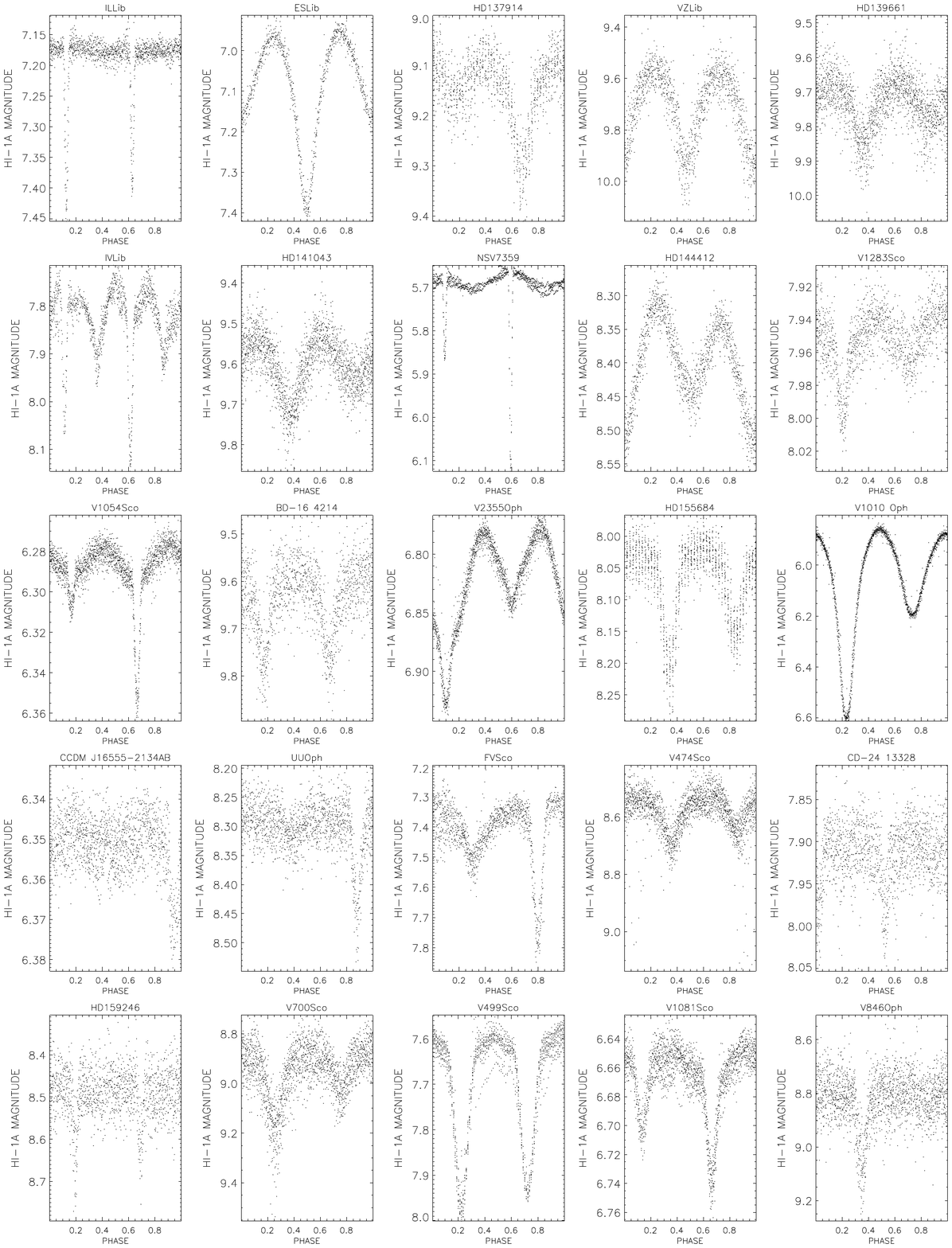}}
\caption{Lightcurves of all EBs in the sample including HD~173770 (see Section~\ref{sec:results}).  For Supporting Information only.}
\label{figB6}
\end{minipage}
\end{figure}

\begin{figure}
\begin{minipage}{16cm}
\centering
\resizebox{15cm}{!}{\includegraphics{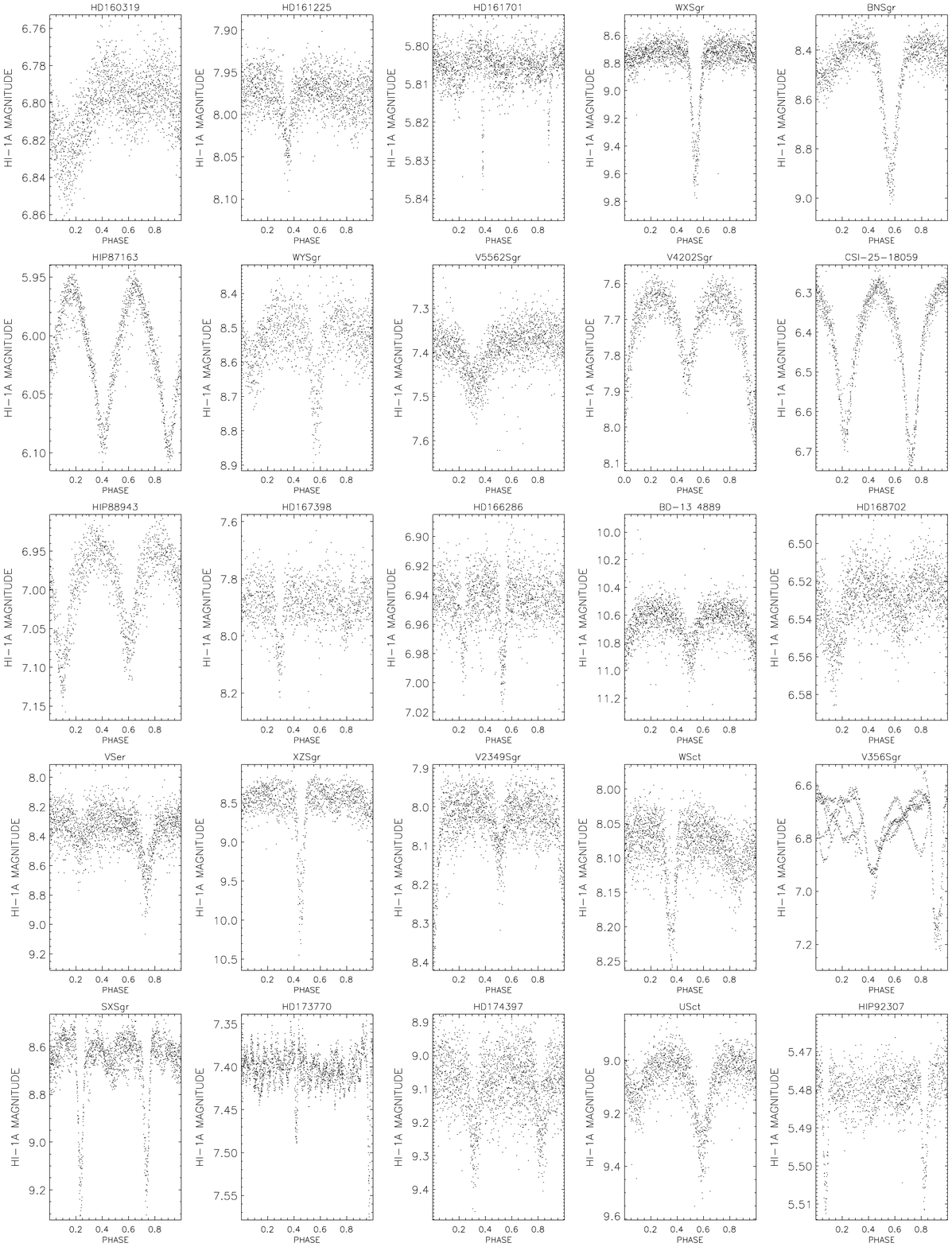}}
\caption{Lightcurves of all EBs in the sample including HD~173770 (see Section~\ref{sec:results}).  For Supporting Information only.}
\label{figB7}
\end{minipage}
\end{figure}

\begin{figure}
\begin{minipage}{16cm}
\centering
\resizebox{15cm}{!}{\includegraphics{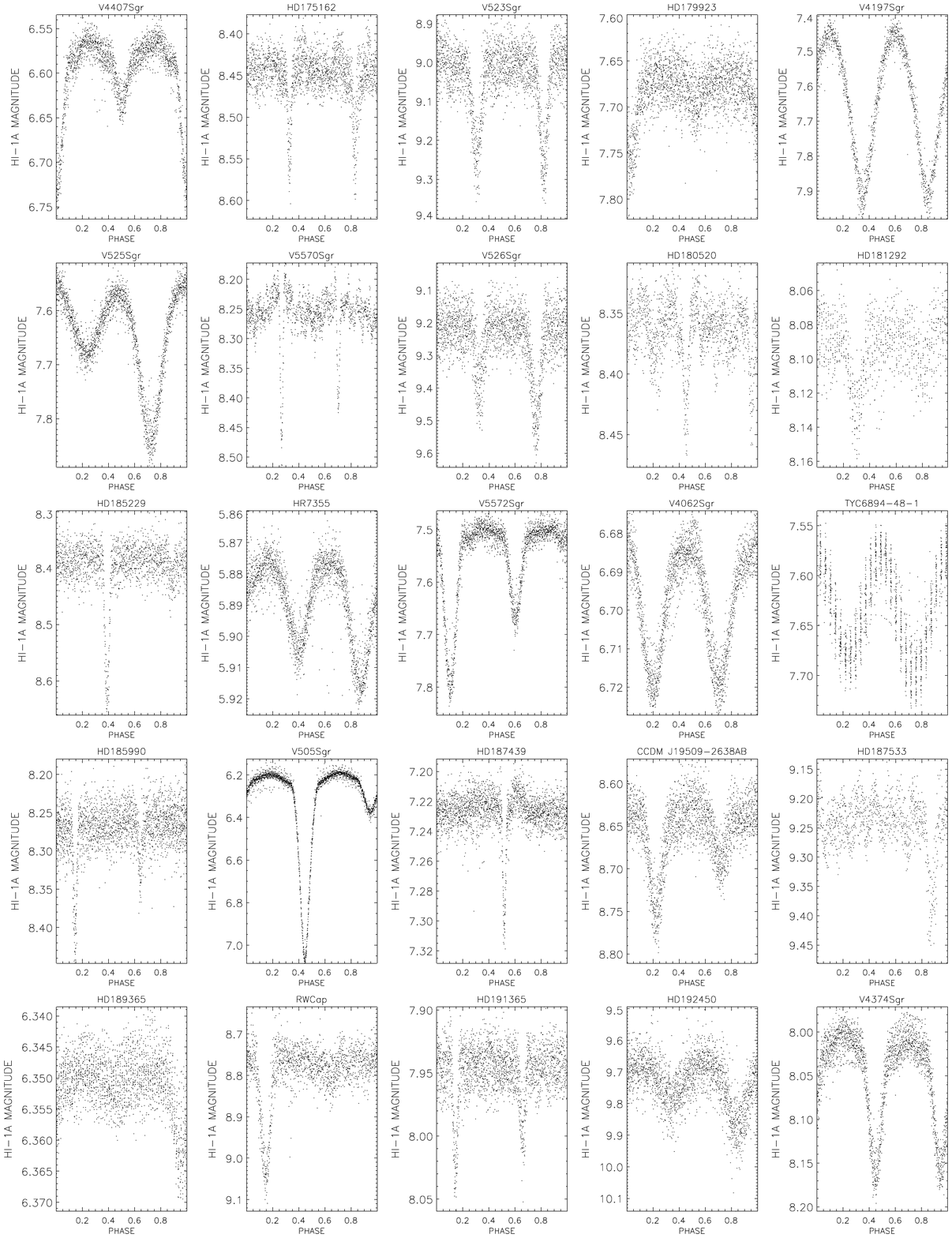}}
\caption{Lightcurves of all EBs in the sample including HD~173770 (see Section~\ref{sec:results}).  For Supporting Information only.}
\label{figB8}
\end{minipage}
\end{figure}

\begin{figure}
\begin{minipage}{16cm}
\centering
\resizebox{15cm}{!}{\includegraphics{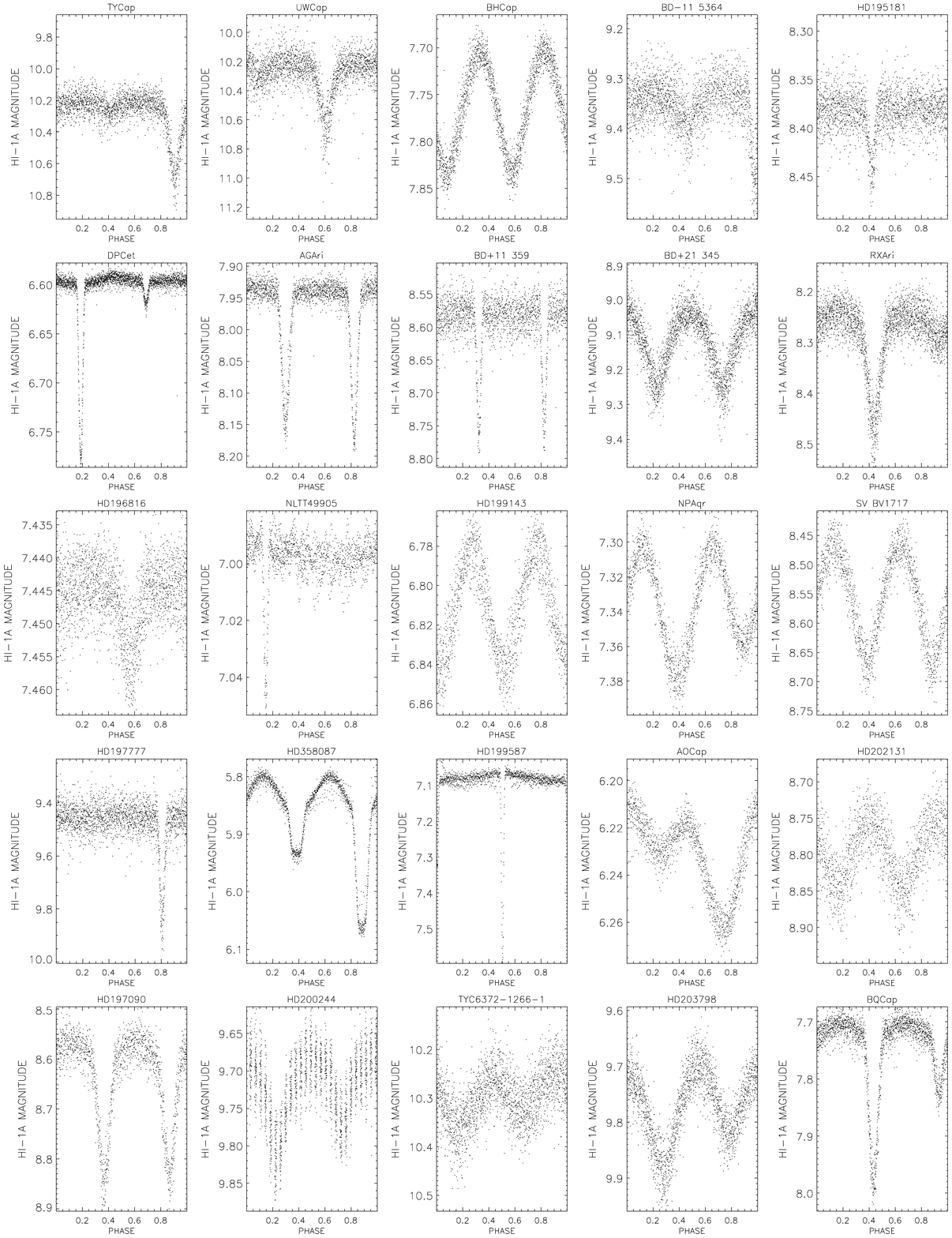}}
\caption{Lightcurves of all EBs in the sample including HD~173770 (see Section~\ref{sec:results}).  For Supporting Information only.}
\label{figB9}
\end{minipage}
\end{figure}

\begin{figure}
\begin{minipage}{16cm}
\centering
\resizebox{15cm}{!}{\includegraphics{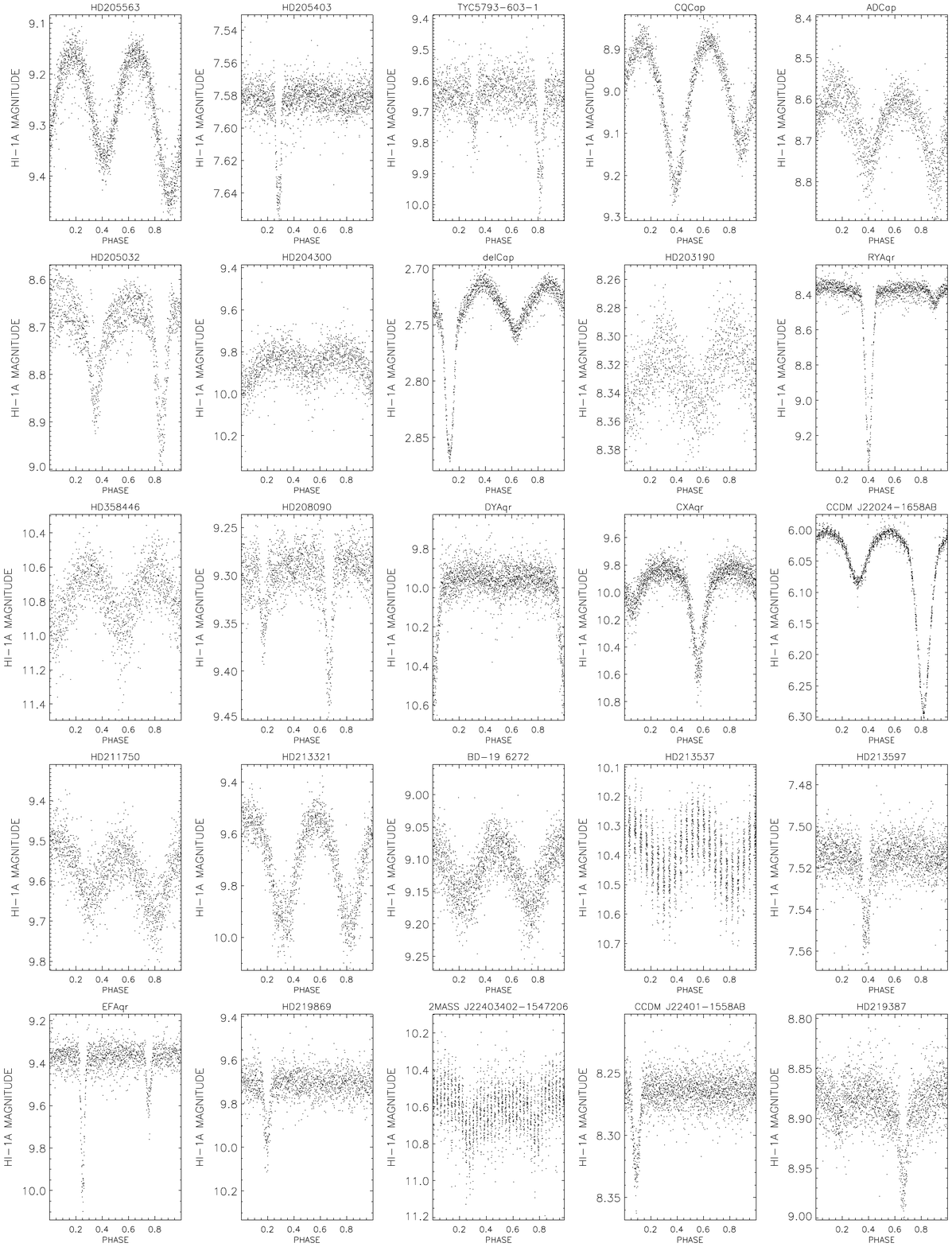}}
\caption{Lightcurves of all EBs in the sample including HD~173770 (see Section~\ref{sec:results}).  For Supporting Information only.}
\label{figB10}
\end{minipage}
\end{figure}

\begin{figure}
\begin{minipage}{16cm}
\centering
\resizebox{15cm}{!}{\includegraphics{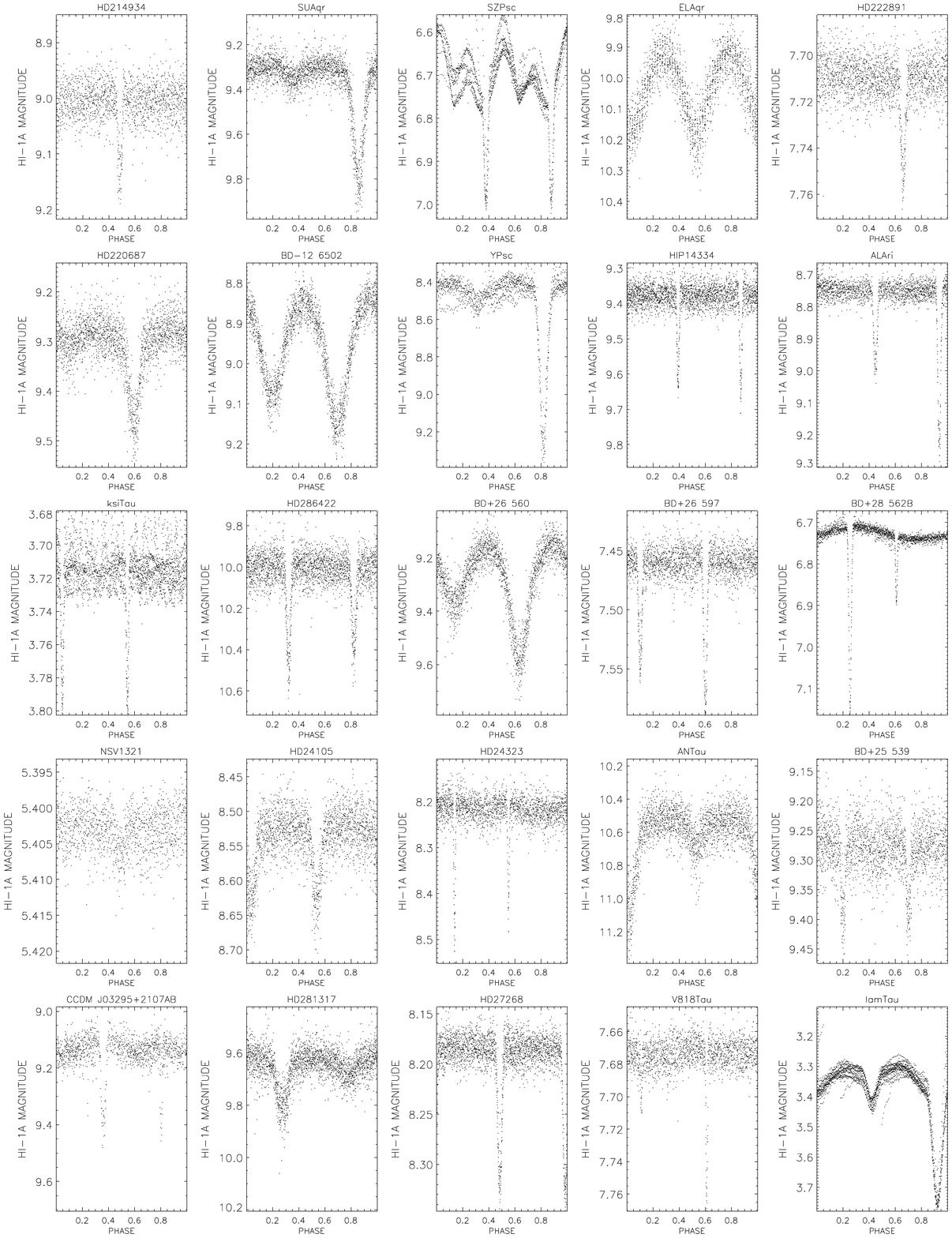}}
\caption{Lightcurves of all EBs in the sample including HD~173770 (see Section~\ref{sec:results}).  For Supporting Information only.}
\label{figB11}
\end{minipage}
\end{figure}

\begin{figure}
\begin{minipage}{16cm}
\centering
\resizebox{15cm}{!}{\includegraphics{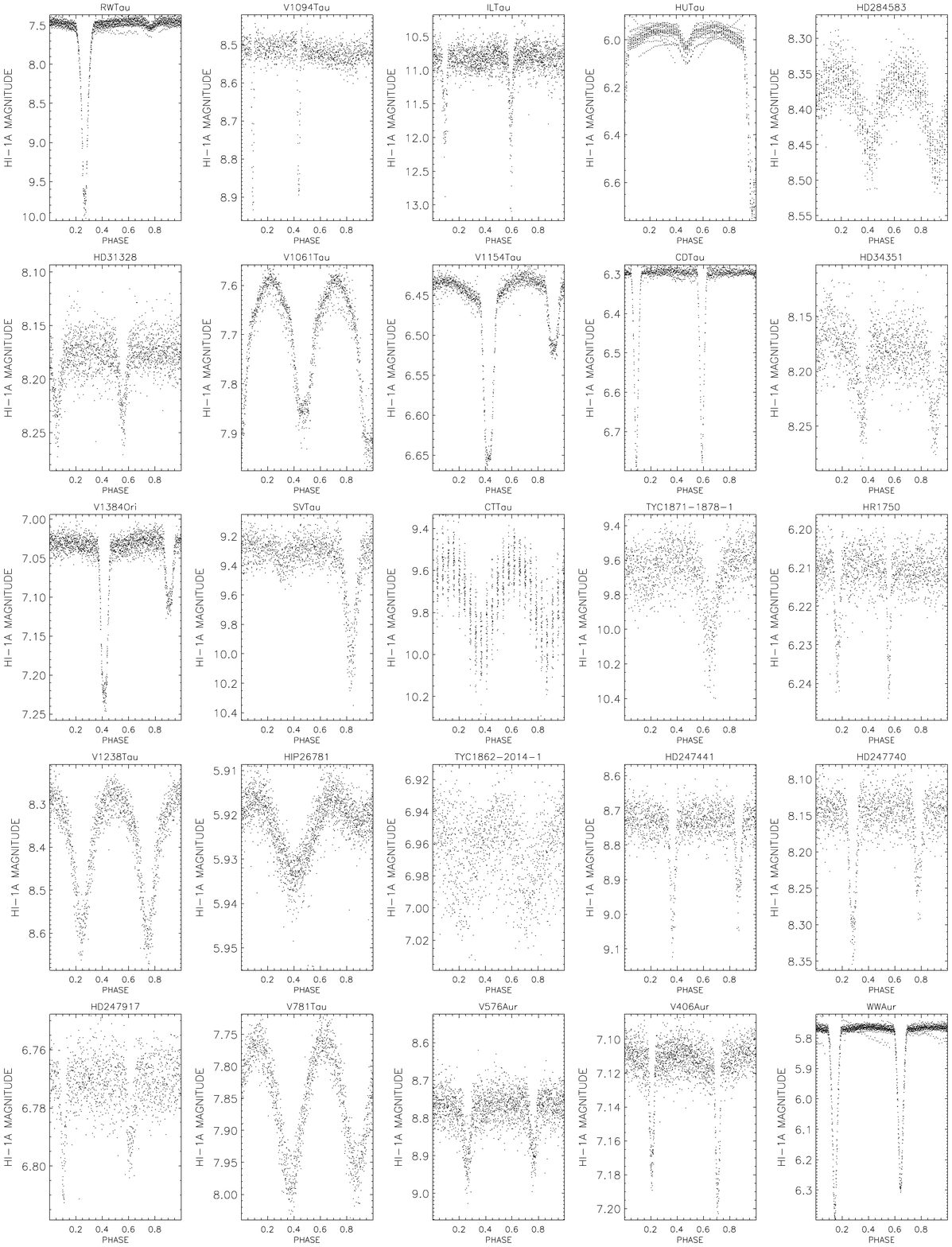}}
\caption{Lightcurves of all EBs in the sample including HD~173770 (see Section~\ref{sec:results}).  For Supporting Information only.}
\label{figB12}
\end{minipage}
\end{figure}

\begin{figure}
\begin{minipage}{16cm}
\centering
\resizebox{15cm}{!}{\includegraphics{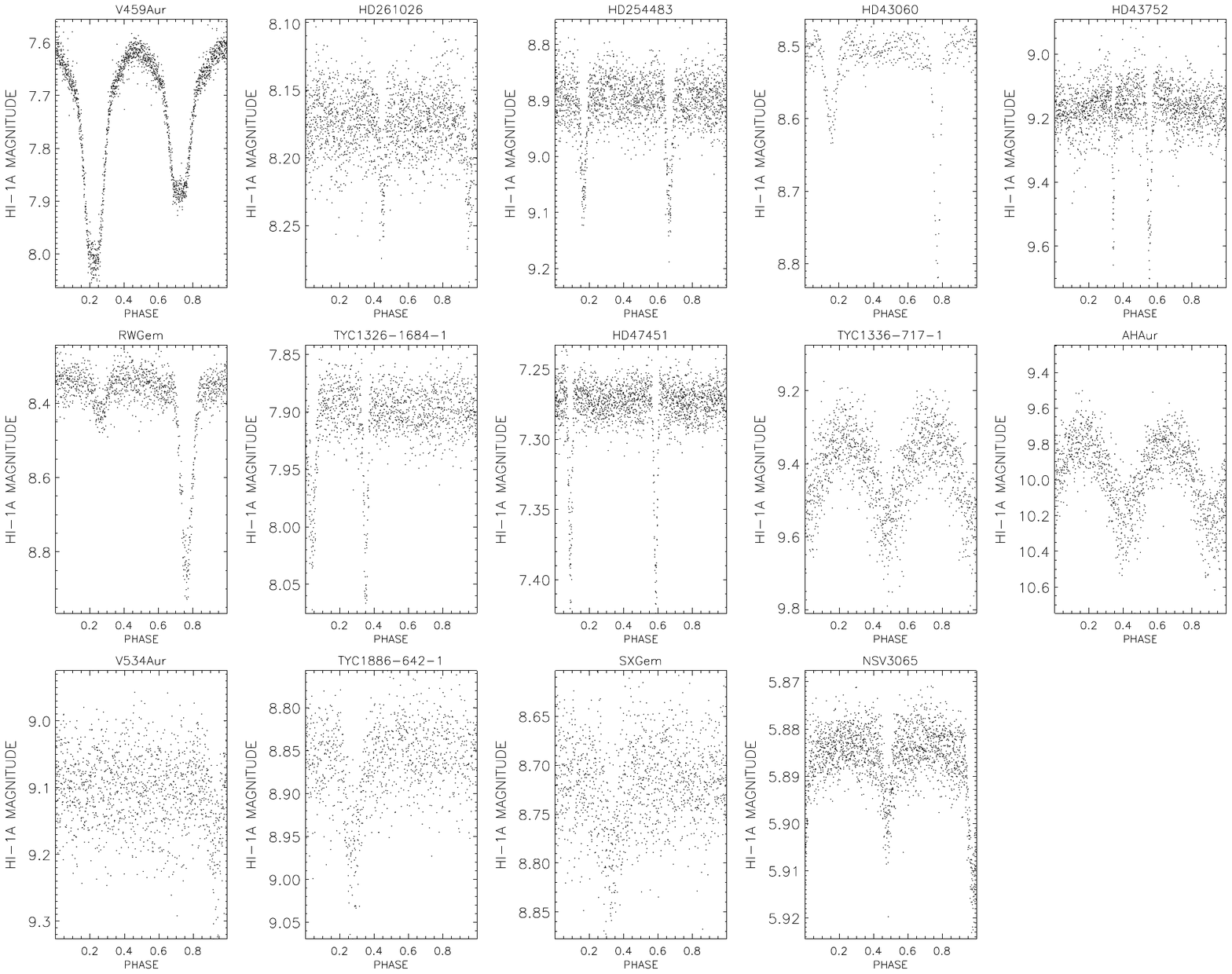}}
\caption{Lightcurves of all EBs in the sample including HD~173770 (see Section~\ref{sec:results}).  For Supporting Information only.}
\label{figB13}
\end{minipage}
\end{figure}

\begin{landscape}
\begin{longtable}{|p{1.0cm}|p{1.0cm}|p{0.9cm}|c|p{1.3cm}|p{0.8cm}|p{1.4cm}|p{1.0cm}|p{1.0cm}|p{1.0cm}|p{0.9cm}|p{1.4cm}|p{1.4cm}|p{0.8cm}|}
\caption{263 eclipsing binaries found from \textit{STEREO}/HI-1A data.} 
\protect\label{tab:1} \\
\hline
RA & DEC & Primary eclipse depth & Period & Weighted mean magnitude & BLS SDE & Stetson index & Peak Lomb-Scargle power & $|e \times \cos \omega|$ & $\pm$ error in $|e \times \cos \omega|$ & Spectral type & Identity & Alternative nearby star & Known EB \\
\hline
58.306 & 30.1915 & 0.15 & 0.79165 & 9.6355 & 4.985 & 0.089959 & 163.27 & 0.014 & 0.05 & A0 & HD281317 & NA & 0 \\
99.8127 & 31.5917 & 0.06 & 4.763 & 8.1252 & 4.0087 & 0.095906 & 83.309 & NA & NA & F5 & HD261026 & AKAur & 1 \\
92.5473 & 30.0289 & 0.15 & 0.9815 & 8.7237 & 4.9764 & 0.10818 & 111.68 & 0.002 & 0.02 & A0 & V576Aur & NA & 1 \\
94.4646 & 32.5046 & 0.07 & 2.7361 & 7.0957 & 5.7669 & 0.12152 & 102.39 & 0.008 & 0.024 & A3 & V406Aur & NA & 1 \\
98.1133 & 32.4549 & 0.55 & 2.5250 & 5.7419 & 6.1559 & 95.241 & 101.2 & 0.03 & 0.04 & A3m & WWAur & NA & 1 \\
98.9069 & 32.5768 & 0.4 & 1.065 & 7.7216 & 4.163 & 4.0988 & 272.15 & 0.00 & 0.036 & B8 & V459Aur & NA & 1 \\
101.605 & 31.4128 & 0.13 & 4.6835 & 8.6636 & 3.1958 & 0.10923 & 189.26 & 0.025 & 0.07 & G5 & HD263268 & HD263212 & 0 \\
33.8366 & 22.5697 & 0.2 & 1.02775 & 8.3924 & 5.9916 & 1.1283 & 133.77 & 0.01 & 0.03 & F2V & RXAri & NA & 1 \\
37.7781 & 22.1983 & 0.25 & 0.3963 & 9.1747 & 3.7634 & 0.1098 & 294.49 & 0.00 & 0.06 & F8 & BD +21 345 & NA & 0 \\
50.8846 & 25.8994 & 0.1 & 1.30555 & 9.2946 & 8.3623 & 0.034856 & 81.708 & NA & NA & F5 & BD +25 539 & NA & 0 \\
52.3671 & 21.1134 & 0.2 & 4.98975 & 9.1 & 4.5718 & 0.1323 & 40.661 & NA & NA & F2 & CCDM J03295 +2107AB & NA & 0 \\
52.2775 & 27.4136 & 0.4 & 1.0 & 9.2641 & 4.1739 & 0.59102 & 268.03 & 0.01 & 0.036 & F2 & BD +26 560 & NA & 1 \\
55.0698 & 26.8486 & 0.12 & 6.114 & 7.4839 & 6.2441 & 0.075213 & 53.906 & 0.00 & 0.02 & B9 & BD +26 597 & NA & 0 \\
55.1605 & 28.7716 & 0.45 & 8.1665 & 6.7556 & 7.38 & 3.8155 & 34.838 & 0.294 & 0.006 & A2V & BD +28 562B & HD22766 & 1 \\
57.0867 & 23.4213 & 0.01 & 2.2663 & 5.3929 & 2.764 & 0.0013492 & 227.86 & NA & NA & B8V & NSV1321 & NA & 0 \\
57.8397 & 25.6354 & 0.15 & 0.63145 & 8.4441 & 4.8421 & 0.073541 & 138.34 & NA & NA & G5 & HD24105 & NA & 0 \\
58.2809 & 25.5354 & 0.3 & 6.53065 & 8.2552 & 4.8853 & 0.21381 & 35.026 & 0.18 & 0.006 & A0 & HD24323 & SXTau & 1 \\
59.0473 & 29.523 & 0.7 & 1.6144 & 10.2486 & 6.14114 & 0.25888 & 172.29 & 0.01 & 0.06 & A3 & ANTau & NA & 1 \\
69.5660 & 20.6847 & 0.75 & 2.0555 & 5.8251 & 7.189 & 100.44 & 69.73 & 0.01 & 0.03 & B8V & HUTau & NA & 1 \\
60.9763 & 28.126 & 2.3 & 2.7778 & 7.9245 & 6.5421 & 35.085 & 73.255 & 0.00 & 0.035 & B8Ve & RWTau & NA & 1 \\
61.8078 & 29.309 & 1.2 & 5.3604 & 10.9617 & 6.4905 & 0.327 & 63.897 & NA & NA & NA & ILTau & NA & 1 \\
63.015 & 21.9474 & 0.4 & 9.0 & 8.5775 & 4.4708 & 0.21727 & 25.498 & 0.305 & 0.02 & G0 & V1094Tau & NA & 1 \\
69.7778 & 22.7121 & 0.1 & 0.7641 & 8.3943 & 4.2479 & 0.05307 & 166.57 & NA & NA & B5 & HD284583 & NA & 0 \\
73.9828 & 27.8354 & 0.06 & 1.2222 & 8.1691 & 4.9429 & 0.022643 & 109.84 & 0.005 & 0.04 & F8 & HD31328 & AQTau & 1 \\
74.7198 & 24.4957 & 0.38 & 0.6944 & 7.6702 & 4.8047 & 24.254 & 199.61 & NA & NA & B5 & V1061Tau & NA & 1 \\
76.4072 & 23.0611 & 0.23 & 1.7778 & 6.5015 & 4.9423 & 1.9962 & 96.445 & 0.02 & 0.02 & B5 & V1154Tau & NA & 1 \\
79.3798 & 20.1318 & 0.5 & 1.71755 & 6.3583 & 6.5897 & 13.857 & 51.274 & NA & NA & F7V & CDTau & NA & 1 \\
79.4109 & 20.5354 & 0.07 & 2.6419 & 8.2016 & 3.5365 & 0.015735 & 120.34 & 0.095 & 0.04 & B9 & HD34351 & NA & 0 \\
80.2472 & 27.9573 & 0.03 & 3.315 & 6.2125 & 5.0512 & 0.0094084 & 42.977 & 0.225 & 0.01 & B9IV & HR1750 & NA & 0 \\
85.3374 & 29.4874 & 0.02 & 0.674 & 5.9259 & 3.4628 & 0.044185 & 143.2 & 0.035 & 0.06 & B8IV & HIP26781 & HD37647 & 0 \\
85.5607 & 22.3714 & 0.3 & 1.1218 & 8.3929 & 3.2723 & 0.62529 & 187.94 & NA & NA & A0 & V1238Tau & NA & 1 \\
85.8421 & 24.1209 & 0.06 & 2.0045 & 6.9761 & 3.1702 & 0.05839 & 163.67 & NA & NA & A & TYC 1862 2014 1 & HD37979 & 0 \\
87.0687 & 20.7695 & 0.28 & 2.43055 & 8.6347 & 4.841 & 0.23099 & 68.82 & 0.005 & 0.04 & NA & HD247740 & HD38693 & 0 \\
86.764 & 28.226 & 0.25 & 4.2676 & 8.7384 & 4.80195 & 0.1336 & 51.331 & 0.00 & 0.02 & A0 & HD247441 & NA & 0 \\
87.3169 & 24.2109 & 0.035 & 2.6897 & 6.7581 & 3.9495 & 0.0052434 & 62.349 & 0.011 & 0.03 & A3 & HD247917 & HD38808 & 0 \\
87.5547 & 26.962 & 0.2 & 0.34508 & 7.8473 & 3.7413 & 0.25811 & 212.95 & 0.005 & 0.03 & G0 & V781Tau & NA & 1 \\
88.0221 & 28.0927 & 0.75 & 2.16665 & 9.35634 & 6.1146 & 0.81875 & 75.385 & 0.025 & 0.100 & B9 & SVTau & NA & 1 \\
89.8347 & 28.0275 & 0.5 & 1.3496 & 9.7615 & 3.8609 & 0.11148 & 83.227 & NA & NA & NA & TYC 1871 1878 1 & NA & 0 \\
89.7362 & 27.0541 & 0.45 & 0.66685 & 9.53182 & 4.096 & 0.18057 & 214.68 & 0.01 & 0.08 & B5 & CTTau & NA & 1 \\
93.4814 & 21.3804 & 0.2 & 3.1505 & 7.9754 & 5.5903 & 0.13942 & 64.155 & 0.365 & 0.03 & NA & TYC 1326 1684 1 & FTOri & 1 \\
90.3669 & 23.141 & 0.5 & 2.8655 & 8.5366 & 5.4542 & 1.2502 & 72.193 & 0.00 & 0.05 & B6V & RWGem & NA & 1 \\
94.8144 & 28.4398 & 0.45 & 10.8922 & 9.1588 & 3.4064 & 0.24165 & 61.223 & 0.575 & 0.005 & A2 & HD43752 & NA & 1 \\
95.0105 & 26.3303 & 0.1 & 2.7483 & 8.794 & 4.413 & 0.02671 & 41.01 & NA & NA & NA & TYC 1886 642 1 & NA & 1 \\
96.0032 & 20.4702 & 0.33 & 0.351 & 9.3721 & 3.5966 & 0.060442 & 172.6 & NA & NA & NA & TYC 1336 717 1 & NA & 0 \\
96.5206 & 27.999 & 0.5 & 0.4934 & 10.025 & 3.4552 & 0.22515 & 183.59 & NA & NA & G1 & AHAur & NA & 1 \\
96.5995 & 27.9456 & 0.1 & 2.1419 & 9.1058 & 3.9544 & 0.020826 & 38.65 & NA & NA & NA & V534Aur & NA & 1 \\
97.0887 & 20.5330 & 0.08 & 1.3668 & 8.6545 & 3.487 & 0.0321 & 33.79 & NA & NA & A0 & SXGem & NA & 1 \\
99.888 & 28.2632 & 0.03 & 1.8797 & 5.8952 & 5.0919 & 0.050436 & 112.29 & 0.005 & 0.04 & B7III & NSV3065 & NA & 0 \\
100.122 & 22.8955 & 0.15 & 2.04755 & 8.401 & 4.387 & 0.08155 & 77.07 & 0.01 & 0.02 & G0 & HD261572 & NA & 0 \\
102.692 & 29.4531 & 0.3 & 0.3581 & 9.5661 & 3.787 & 0.34595 & 206.45 & NA & NA & G0 & QWGem & NA & 1 \\
104.411 & 20.8924 & 0.7 & 1.3913 & 9.6675 & 4.403 & 0.145 & 44.64 & NA & NA & F6V & ALGem & NA & 1 \\
105.523 & 21.7981 & 0.09 & 1.7678 & 7.7499 & 5.9235 & 0.0376 & 70.74 & 0.005 & 0.04 & B9 & V335Gem & NA & 1 \\
109.714 & 29.1009 & 0.33 & 2.88035 & 8.3468 & 6.5489 & 2.2977 & 93.296 & NA & NA & F2 & V339Gem & NA & 1 \\
117.254 & 28.574 & 0.045 & 1.425 & 7.4336 & 5.884 & 0.02342 & 54.56 & 0.01 & 0.02 & F0 & HD63238 & STF 1144AB & 0 \\
125.384 & 21.5953 & 0.06 & 1.6054 & 8.5395 & 5.9816 & 0.006909 & 58.784 & NA & NA & A5 & HD70205 & NA & 0 \\
131.067 & 25.0757 & 0.16 & 1.19035 & 8.8014 & 5.6777 & 0.06538 & 109.62 & NA & NA & NA & TYC 1945 2078 1 & TYC 1945 1438 1 & 0 \\
133.458 & 26.9132 & 0.025 & 0.5342 & 7.8157 & 4.072 & 0.05791 & 95.53 & 0.015 & 0.200 & G8V & IICnc & NA & 0 \\
138.547 & 20.7034 & 0.05 & 0.6855 & 8.837 & 3.428 & 0.0020536 & 105.924 & NA & NA & G5 & BD +21 1994 & BD +21 1994p & 0 \\
16.3016 & 12.8321 & 0.11 & 1.62785 & 8.933 & 5.922 & 0.01696 & 114.13 & 0.015 & 0.05 & G0 & CCDM J01052 +1250AB & BD +12 131C & 0 \\
26.2229 & 19.8568 & 0.05 & 0.8463 & 8.0935 & 6.019 & 0.01728 & 106.64 & NA & NA & F0 & EUPsc & NA & 1 \\
39.128 & 12.1424 & 0.20 & 1.80245 & 8.6189 & 6.2292 & 0.059477 & 52.623 & NA & NA & F8 & BD +11 359 & NA & 1 \\
36.6123 & 12.8988 & 0.25 & 1.9631 & 7.9653 & 5.3368 & 0.32205 & 98.264 & 0.055 & 0.015 & B9 & AGAri & NA & 1 \\
40.6514 & 12.7355 & 0.45 & 3.7474 & 8.7802 & 6.514 & 0.2791 & 36.44 & 0.02 & 0.02 & F8 & ALAri & NA & 1 \\
59.8049 & 12.6382 & 0.45 & 2.32225 & 10.115 & 5.8665 & 0.15452 & 68.537 & NA & NA & A2 & HD286422 & NA & 0 \\
60.1701 & 12.4903 & 0.4 & 3.9529 & 3.4625 & 5.8633 & 1374.4 & 140.8 & 0.005 & 0.02 & B3V & lamTau & NA & 1 \\
64.7172 & 12.9248 & 0.13 & 2.06345 & 8.1948 & 6.2857 & 0.07836 & 72.31 & NA & NA & A0 & HD27268 & NA & 0 \\
64.4123 & 16.9479 & 0.08 & 5.6093 & 7.6616 & 5.5192 & 0.01604 & 41.01 & 0.005 & 0.01 & G8V & V818Tau & NA & 1 \\
89.2108 & 16.3551 & 0.19 & 3.7802 & 7.0654 & 5.9208 & 0.816 & 97.582 & 0.015 & 0.02 & B2V & V1384Ori & NA & 1 \\
94.2841 & 14.4566 & 0.20 & 1.51305 & 8.9091 & 5.7367 & 0.16955 & 68.938 & NA & NA & F8IV & HD254483 & GSC 00743 01238 & 0 \\
93.6901 & 13.2113 & 0.30 & 2.73 & 8.5239 & 6.3801 & 0.047377 & 42.956 & 0.23 & 0.02 & B3V & HD43060 & NA & 0 \\
99.8207 & 13.6861 & 0.13 & 4.55025 & 7.2869 & 7.6666 & 0.22473 & 46.076 & NA & NA & K0 & HD47451 & NA & 0 \\
102.069 & 14.5924 & 0.30 & 1.6199 & 7.4618 & 5.1785 & 2.5358 & 164.603 & 0.015 & 0.04 & A0 & QTGem & NA & 1 \\
110.796 & 15.9870 & 0.12 & 1.69445 & 8.6423 & 7.2193 & 0.072526 & 71.826 & NA & NA & A7V & BD +16 1463 & NA & 0 \\
114.053 & 16.9029 & 0.15 & 2.80005 & 8.5922 & 5.9309 & 0.028149 & 59.059 & NA & NA & K2 & BD +17 1620a & TXGem & 1 \\
121.8 & 10.147 & 0.07 & 1.8852 & 7.9903 & 5.1821 & 0.089832 & 276.58 & 0.015 & 0.04 & F0 & HD67206 & NA & 0 \\
124.75 & 19.976 & 0.25 & 5.3899 & 8.5718 & 6.4434 & 0.11706 & 35.476 & 0.014 & 0.01 & A3 & HD69735 & NA & 0 \\
125.993 & 17.1141 & 0.25 & 3.5792 & 8.9663 & 5.8786 & 0.054716 & 54.377 & NA & NA & A2 & BD +17 1832 & NA & 0 \\
133.232 & 14.8165 & 0.35 & 1.8554 & 8.7986 & 6.7368 & 0.29514 & 48.61 & NA & NA & A0 & BD +15 1927 & Cl* NGC 2632 KP 102662 & 0 \\
139.409 & 16.705 & 0.27 & 3.65 & 8.2735 & 5.3567 & 0.19792 & 69.855 & 0.08 & 0.03 & A0 & GOCnc & NA & 1 \\
168.984 & 14.7035 & 0.3 & 4.4507 & 9.0733 & 5.7337 & 0.1575 & 52.187 & 0.00 & 0.015 & NA & TYC 861 257 1 & NA & 0 \\
173.541 & 11.0239 & 0.032 & 2.2096 & 6.3907 & 6.3079 & 0.11581 & 61.79 & 0.009 & 0.02 & A2m & HR4454 & NA & 0 \\
0.765435 & 8.26129 & 0.03 & 2.2604 & 7.4773 & 5.387 & 0.010015 & 57.588 & NA & NA & F0 & HIP 247 & HIP 248 & 0 \\
1.36379 & 4.80649 & 0.4 & 2.1027 & 10.994 & 3.267 & 0.0691 & 61.1486 & NA & NA & NA & TYC 4 517 1 & NA & 0 \\
3.43064 & 6.9384 & 0.3 & 0.951 & 8.701 & 3.789 & 0.398 & 223.12 & 0.015 & 0.03 & A5 & HD932 & NA & 0 \\
4.47026 & 7.86604 & 0.04 & 3.86045 & 7.3076 & 4.012 & 0.01586 & 206.115 & NA & NA & A5 & BXPsc & NA & 1 \\
3.74693 & 8.81828 & 0.032 & 0.8417 & 5.565 & 2.976 & 0.0978 & 314.956 & 0.015 & 0.03 & A7 & BD +08 19B & UUPsc & 1 \\
2.19967 & 5.47131 & 0.3 & 2.3999 & 10.246 & 3.619 & 0.2143 & 59.166 & 0.015 & 0.03 & NA & TYC 7 613 1 & NA & 0 \\
19.2297 & 6.8117 & 0.7 & 0.8611 & 8.4686 & 6.1005 & 1.0814 & 100.71 & 0.015 & 0.02 & G5 & UVPsc & NA & 1 \\
28.7275 & 3.78465 & 0.05 & 1.2959 & 8.4435 & 5.5528 & 0.0033295 & 63.903 & NA & NA & NA & BD +03 263p & BD +03 263s & 0 \\
32.464 & 3.76935 & 0.18 & 2.3682 & 6.6093 & 8.0387 & 0.61444 & 57.542 & 0.005 & 0.015 & A2 & DPCet & NA & 1 \\
46.2098 & 8.11422 & 0.29 & 5.9971 & 9.3981 & 6.365 & 0.050353 & 41.737 & 0.13 & 0.01 & F8 & HIP 14334 & NA & 0 \\
51.7923 & 9.73267 & 0.07 & 3.5732 & 3.7049 & 7.3388 & 10.227 & 38.366 & NA & NA & B9Vn & ksiTau & NA & 1 \\
127.977 & 9.81409 & 0.06 & 22.013 & 6.6731 & 4.5847 & 0.083666 & 44.264 & 0.391 & 0.006 & B9p & HD72208 & NA & 0 \\
133.069 & 9.08854 & 0.4 & 5.5615 & 9.1992 & 3.5363 & 9.2918 & 64.636 & NA & NA & A2 & TUCnc & NA & 1 \\
143.758 & 4.75613 & 0.3 & 3.0095 & 9.009 & 6.2857 & 286470 & 106.75 & 0.005 & 0.04 & A2 & V358Hya & NA & 1 \\
155.541 & 6.21829 & 0.08 & 3.07835 & 8.3795 & 3.1369 & 0.0061672 & 42.305 & NA & NA & F8 & HD89849 & AG +06 1309 & 0 \\
158.759 & 8.65043 & 0.07 & 2.445 & 5.5845 & 5.2227 & 0.14123 & 65.099 & 0.025 & 0.03 & A2V & TXLeo & NA & 1 \\
162.333 & 2.57512 & 0.5 & 8.0319 & 10.053 & 6.7082 & 0.017853 & 32.705 & 0.183 & 0.01 & NA & TYC 257 44 1 & NA & 0 \\
162.738 & 3.59068 & 0.03 & 0.4325 & 7.0637 & 4.025 & 0.002475 & 125.72 & NA & NA & A2 & HD93982 & NA & 0 \\
165.545 & 9.89519 & 0.42 & 0.3658 & 8.6466 & 3.935 & 0.4847 & 145.74 & NA & NA & F8Vn & AMLeo & NA & 1 \\
166.271 & 5.15179 & 0.32 & 0.4306 & 8.5874 & 3.2822 & 0.37463 & 161.93 & 0.00 & 0.04 & F8V & APLeo & NA & 1 \\
168.070 & 1.31822 & 0.55 & 0.3116 & 10.391 & 3.703 & 0.0869 & 183.6 & 0.00 & 0.06 & NA & HILeo & NA & 1 \\
168.188 & 0.348007 & 0.47 & 6.7287 & 8.0948 & 7.2493 & 1.02896 & 28.736 & 0.004 & 0.006 & F8 & FMLeo & NA & 1 \\
169.662 & 2.82139 & 0.1 & 9.1784 & 8.959 & 5.097 & 0.0075 & 26.6756 & 0.015 & 0.01 & G0 & BD +03 2482 & BD +03 2483A & 0 \\
169.789 & 1.95757 & 1.1 & 1.29785 & 11.543 & 5.225 & 0.211 & 81.82 & NA & NA & F8 & BD +02 2410 & NA & 0 \\
179.134 & 7.29747 & 0.75 & 2.2392 & 8.937 & 5.7896 & 0.81757 & 59.004 & 0.005 & 0.006 & K2 & HD103694 & NA & 0 \\
198.095 & 2.65383 & 0.06 & 1.1318 & 8.0144 & 4.2689 & 0.17505 & 296.99 & 0.00 & 0.06 & F2 & KZVir & NA & 1 \\
202.446 & 1.09535 & 0.27 & 0.7756 & 6.3652 & 4.5741 & 12.042 & 273.66 & 0.015 & 0.05 & A2 & FOVir & CCDM J13298 +0106AB & 1 \\
338.136 & 1.58245 & 0.025 & 2.4238 & 7.493 & 6.3662 & 0.29712 & 79.855 & NA & NA & F0 & HD213597 & HD213598 & 0 \\
348.349 & 2.67544 & 0.27 & 3.96565 & 6.733 & 6.4122 & 7.8083 & 162.97 & NA & NA & K1 IV-V & SZPsc & NA & 1 \\
353.606 & 7.9246 & 0.9 & 3.7657 & 8.636 & 5.285 & 4.542 & 65.128 & NA & NA & A3V & YPsc & NA & 1 \\
9.2297 & -5.87404 & 0.35 & 3.4885 & 9.4566 & 6.3841 & 0.23564 & 138.12 & 0.005 & 0.08 & A5 & HD3399 & NA & 1 \\
13.9303 & -2.09399 & 0.4 & 0.5224 & 10.036 & 3.9492 & 0.18227 & 255.32 & 0.025 & 0.05 & A5 & VVCet & NA & 1 \\
10.2136 & -0.962273 & 0.3 & 1.6405 & 9.7072 & 5.9898 & 0.038305 & 93.877 & 0.005 & 0.03 & F8 & HD3820 & FASTT28 & 0 \\
11.5631 & -0.0386658 & 0.45 & 1.6900 & 9.6328 & 7.2307 & 0.093337 & 59.875 & NA & NA & NA & TYC 4673 353 1 & SDSS J004617.88 -000130.3 & 0 \\
160.922 & -1.09624 & 0.15 & 0.77895 & 9.9023 & 3.3872 & 0.039807 & 160.96 & 0.035 & 0.08 & NA & TYC 4913 903 1 & NA & 0 \\
162.624 & -2.6953 & 0.3 & 0.4434 & 8.5033 & 3.3536 & 0.50124 & 268.8 & 0.015 & 0.02 & F9.5V & VYSex & NA & 1 \\
164.775 & -3.54405 & 0.3 & 1.0713 & 9.5849 & 4.3149 & 0.17634 & 261.69 & 0.005 & 0.05 & NA & BD -02 3261 & NA & 0 \\
166.176 & -2.97221 & 0.075 & 1.0847 & 8.591 & 4.6747 & 0.010763 & 207.77 & 0.025 & 0.140 & F2 & HRLeo & NA & 1 \\
169.152 & -1.90306 & 0.3 & 0.6288 & 9.8227 & 4.279 & 0.15639 & 218.56 & 0.015 & 0.07 & NA & BD -01 2503 & NA & 0 \\
171.997 & -1.92144 & 0.18 & 1.0143 & 8.8186 & 6.0283 & 0.04405 & 83.213 & 0.005 & 0.100 & F2 & HD99666 & BD -01 2527B & 0 \\
173.589 & -5.5271 & 0.06 & 2.5670 & 6.5105 & 7.6481 & 0.06868 & 140.77 & NA & NA & A2 & HD100565 & NA & 0 \\
181.587 & -4.91402 & 0.4 & 2.7113 & 10.121 & 6.0989 & 0.036731 & 31.688 & NA & NA & NA & BD -04 3206 & NA & 0 \\
192.411 & -6.07913 & 0.5 & 1.3086 & 9.1087 & 6.6385 & 0.23859 & 55.119 & 0.015 & 0.024 & G5 & IMVir & NA & 1 \\
197.118 & -2.68878 & 0.2 & 2.7323 & 7.4267 & 6.0494 & 0.57602 & 93.207 & 0.001 & 0.009 & F8 & BD -01 2777B & HYVir & 1 \\
208.716 & -2.38036 & 0.42 & 6.6950 & 8.4903 & 6.0823 & 0.31388 & 24.374 & 0.045 & 0.01 & F8 & HD121322 & NA & 0 \\
220.288 & -5.09566 & 0.18 & 0.3846 & 9.4452 & 2.8371 & 0.0097679 & 66.192 & 0.015 & 0.150 & F8 & BD -04 3732 & NA & 0 \\
225.243 & -8.51894 & 0.85 & 2.3273 & 4.9170 & 6.3210 & 536.33 & 70.608 & 0.005 & 0.10 & B9.5V & delLib & NA & 1 \\
323.766 & -3.7349 & 0.06 & 2.4449 & 7.5587 & 6.7961 & 0.053739 & 48.11 & NA & NA & F5 & HD205403 & NA & 0 \\
328.78 & -9.79926 & 0.25 & 1.8636 & 9.6842 & 5.7501 & 0.053637 & 51.422 & 0.007 & 0.02 & K0V & TYC 5793 603 1 & NA & 0 \\
334.768 & -2.64167 & 0.5 & 2.1597 & 9.9316 & 6.7132 & 0.31543 & 90.597 & 0.04 & 0.12 & A0 & DYAqr & NA & 1 \\
338.933 & -0.692436 & 0.7 & 0.556 & 10.003 & 4.5452 & 1.5886 & 210.04 & 0.015 & 0.05 & F2pv & CXAqr & FASTT1583 & 1 \\
345.33 & -6.4376 & 0.6 & 2.8536 & 9.4475 & 5.8903 & 5.8698 & 44.337 & 0.01 & 0.02 & G0 & EFAqr & NA & 1 \\
349.842 & -8.87043 & 0.25 & 3.03095 & 9.8713 & 5.5423 & 0.03644 & 55.451 & NA & NA & F8 & HD219869 & G273 -14 & 0 \\
356.826 & -8.08669 & 0.28 & 0.4814 & 9.8908 & 4.3021 & 0.072444 & 248.93 & 0.015 & 0.07 & F & ELAqr & NA & 1 \\
356.162 & -8.84879 & 0.04 & 1.59495 & 7.7152 & 6.0264 & 0.0062216 & 57.745 & NA & NA & F8 & HD222891 & 2MASS J23443838 -0851082 & 0 \\
190.304 & -13.0246 & 0.13 & 3.14445 & 4.8099 & 5.0159 & 15.226 & 148.89 & 0.04 & 0.03 & NA & BD-12 3675 & VVCrv & 1 \\
196.428 & -12.37 & 0.1 & 1.581 & 8.3555 & 4.8238 & 0.025355 & 97.58 & 0.005 & 0.04 & F8 & HD113718 & NA & 0 \\
197.823 & -11.1059 & 0.22 & 2.99275 & 8.789 & 6.398 & 0.074939 & 46.158 & NA & NA & F5 & V338Vir & NA & 1 \\
198.836 & -17.4714 & 2.2 & 1.8108 & 8.8216 & 6.723 & 10.433 & 85.636 & 0.00 & 0.03 & A8 IV/V & UWVir & NA & 1 \\
201.669 & -16.1046 & 1.0 & 2.5486 & 9.5171 & 6.2785 & 2.1003 & 85.011 & 0.005 & 0.025 & A8V & BDVir & NA & 1 \\
203.192 & -17.759 & 0.14 & 0.4094 & 7.8799 & 3.4011 & 0.22645 & 277.84 & 0.025 & 0.03 & F6V & LVVir & NA & 1 \\
205.697 & -18.8854 & 0.2 & 0.4090 & 9.263 & 3.2808 & 0.050989 & 227.88 & 0.005 & 0.04 & NA & BD -18 3667 & PPM 717337 & 0 \\
211.968 & -11.1521 & 0.7 & 2.3347 & 8.4074 & 6.3465 & 1.9933 & 38.449 & NA & NA & F5 & DMVir & NA & 1 \\
211.626 & -15.5078 & 0.3 & 0.7496 & 8.8799 & 5.0803 & 0.17877 & 79.244 & 0.015 & 0.11 & A2III & FQVir & NA & 1 \\
211.207 & -18.3606 & 0.25 & 1.6936 & 8.7012 & 5.7877 & 0.14933 & 74.205 & 0.005 & 0.06 & NA & BD -17 3996 & AKVir & 1 \\
212.364 & -15.5816 & 0.35 & 0.7493 & 9.5831 & 4.2051 & 0.19179 & 121.97 & 0.025 & 0.06 & F5V & CXVir & NA & 1 \\
227.236 & -11.7905 & 0.25 & 2.88475 & 7.2047 & 5.8387 & 0.95864 & 35.836 & NA & NA & F2 & ILLib & NA & 1 \\
229.203 & -13.0392 & 0.4 & 0.883 & 7.0607 & 3.9639 & 5.2944 & 119.92 & 0.005 & 0.04 & A3IV & ESLib & NA & 1 \\
232.32 & -14.4351 & 0.2 & 1.27605 & 9.1309 & 5.2831 & 0.071499 & 103.16 & 0.005 & 0.14 & A1IV & HD137914 & NA & 0 \\
232.966 & -15.6862 & 0.35 & 0.3583 & 9.6686 & 3.2879 & 0.12395 & 169.29 & 0.015 & 0.05 & F5 & VZLib & NA & 1 \\
234.948 & -12.8062 & 0.2 & 0.797 & 9.726 & 3.6692 & 0.043077 & 184.94 & 0.015 & 0.08 & A1/A2 III/IV & HD139661 & NA & 0 \\
234.306 & -18.335 & 0.32 & 6.8617 & 7.9329 & 5.3866 & 0.94108 & 122.83 & 0.004 & 0.02 & K1III & IVLib & BD -17 4379B & 1 \\
236.792 & -10.6215 & 0.2 & 1.0944 & 9.5769 & 3.4825 & 0.083535 & 241.16 & 0.005 & 0.075 & A2 & HD141043 & NA & 0 \\
241.475 & -16.7817 & 0.15 & 0.68175 & 9.5983 & 4.4546 & 0.032861 & 130.9 & NA & NA & NA & BD -16 4214 & NA & 0 \\
241.605 & -18.6567 & 0.2 & 0.9483 & 8.3735 & 3.7872 & 0.13228 & 170.33 & 0.05 & 0.035 & A2V & HD144412 & NA & 0 \\
242.071 & -17.1348 & 0.05 & 1.75875 & 7.9512 & 3.9072 & 0.0066278 & 135.87 & 0.025 & 0.08 & A0V & V1283Sco & NA & 1 \\
245.118 & -12.911 & 0.06 & 2.7191 & 6.2926 & 6.4767 & 0.25589 & 161.32 & 0.01 & 0.02 & F0 III/IV & V1054Sco & NA & 1 \\
252.365 & -15.668 & 0.7 & 0.6614 & 6.0757 & 4.5305 & 117.58 & 256.26 & 0.005 & 0.03 & A3V & V1010Oph & NA & 1 \\
252.731 & -16.5469 & 0.15 & 2.1077 & 6.7612 & 5.6012 & 0.6219 & 208.46 & 0.00 & 0.025 & B5V & V2355Oph & NA & 1 \\
258.477 & -18.2188 & 0.17 & 1.1951 & 8.0611 & 4.6524 & 0.20507 & 160.37 & 0.005 & 0.035 & A9V & HD155684 & NA & 0 \\
263.582 & -18.5839 & 0.15 & 2.6282 & 8.5106 & 3.57 & 0.14658 & 53.141 & NA & NA & B9III & HD159246 & NA & 0 \\
266.238 & -13.5493 & 0.05 & 2.35315 & 7.9179 & 3.921 & 0.21308 & 109.84 & NA & NA & B9V & HD161225 & NA & 0 \\
266.903 & -14.7258 & 0.025 & 12.451 & 5.8048 & 5.9658 & 1.5153 & 42.992 & 0.0 & 0.012 & B9V & HD161701 & NA & 0 \\
269.851 & -17.3986 & 0.85 & 2.1293 & 8.7559 & 5.646 & 1.1945 & 95.882 & 0.035 & 0.07 & A2 III/IV & WXSgr & NA & 1 \\
273.948 & -18.4224 & 0.2 & 4.6717 & 8.0041 & 5.6725 & 0.07271 & 34.117 & 0.01 & 0.03 & B5III & HD167398 & HD167412 & 0 \\
272.64 & -16.7493 & 0.06 & 7.8075 & 6.8742 & 5.1733 & 0.2302 & 50.363 & 0.393 & 0.02 & B1II & HD166286 & BD -16 4736P & 0 \\
273.655 & -13.4002 & 0.3 & 0.6996 & 9.8555 & 4.0551 & 0.5291 & 132.89 & NA & NA & NA & BD -13 4889 & NA & 0 \\
275.454 & -16.3246 & 0.035 & 1.3827 & 6.4552 & 4.1802 & 0.1946 & 109.19 & 0.005 & 0.08 & B9V & HD168702 & V4390Sgr & 1 \\
274.259 & -15.5087 & 0.3 & 3.4535 & 8.8313 & 4.3047 & 0.19809 & 115.05 & 0.025 & 0.04 & B2/B3 III & VSer & NA & 1 \\
277.105 & -16.7012 & 0.3 & 3.4085 & 7.9229 & 5.3509 & 0.37642 & 107.49 & 0.00 & 0.025 & B1/B2 Ib & V2349Sgr & NA & 1 \\
276.160 & -13.6748 & 0.16 & 10.2706 & 8.0667 & 4.091 & 0.32749 & 124.63 & 0.025 & 0.04 & B3n & WSct & NA & 1 \\
282.204 & -18.6322 & 0.03 & 4.7907 & 5.4828 & 3.9456 & 0.072968 & 44.876 & 0.500 & 0.02 & A2V & HIP 92307 & CCDM J18488 -1838AB & 0 \\
282.714 & -13.9092 & 0.2 & 1.57295 & 8.9156 & 3.3882 & 0.17838 & 93.953 & NA & NA & A9 & HD174397 & NA & 0 \\
283.613 & -12.6098 & 0.3 & 0.95835 & 8.9229 & 4.6523 & 0.279 & 162.2 & 0.005 & 0.08 & A9 & USct & NA & 1 \\
286.491 & -19.4813 & 0.45 & 0.7148 & 7.9042 & 3.3453 & 3.9489 & 215.73 & 0.005 & 0.025 & B2/B3 V & V4197Sgr & NA & 1 \\
288.756 & -15.9766 & 0.08 & 0.875 & 7.7244 & 4.64 & 0.28778 & 132.32 & 0.015 & 0.06 & B9 IV/V & HD179923 & BD -16 5215B & 0 \\
294.679 & -17.0942 & 0.23 & 3.7186 & 8.3576 & 5.9003 & 0.12296 & 38.655 & 0.01 & 0.04 & B8V & HD185229 & NA & 0 \\
295.571 & -14.9705 & 0.125 & 3.7067 & 8.3026 & 5.8637 & 0.043614 & 54.159 & 0.002 & 0.02 & A0V & HD185990 & NA & 1 \\
298.277 & -14.6032 & 0.8 & 1.1829 & 6.4019 & 5.2705 & 94.422 & 155.64 & 0.005 & 0.015 & A1V & V505Sgr & NA & 1 \\
297.611 & -13.9239 & 0.08 & 6.074 & 7.2224 & 6.2693 & 0.066809 & 45.529 & NA & NA & B6III & HD187439 & NA & 0 \\
304.484 & -17.6731 & 0.25 & 3.3921 & 8.8321 & 5.6064 & 0.2062 & 56.448 & NA & NA & A1 III/IV & RWCap & NA & 1 \\
302.504 & -13.0464 & 0.09 & 5.5376 & 7.7802 & 5.7766 & 7.3784 & 53.202 & 0.03 & 0.02 & F5V & HD191365 & HD191430 & 0 \\
303.839 & -12.6246 & 0.2 & 0.6364 & 9.4584 & 4.5911 & 12.766 & 222.62 & 0.025 & 0.08 & A9V & HD192450 & NA & 0 \\
306.124 & -12.9654 & 0.45 & 1.4235 & 9.5064 & 5.8738 & 50.299 & 117.51 & 0.01 & 0.035 & A5III & TYCap & NA & 1 \\
305.214 & -10.5152 & 0.4 & 0.69955 & 9.9763 & 5.1104 & 3.0141 & 125.17 & 0.015 & 0.07 & A5 & UWCap & NA & 1 \\
306.785 & -11.5473 & 0.13 & 1.0798 & 7.6587 & 3.7731 & 0.46354 & 327.51 & 0.01 & 0.035 & F0 & BHCap & NA & 1 \\
308.604 & -10.6828 & 0.18 & 7.346 & 9.2075 & 5.3282 & 0.0081122 & 137.82 & 0.00 & 0.05 & K1III & BD -11 5364 & NA & 0 \\
307.511 & -10.0758 & 0.05 & 1.9980 & 8.3212 & 5.9416 & 2.8639 & 59.2 & NA & NA & F8 & HD195181 & NA & 0 \\
310.453 & -13.9043 & 0.25 & 0.5124 & 8.7696 & 4.7275 & 4.0739 & 157.1 & NA & NA & F3V & HD197090 & NA & 0 \\
313.949 & -17.1142 & 0.08 & 1.0748 & 6.8846 & 3.375 & 0.13401 & 219.73 & 0.025 & 0.06 & F7V & HD199143 & AZCap & 1 \\
312.829 & -13.9244 & 0.07 & 0.807 & 7.3746 & 3.52 & 0.20018 & 196.77 & 0.015 & 0.05 & F0V & NPAqr & NA & 1 \\
312.430 & -13.1265 & 0.2 & 0.5861 & 8.5199 & 3.6104 & 0.45042 & 203.34 & 0.015 & 0.04 & A5 IV/V & SV* BV1717 & OOAqr & 1 \\
311.534 & -11.3438 & 0.4 & 2.52715 & 9.2787 & 7.08 & 0.66414 & 49.435 & NA & NA & G0 & HD197777 & NA & 0 \\
314.707 & -14.5001 & 0.26 & 1.5755 & 5.807 & 4.8596 & 8.4242 & 158.54 & 0.015 & 0.06 & K0 & HD358087 & DVAqr & 1 \\
314.648 & -13.3685 & 0.5 & 5.7826 & 7.0746 & 7.2661 & 1.6567 & 24.215 & 0.0 & 0.014 & G2/G3 V & HD199587 & NA & 1 \\
314.901 & -19.0353 & 0.05 & 2.2413 & 6.2104 & 3.5554 & 0.24023 & 116.97 & 0.065 & 0.14 & Ap & AOCap & NA & 0 \\
318.577 & -13.8896 & 0.1 & 0.4191 & 8.8622 & 3.06 & 0.065375 & 187.91 & 0.025 & 0.08 & F6IV & HD202131 & HD202203 & 0 \\
320.28 & -11.6867 & 0.05 & 0.5627 & 8.2740 & 3.5015 & 0.030002 & 161.54 & 0.025 & 0.06 & F3V & HD203190 & HD203161 & 0 \\
320.067 & -10.8023 & 0.9 & 1.9666 & 8.4764 & 7.4309 & 3.1889 & 57.605 & 0.01 & 0.03 & A3 & RYAqr & NA & 1 \\
320.872 & -11.1275 & 0.33 & 0.6617 & 10.504 & 3.1758 & 0.34573 & 185.49 & 0.005 & 0.20 & F5 & HD358446 & NA & 0 \\
321.636 & -17.8785 & 0.35 & 0.6908 & 8.5995 & 3.7531 & 0.35279 & 205.59 & 0.015 & 0.04 & A9V & CQCap & NA & 1 \\
324.954 & -16.0058 & 0.21 & 2.959 & 8.8822 & 3.9324 & 0.41874 & 188.31 & 0.015 & 0.05 & K0 IV/V & ADCap & NA & 1 \\
323.237 & -17.564 & 0.3 & 6.7156 & 8.3946 & 4.9643 & 0.49337 & 133.77 & 0.005 & 0.03 & K0III & HD205032 & NA & 0 \\
322.049 & -19.1004 & 0.15 & 0.4773 & 9.6387 & 3.4737 & 0.020116 & 151.55 & 0.025 & 0.15 & F8 & HD204300 & BD -19 6102B & 0 \\
326.76 & -16.1266 & 0.15 & 1.0228 & 2.8337 & 5.4122 & 107.4 & 128.8 & 0.005 & 0.04 & A6m & delCap & NA & 1 \\
330.609 & -16.9648 & 0.28 & 0.9450 & 6.0641 & 5.6883 & 5.5757 & 133.9 & 0.005 & 0.045 & NA & CCDM J22024 -1658AB & DXAqr & 1 \\
334.908 & -14.3124 & 0.2 & 0.8846 & 9.503 & 3.8571 & 0.03637 & 172.39 & 0.005 & 0.1 & F0V & HD211750 & NA & 0 \\
337.687 & -13.5436 & 0.42 & 0.4076 & 9.7315 & 4.0316 & 0.17653 & 218.58 & 0.025 & 0.08 & F6/F7V & HD213321 & NA & 0 \\
336.434 & -18.693 & 0.1 & 0.7357 & 9.154 & 4.0186 & 0.038585 & 279.07 & 0.025 & 0.06 & NA & BD -19 6272 & HD212561 & 0 \\
338.118 & -18.9711 & 0.27 & 0.6389 & 10.399 & 4.0963 & 0.032927 & 242.77 & 0.025 & 0.07 & A2 & HD213537 & NA & 0 \\
340.478 & -12.4 & 0.15 & 3.3953 & 9.0714 & 6.6144 & 0.011307 & 47.59 & NA & NA & G0V & HD214934 & NA & 0 \\
340.034 & -15.9588 & 0.055 & 1.88275 & 8.2237 & 6.7596 & 0.0072071 & 49.433 & NA & NA & F3V & CCDM J22401 -1558AB & NA & 0 \\
343.016 & -12.9455 & 0.5 & 1.0447 & 9.4601 & 5.5955 & 0.42465 & 114.3 & 0.025 & 0.12 & A2IV & SUAqr & NA & 1 \\
340.142 & -15.7891 & 0.25 & 1.0273 & 10.569 & 4.3901 & 0.048501 & 124.39 & 0.015 & 0.08 & K2V & 2MASS J22403402 -1547206 & NA & 1 \\
348.863 & -12.3715 & 0.1 & 3.3025 & 8.8777 & 4.1945 & 0.01125 & 112.8 & 0.025 & 0.08 & A3II & HD219387 & NA & 0 \\
351.449 & -11.6099 & 0.15 & 1.5943 & 9.3253 & 4.3647 & 0.062115 & 112.25 & 0.015 & 0.08 & A2III & HD220687 & NA & 1 \\
352.508 & -11.9084 & 0.3 & 0.4455 & 8.8916 & 3.4634 & 2.3354 & 285.5 & 0.015 & 0.06 & F8V & BD-12-6502 & HD221165 & 1 \\
239.419 & -20.9831 & 0.43 & 9.1999 & 5.7287 & 5.5398 & 12.261 & 28.862 & 0.0 & 0.006 & B3V & NSV 7359 & NA & 0 \\
253.886 & -21.5695 & 0.02 & 3.7399 & 6.3529 & 3.7192 & 0.0059572 & 85.297 & NA & NA & B9III & CCDM J16555 -2134AB & NA & 0 \\
254.344 & -25.7996 & 0.2 & 8.793 & 8.2323 & 3.5348 & 0.10286 & 66.112 & 0.0 & 0.02 & A1IV & UUOph & NA & 1 \\
261.295 & -24.6052 & 0.09 & 1.6942 & 7.8548 & 3.8281 & 0.058072 & 57.763 & NA & NA & B9 & CD -24 13328 & NA & 0 \\
264.92 & -28.8534 & 0.33 & 3.12695 & 8.7633 & 3.9533 & 0.25973 & 67.086 & NA & NA & A2 & V846Oph & NA & 1 \\
265.099 & -28.9232 & 0.045 & 0.9215 & 6.7874 & 4.5101 & 0.14926 & 141.9 & 0.035 & 0.10 & B3Vne & HD160319 & NA & 0 \\
266.772 & -28.1499 & 0.5 & 2.5197 & 8.5183 & 4.8781 & 1.7697 & 120.32 & 0.035 & 0.06 & F3V & BNSgr & NA & 1 \\
267.116 & -26.9749 & 0.14 & 2.6187 & 6.034 & 4.6453 & 1.7625 & 190.21 & 0.005 & 0.05 & B4IVe & HIP87163 & NA & 1 \\
270.245 & -23.0322 & 0.3 & 4.6712 & 8.3064 & 3.4461 & 0.42562 & 96.239 & 0.035 & 0.08 & B4III & WYSgr & NA & 1 \\
270.702 & -29.3766 & 0.12 & 1.5194 & 7.2747 & 5.2504 & 0.43214 & 138.64 & NA & NA & B3II & V5562Sgr & NA & 1 \\
270.865 & -22.6462 & 0.35 & 1.3193 & 7.6687 & 5.0436 & 1.1983 & 154.22 & 0.0 & 0.06 & B3nn & V4202Sgr & NA & 1 \\
272.255 & -25.4921 & 0.42 & 2.248 & 6.4448 & 3.7614 & 9.05 & 176.97 & 0.005 & 0.06 & B9IV & CSI -25 -18059 & V3792Sgr & 1 \\
272.363 & -24.0186 & 0.15 & 1.7442 & 6.9113 & 4.3734 & 0.71551 & 157.86 & 0.005 & 0.07 & B6III & HIP88943 & NA & 0 \\
275.577 & -25.2639 & 1.5 & 3.2755 & 8.537 & 6.8388 & 1.6201 & 38.873 & NA & NA & A2/A3 IV & XZSgr & NA & 1 \\
281.968 & -20.2745 & 0.5 & 4.44825 & 6.7588 & 7.8399 & 34.33 & 97.941 & NA & NA & B9III & V356Sgr & NA & 1 \\
285.728 & -29.1427 & 0.25 & 1.1619 & 9.0931 & 4.7485 & 0.37938 & 115.62 & NA & NA & A4 II/III & V523Sgr & NA & 1 \\
287.737 & -25.911 & 0.28 & 11.7899 & 8.2783 & 3.5682 & 0.26594 & 28.664 & 0.134 & 0.007 & B9V & V5570Sgr & NA & 1 \\
289.464 & -24.7061 & 0.1 & 4.8142 & 8.4851 & 3.7519 & 0.30322 & 75.183 & 0.002 & 0.02 & A6IV & HD180520 & NA & 0 \\
290.228 & -23.8356 & 0.075 & 1.6669 & 8.1335 & 3.97 & 6.1353 & 84.66 & 0.025 & 0.1 & F0V & HD181292 & NA & 0 \\
291.126 & -27.8659 & 0.035 & 0.5214 & 5.8645 & 4.098 & 0.24242 & 275.79 & 0.065 & 0.08 & B2Vn & HR7355 & HD182120 & 0 \\
293.013 & -28.2124 & 0.28 & 1.4398 & 7.4899 & 4.7352 & 2.4628 & 194.82 & 0.005 & 0.03 & B8V & V5572Sgr & NA & 1 \\
294.734 & -28.6072 & 0.035 & 3.4753 & 6.6342 & 3.7624 & 0.15006 & 291.57 & 0.025 & 0.06 & B9 & V4062Sgr & NA & 0 \\
295.664 & -24.4329 & 0.125 & 0.7224 & 7.465 & 3.9348 & 2.616 & 198.95 & 0.005 & 0.1 & NA & TYC 6894 48 1 & V4063Sgr & 0 \\
297.735 & -26.6387 & 0.13 & 1.1715 & 8.4906 & 4.7148 & 0.211 & 163.02 & 0.005 & 0.05 & A5III & CCDM J19509 -2638AB & NA & 0 \\
297.803 & -20.5028 & 0.18 & 3.5036 & 9.4256 & 3.3937 & 1.011 & 34.81 & NA & NA & F5V & HD187533 & NA & 1 \\
300.15 & -28.5864 & 0.015 & 1.6652 & 6.3824 & 4.8303 & 0.063667 & 69.175 & NA & NA & K0III & HD189365 & V1173Sgr & 1 \\
304.496 & -28.1331 & 0.13 & 1.177 & 8.001 & 4.1505 & 0.45996 & 239.69 & 0.015 & 0.035 & G2III & V4374Sgr & NA & 1 \\
312.091 & -22.7407 & 0.055 & 5.15195 & 6.9182 & 5.5004 & 0.017026 & 24.025 & NA & NA & F7V & NLTT 49905 & NA & 0 \\
310.147 & -27.5866 & 0.013 & 1.3399 & 7.4484 & 3.9504 & 0.077899 & 75.090 & 0.025 & 0.14 & A4III & HD196816 & NA & 0 \\
315.743 & -22.9336 & 0.15 & 0.7226 & 9.5378 & 4.6812 & 0.074199 & 188.17 & 0.035 & 0.15 & G8V & HD200244 & NA & 0 \\
321.889 & -20.9133 & 0.15 & 0.48225 & 9.9485 & 4.3722 & 0.041294 & 163.15 & 0.025 & 0.12 & G8V & TYC 6372 1266 1 & NA & 1 \\
321.236 & -21.6525 & 0.18 & 0.7156 & 9.5101 & 3.7546 & 0.11808 & 219.69 & 0.025 & 0.08 & F4V & HD203798 & NA & 0 \\
324.401 & -20.4319 & 0.25 & 1.4741 & 7.7244 & 6.7401 & 1.0809 & 155.22 & 0.005 & 0.03 & F3V & BQCap & NA & 1 \\
324.153 & -22.2997 & 0.27 & 0.5440 & 9.2615 & 4.1631 & 0.20773 & 287.69 & 0.02 & 0.055 & F5V & HD205563 & NA & 0 \\
328.575 & -22.9645 & 0.13 & 2.4465 & 9.3474 & 7.3198 & 0.053594 & 74.716 & 0.004 & 0.03 & Fm & HD208090 & NA & 0 \\
258.437 & -32.8526 & 0.4 & 5.7282 & 7.5473 & 3.9281 & 2.9865 & 120.46 & 0.005 & 0.04 & B6V & FVSco & NA & 1 \\
259.714 & -31.583 & 0.19 & 1.6201 & 8.6409 & 4.8226 & 0.4785 & 152.00 & 0.005 & 0.05 & B3 & V474Sco & NA & 1 \\
262.304 & -31.5343 & 0.075 & 2.5136 & 6.6946 & 5.2055 & 0.39186 & 149.19 & 0.065 & 0.04 & O9.5V & V1081Sco & NA & 1 \\
262.805 & -31.3785 & 0.3 & 1.17345 & 8.8562 & 5.3651 & 0.66093 & 144.63 & 0.005 & 0.06 & B5 & V700Sco & NA & 1 \\
262.26 & -33.0045 & 0.38 & 1.1667 & 7.6921 & 4.541 & 3.2044 & 164.56 & 0.005 & 0.03 & B1III & V499Sco & NA & 1 \\
281.522 & -30.489 & 0.6 & 4.15415 & 8.7902 & 4.9901 & 1.4934 & 98.102 & NA & NA & A9V & SXSgr & NA & 1 \\
283.174 & -30.7341 & 0.16 & 1.4517 & 6.4872 & 4.4953 & 2.1661 & 186.67 & 0.01 & 0.03 & B8V & V4407Sgr & NA & 1 \\
283.886 & -32.0235 & 0.11 & 2.8124 & 8.4159 & 6.3134 & 0.14475 & 76.435 & NA & NA & B9V & HD175162 & NA & 0 \\
286.806 & -30.1605 & 0.3 & 0.7051 & 7.5461 & 4.3497 & 3.0389 & 234.00 & 0.005 & 0.065 & A6V & V525Sgr & NA & 1 \\
287.063 & -31.3486 & 0.3 & 1.9193 & 9.2647 & 4.9514 & 0.30954 & 95.852 & 0.13 & 0.04 & A0V & V526Sgr & NA & 1 \\
\hline
\end{longtable}
\end{landscape}

\twocolumn